\documentclass[aps,prd,tightenlines,nofootinbib,reprint,floatfix]{revtex4-1}

\usepackage{graphicx}
\usepackage{dcolumn}
\usepackage{bm}
\usepackage{amsmath}
\usepackage{multirow}
\usepackage{longtable}
\usepackage{dcolumn}

\begin{document}

\newcommand{\mevcc}{\!\mathrm{MeV}\!/c^2}
\newcommand{\mevc}{\!\mathrm{MeV}/\!c}
\newcommand{\mev}{\!\mathrm{MeV}}
\newcommand{\gevcc}{\!\mathrm{GeV}/\!c^2}
\newcommand{\gevc}{\!\mathrm{GeV}/\!c}
\newcommand{\gev}{\!\mathrm{GeV}}

\title{A Comprehensive Study of the Radiative Decays of $J/\psi$ and $\psi(2S)$ to Pseudoscalar Meson Pairs, and Search for Glueballs}

\author{S.~Dobbs}
\author{A.~Tomaradze}
\author{T.~Xiao}
\author{Kamal~K.~Seth}
\affiliation{Northwestern University, Evanston, Illinois 60208, USA}

\date{\today}

\begin{abstract}
Using 53~pb$^{-1}$ of $e^+e^-$ annihilation data taken at $\sqrt{s}=3.686$~GeV, a comprehensive study has been made of the radiative decays of samples of 5.1~million~$J/\psi$ and 24.5~million~$\psi(2S)$ into pairs of pseudoscalar mesons, $\pi^+\pi^-$, $\pi^0\pi^0$, $K^+K^-$, $K_S^0K_S^0$, and $\eta\eta$.  Product branching fractions for the radiative decays of $J/\psi$ and $\psi(2S)$ to scalar resonances $f_0(1370,1500,1710,2100,~\text{and}~2200)$, and tensor resonances $f_2(1270,1525,~\text{and}~2230)$ have been determined, and are discussed in relation to predicted glueballs.  For $\psi(2S)$ radiative decays the search for glueballs has been extended to masses between 2.5~GeV and 3.3~GeV.
\end{abstract}

\maketitle

\section{Introduction}

Quantum Chromodynamics (QCD) is presently the accepted theory of strong interactions.  Its predictions for hadronic phenomena at high energies have received remarkable experimental confirmation, and it is an integral part of the Standard Model of particle physics.  In contrast, at lower energies the success of QCD has been comparatively limited.  This is mainly due to the fact that while at high energies reliable predictions can be made in perturbative QCD (pQCD) models, at low energies pQCD predictions are unreliable, and lattice calculations with the required precision are in many cases not yet available.

In 1972, Fritzsch and Gell-Mann~\cite{fgm} pointed out that ``if the quark-gluon field theory indeed yields a correct description of strong interactions, there must exist glue states in the hadron spectrum'', and Fritzsch and Minkowski~\cite{fm} presented a detailed discussion of the phenomenology of the spectrum of what they called ``glue-states''.  The ``glue-states'', which we here call ``glueballs'', are bosons with baryon number and isospin equal to zero, and they are $SU(3)$ flavor singlets ``to the extent to which $SU(3)$ breaking effect can be neglected''~\cite{fm}.

Ever since 1985, when $\eta(1440)$ was claimed as the first glueball discovered~\cite{eta1440}, numerous experimental searches for glueballs have been reported, and numerous claims and counterclaims abound.  Many reviews of QCD exotics, including glueballs, have been published.   Two of the latest comprehensive reviews are by Klempt and Zaitsev~\cite{review} and Ochs~\cite{review2}.

Fritzsch and Minkowski~\cite{fm} have detailed color-singlet states of two and three gluons allowed by conservation rules, which are identical to those for states of two and three photons.  The two-gluon states have charge conjugation, $C=+1$, while the three gluon states can have $C=+1$ or $-1$.  The lowest mass states have S--waves between the gluons, and are two-gluon states with $J^{PC} = 0^{++}$, $2^{++}$, and $0^{-+}$, and three-gluon states with $J^{PC} = 1^{++}$, $1^{+-}$, $1^{--}$, and $3^{--}$.

\begin{figure}
\begin{center}
\includegraphics[width=3.5in]{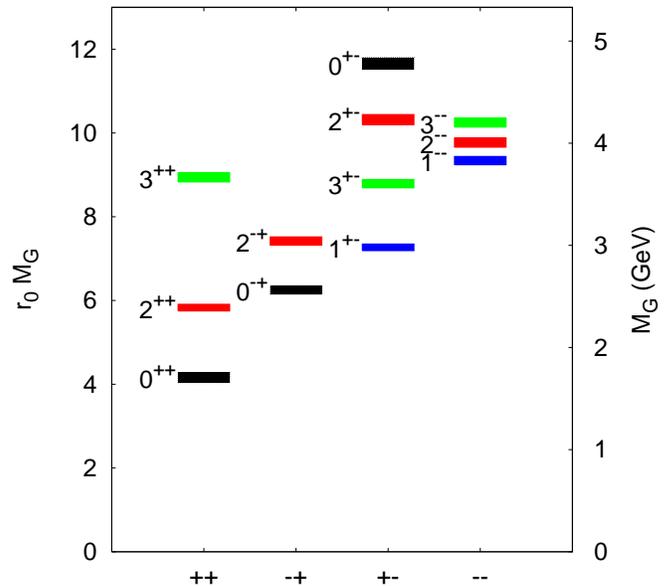}
\end{center}

\caption{Quenched lattice predictions of glueball spectrum by Chen et al.~\cite{lattice4}.  The states are grouped by their parity and charge conjugation indicated in the abscissa.}
\label{lqcd-prediction}
\end{figure}

Predictions for glueball spectra have been made in bag models~\cite{bagmodel}, QCD-based potential models~\cite{potential}, QCD sum rules~\cite{sumrule}, and lattice gauge calculations~\cite{lattice1,lattice2,lattice3,lattice4,lattice5}.  Lattice gauge calculations have supplanted most other predictions.  The lattice calculations of Refs.~10,11, and 12 were done in the quenched approximation with no quarks, and their predictions are in general agreement with each other.  The first unquenched lattice calculations with $2+1$ flavor sea-quarks have been reported by the UKQCD Collaboration~\cite{lattice5}.  They obtain results for the masses of the $0^{++}$, $0^{-+}$, and $2^{++}$ glueballs which are in good agreement with the lattice predictions of Refs.~10 and 12 in the quenched approximation,  and conclude that they ``find no evidence of strong unquenching effects''.  Thus, the results of Ref.~12, reproduced in Fig.~\ref{lqcd-prediction}, can be considered as representative of lattice predictions so far.  The order of the lowest lying glueballs is as anticipated by Fritzsch and Minkowski~\cite{fm}, with the prediction for the first scalar ($0^{++}$) and tensor ($2^{++}$) glueball masses are $G(0_1^{++})=1710(50)(80)$~MeV, $G(2_1^{++})=2390(30)(120)$~MeV.  These are followed by five heavier glueballs with masses less than $M(\psi(2S))$.  These are $0^{-+}(2560(35)(120)~\text{MeV})$, $1^{+-}(2980(30)(140)~\text{MeV})$, $2^{-+}(3040(40)(150)~\text{MeV})$, $3^{+-}(3600(40)(170)~\text{MeV})$, and $3^{++}(3670(50)(150)~\text{MeV})$.  In the present paper, our glueball search is limited to masses less than 3.3~GeV, below the masses of the $\chi_{cJ}$ resonances of charmonium.

Early searches for glueballs were misled by the wishful expectation of ``pure'' glueballs.  With the realization that the predicted glueballs have masses close to those of normal $q\bar{q}$ mesons with the same quantum numbers with which they must mix, the more realistic searches aspire to identify states which contain substantial glueball content in their wave functions.  Searches for such states have been based mainly on three expectations.  These are: (a) for a given $J^{PC}$, the glueball state is ``supernumary'', i.e. a state in excess of the expected number of $q\bar{q}$ states, e.g., the nonet of $0^{++}$ $SU(3)$ states, (b) a pure glueball state does not decay preferentially to states containing up, down, or strange quarks, i.e., ``pure'' glueball decays to pseudoscalar pairs should have equal branching fractions for each available charged state.  Thus decays to $\pi\pi$, $KK$, $\eta\eta$, $\eta'\eta'$ and $\eta\eta'$, should have branching fractions proportional to $3:4:1:1:0$, and (c) certain decays are particularly favorable for populating glueballs, e.g., $p\bar{p}$ annihilation, and radiative decays of vector states of quarkonia, $J/\psi,\psi',\Upsilon$, to $\gamma gg \ldots$, where $g$ stands for gluons. Numerous experimental attempts have been made to exploit these expectations, but they have been generally limited in their scope.  

An extensive review of searches for glueballs in $p\bar{p}$ annihilations, central production, and hadronic and radiative decays of vector states of charmonium and bottomonium has been made by Klempt and Zaitsev~\cite{review}.  Radiative decays of $J/\psi$ have been studied by Crystal~Ball~\cite{cb}, Mark~III~\cite{markiii,markiii2}, DM2~\cite{dm2}, and BES~\cite{besii,besii2,besii3}. Among the latest of these are BES measurements of $J/\psi\to \gamma(\pi^+\pi^-,\pi^0\pi^0)$~\cite{besii} and $J/\psi\to\gamma(K^+K^-,K_S^0K_S^0)$~\cite{besii3} with 58~million directly produced $J/\psi$ in $e^+e^-\to J/\psi$.  The invariant mass spectra in the decays of these directly produced $J/\psi$ contain large backgrounds due to ISR production of $\pi^+\pi^-\pi^0$ and $K\overline{K}\pi^0$ with one missing photon which simulate $\gamma\pi^+\pi^-$ and $\gamma K^+K^-$ decays.  More recently, BES~III has reported a study of $J/\psi\to\gamma\eta\eta$ with $2.28\times10^8$ directly produced $J/\psi$~\cite{besiii}.  No studies of radiative decays of $\psi(2S)$ have been published, and only an unpublished report with 14~million~$\psi(2S)$ by BES exists~\cite{besii4}.

In the present paper we present results of a study of the radiative decays of the charmonium resonances, 5.1~million $J/\psi$, and 24.5~million $\psi(2S)$ resonances populated in $e^+e^-$ annihilation, and decaying into pairs of pseudoscalar, $\pi\pi$, $KK$, and $\eta\eta$.  We determine the masses, widths, and product branching fractions of the resonances observed, and discuss their possible relations to glueballs.
Since the $J/\psi$ used in the present investigation are produced by the exclusive decay $\psi(2S)\to\pi^+\pi^-J/\psi$, the invariant mass spectra we observe from the $J/\psi$ radiative decays are free from the ISR-produced backgrounds.  The invariant mass spectra for $\psi(2S)$ decays contain, of course, background contribution from ISR.

The paper is organized as follows.  In Sec.~II we describe the event selection criteria.  In Sec.~III we describe Monte Carlo event simulations and use them to determine instrumental resolutions and efficiencies. In Sec.~IV we present the Dalitz plots for $J/\psi$ and $\psi(2S)$ radiative decays to pseudoscalar pairs.  In Sec.~V.A we present the $\pi\pi$ and $K\overline{K}$ invariant mass distributions for the radiative decays of $J/\psi$, and discuss the resonances observed in them.  In Sec.~V.B we present the similar mass distributions and discussion for the radiative decays of $\psi(2S)$.  In Sec.~V.C we present the results of our search for the putative tensor glueball $\xi(2230)$.  In Sec.~VI we summarize and compare our results for $J/\psi$ and $\psi(2S)$ radiative decays to pseudoscalar pairs for the observed resonances with masses below 2.5~GeV.  
In Sec.~VII.A we present the invariant mass spectra for $\psi(2S)$ radiative decays to $\pi\pi$, $K\overline{K}$, and $\eta\eta$ decays in search for potential glueball candidates in the mass range $2.5-3.3$~GeV, and establish upper limits for any resonance enhancements. 
We observe strong decays of $\chi_{c0}(0^{++})$ and $\chi_{c2}(2^{++})$ resonances of charmonium in our data for $\psi(2S)$ radiative decays.  Because the decays of these resonances are necessarily mediated by two gluons, just as the decays of two-gluon glueballs, it is instructive to examine and compare their characteristics.  We do so in Sec.~VII.B in terms of the branching fractions measured using the same data as we use in the present paper, and published as Ref.~\cite{cleoprev}.
In Sec.~VIII we discuss systematic uncertainties in our results.  In Sec.~IX we summarize the results of the present investigation for the decays of known and proposed resonances into pseudoscalar pairs, and their implications for scalar and tensor glueball candidates.

\begin{table}[!tb]
\caption{Summary of the requirements for the number of reconstructed particles for the reactions investigated in this paper.}
\begin{ruledtabular}
\begin{tabular}{rlccl}
\multicolumn{2}{l}{Decay mode} & \# tracks & \# showers & Other \\
\hline
$\psi(2S)\to$ & $\gamma\pi^+\pi^-$ & 2 & $\ge1$ & \\
 & $\gamma K^+K^-$ & 2 & $\ge1$ & \\
 & $\gamma\pi^0\pi^0$ & 0 & $\ge5$ & 2  $\pi^0$ \\
 & $\gamma\eta\eta$ & 0 & $\ge5$ & 2  $\eta$ \\
 & $\gamma K_S^0K_S^0$ & $\ge4$ & $\ge1$ & 2  $K_S^0$ \\
\hline
$J/\psi\to$ & $\gamma\pi^+\pi^-$ & 4 & $\ge1$ & \\
 & $\gamma K^+K^-$ & 4 & $\ge1$ & \\
 & $\gamma\pi^0\pi^0$ & 2 & $\ge5$ & 2  $\pi^0$ \\
 & $\gamma\eta\eta$ & 2 & $\ge5$ & 2  $\eta$ \\
 & $\gamma K_S^0K_S^0$ & $\ge6$ & $\ge1$ & 2  $K_S^0$ \\
\end{tabular}
\end{ruledtabular}
\label{tbl:multiplicites}
\end{table}

\begin{figure}[!tb]
\begin{center}
\includegraphics[width=3.in]{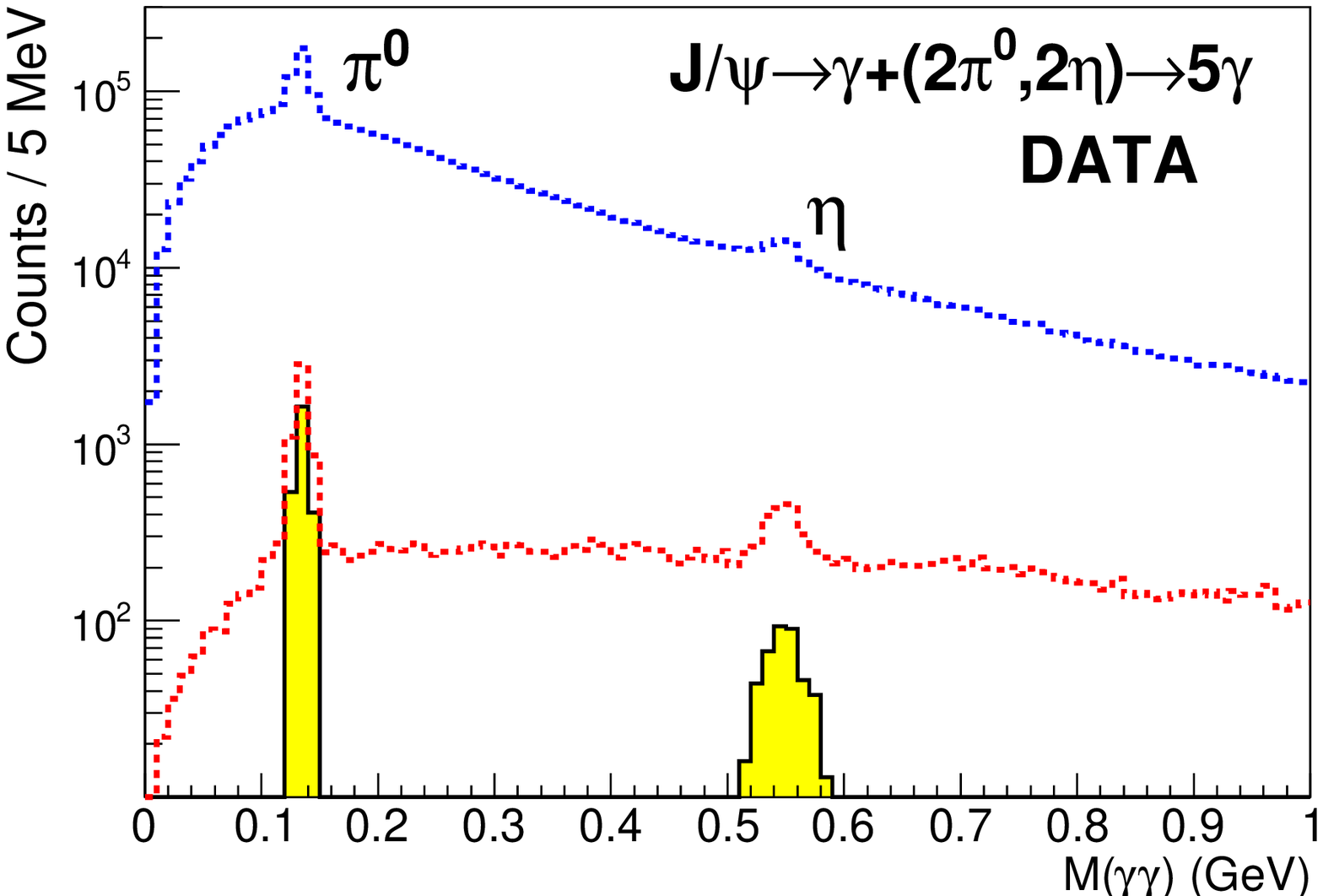}
\includegraphics[width=3.in]{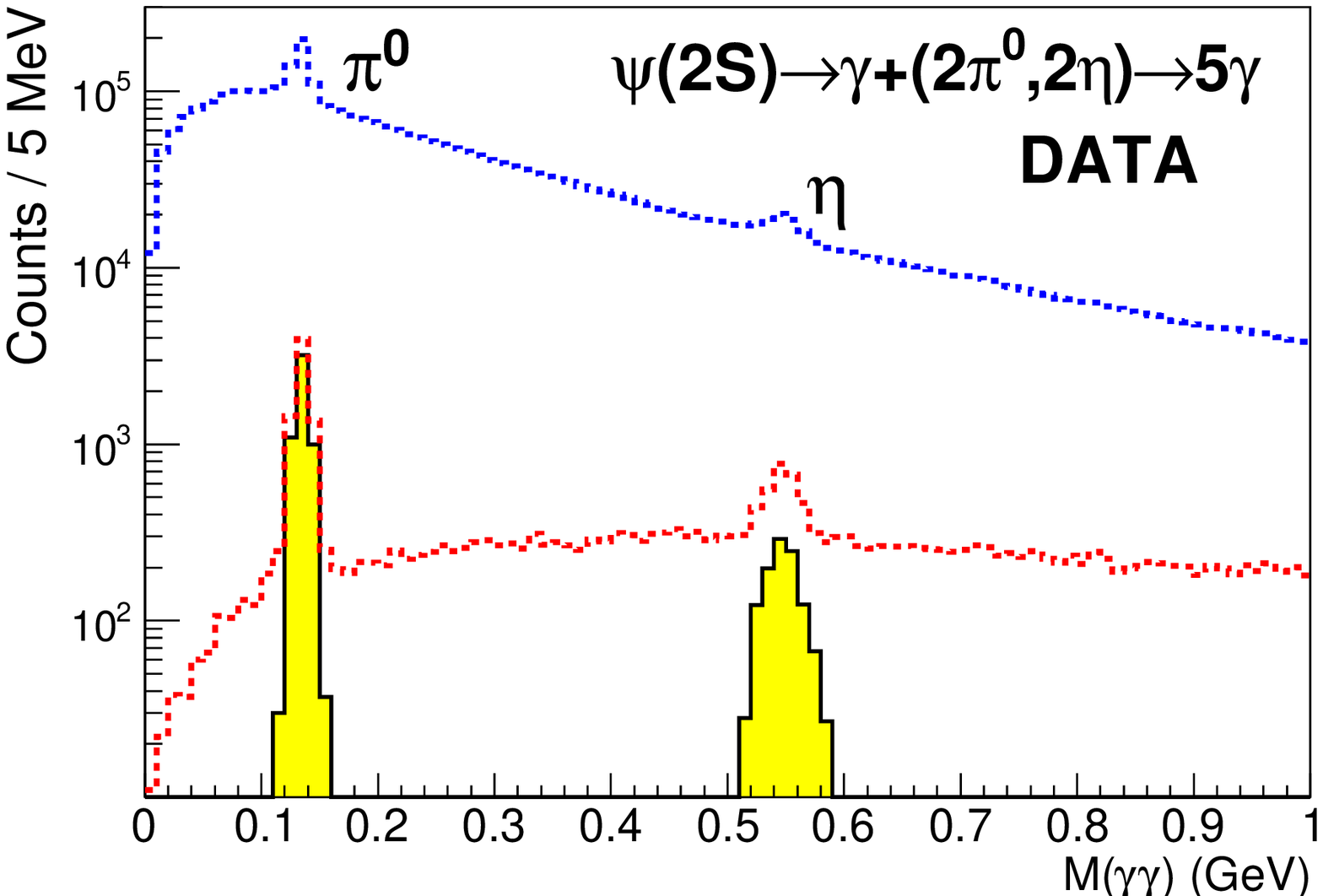}
\end{center}

\caption{Two--photon invariant mass distributions for radiative decays of (top panel) $J/\psi$ and (bottom panel) $\psi(2S)$, with at least 5 photons in the final state.  In both panels, the top histograms show invariant mass distribution for all two--photon combinations in the selected events.  The middle histograms show distributions for all two--photon combinations in the events that pass the 4C energy-momentum kinematic fit with $\chi^{2}_{4C}<25$.  The shaded solid histograms show the distributions of events which contain two $\pi^0$ or two $\eta$ candidates which best fit $\pi^0$ and $\eta$ masses.  Signals from $\pi^0\to\gamma\gamma$ and $\eta\to\gamma\gamma$ decays are clearly visible in all three histograms.} 
\label{fig:ggmass}
\end{figure}

\begin{figure}[!tb]
\begin{center}
\includegraphics[width=3.in]{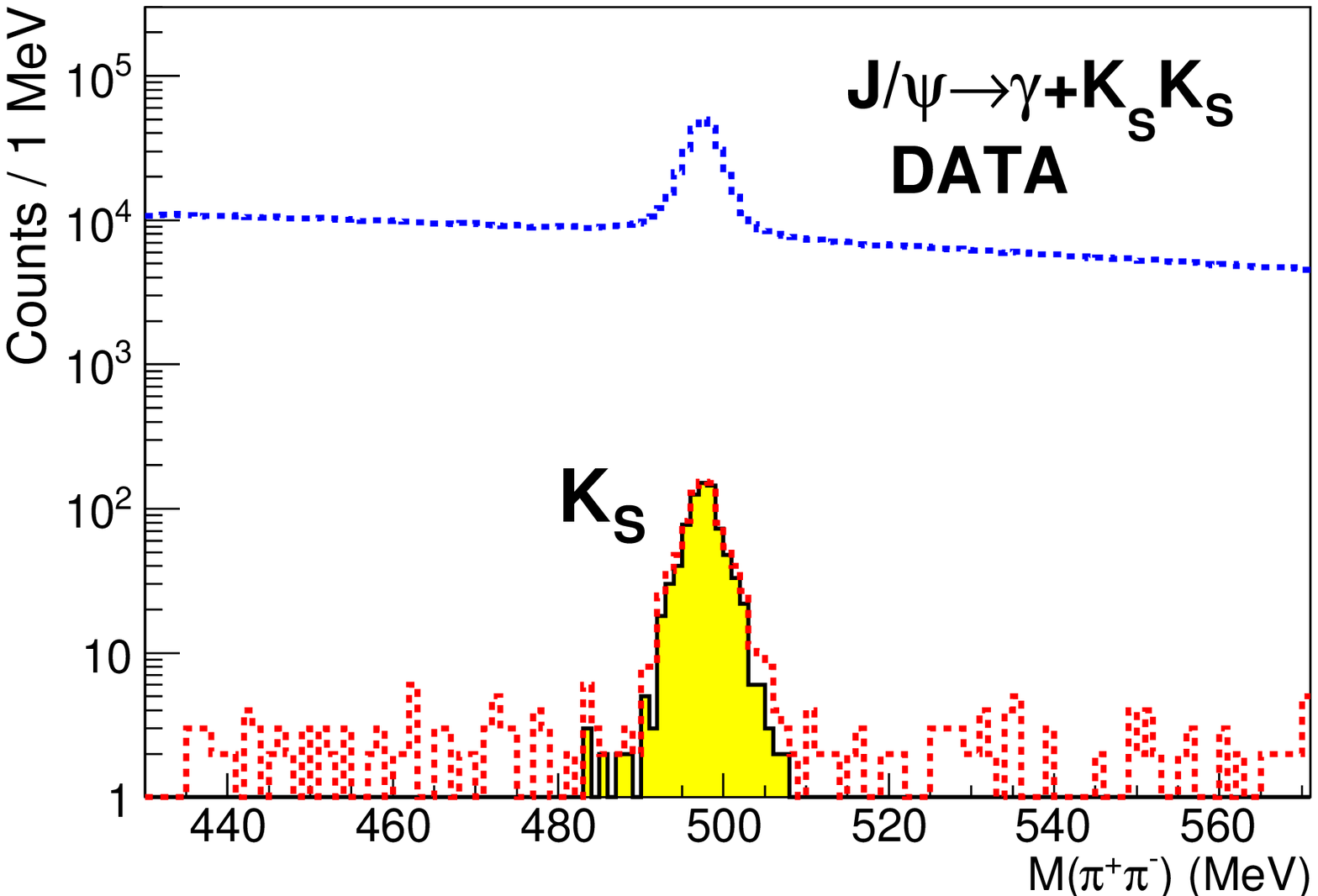}
\includegraphics[width=3.in]{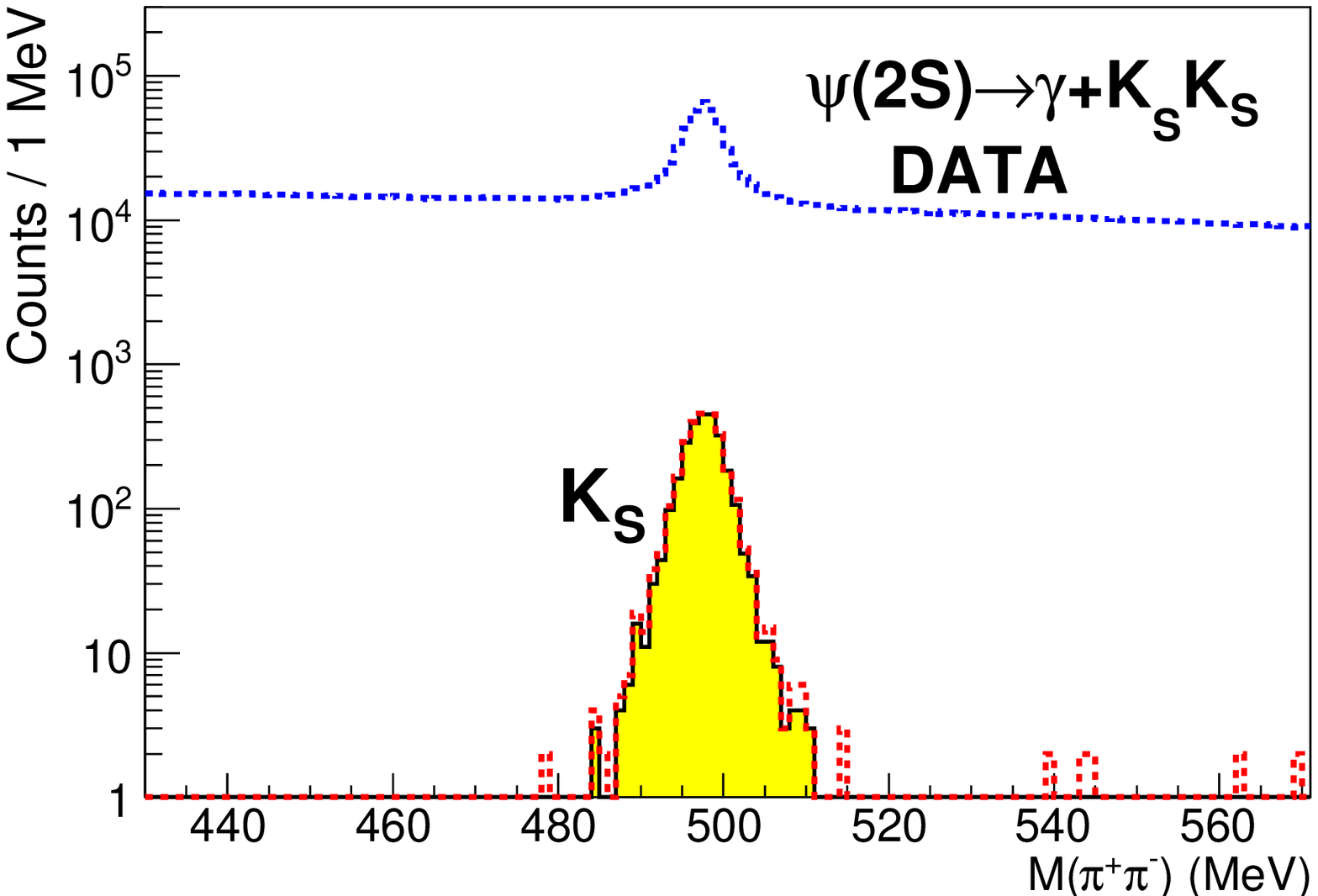}
\end{center}

\caption{Two--pion invariant mass distributions for radiative decays of (top panel) $J/\psi$ and (bottom panel) $\psi(2S)$ with at least two $K_S^0\to\pi^+\pi^-$ candidates reconstructed.  In both panels the top histograms show all $K_S^0\to\pi^+\pi^-$ candidates in the selected events.  The dashed histograms show distributions for all $K_S^0\to\pi^+\pi^-$ candidates in the events that pass the 4C energy-momentum kinematic fit with $\chi^{2}_{4C}<25$.  The shaded histograms show the distribution for the best two $K_S^0$ candidates in the event, as determined by the kinematic fit. The mass peaks corresponding to the selected $K_S^0$ candidates are clearly visible in all three histograms.}
\label{fig:pipimass}
\end{figure}

\section{Event Selections}

We use 53~pb$^{-1}$ of $e^+e^-$ annihilation data taken at $\sqrt{s}=3.686$~GeV at the CESR $e^+e^-$ collider at Cornell University.  It corresponds to 24.5~million $\psi(2S)$ produced, and 5.1~million $J/\psi$ tagged by the decay $\psi(2S)\to\pi^+\pi^-J/\psi$.  We also use 21~pb$^{-1}$ of $e^+e^-$ annihilation data taken at $\sqrt{s}=3.672$~GeV to study backgrounds from off-resonance decays.  The decay products were analyzed by the CLEO-c detector, which has been described elsewhere in detail~\cite{cleocdetector}.  Briefly, it consists of a CsI(Tl) electromagnetic calorimeter, an inner vertex drift chamber, a central drift chamber, and a ring imaging Cherenkov (RICH) detector, all inside a superconducting solenoid magnet providing a nominal 1.0 Tesla magnetic field.  The acceptance for charged and neutral particles is $|\cos\theta|<0.93$. Charged particle resolution is $\sigma_p/p = 0.6\%$~@~1~GeV/$c$.  Photon resolution is $\sigma_E/E=2.2\%$~@~1~GeV, and $5\%$~@~100~MeV.  The detector response was studied using a GEANT-based Monte Carlo (MC) simulation including radiation corrections~\cite{GEANTMC}. 

The particle multiplicity requirements for the various decay modes are given in Table~\ref{tbl:multiplicites}.  
 Charged particle tracks were accepted in the full range of the drift chamber's acceptance, $|\cos\theta|<0.93$ ($\theta$ is the polar angle with respect to the positron beam), and were selected by standard CLEO quality requirements. The total charge in the event was required to be zero. 
 Photon candidates were accepted in the ``Good Barrel'' and ``Good Endcap regions'' of the calorimeter  ($|\cos\theta|<0.81$ and $|\cos\theta|=0.85-0.93$, respectively), and were required to contain none of the few noisy calorimeter cells, and to have a transverse energy deposition that was consistent with that of an electromagnetic shower.

The $\pi^0$ and $\eta$ were reconstructed in their decay to two photons.  Pairs of photon candidates were kinematically fitted to the known $\pi^0$ and $\eta$ masses, and the pairs were accepted if they had a mass $\le3\sigma$ of the particle mass.  Invariant mass distributions at different stages of these selections are shown in Fig.~\ref{fig:ggmass}.

The $K_S^0$ were reconstructed in their decay to two charged pions.  Pairs of oppositely charged pions were kinematically fitted to originate from a common vertex, and this common vertex was required to be displaced by $\ge3\sigma$ from the $e^+e^-$ interaction point. Invariant mass distributions at different stages of these selections are shown in Fig.~\ref{fig:pipimass}.

\begin{figure}[!tb]
\begin{center}
\includegraphics[width=2.8in]{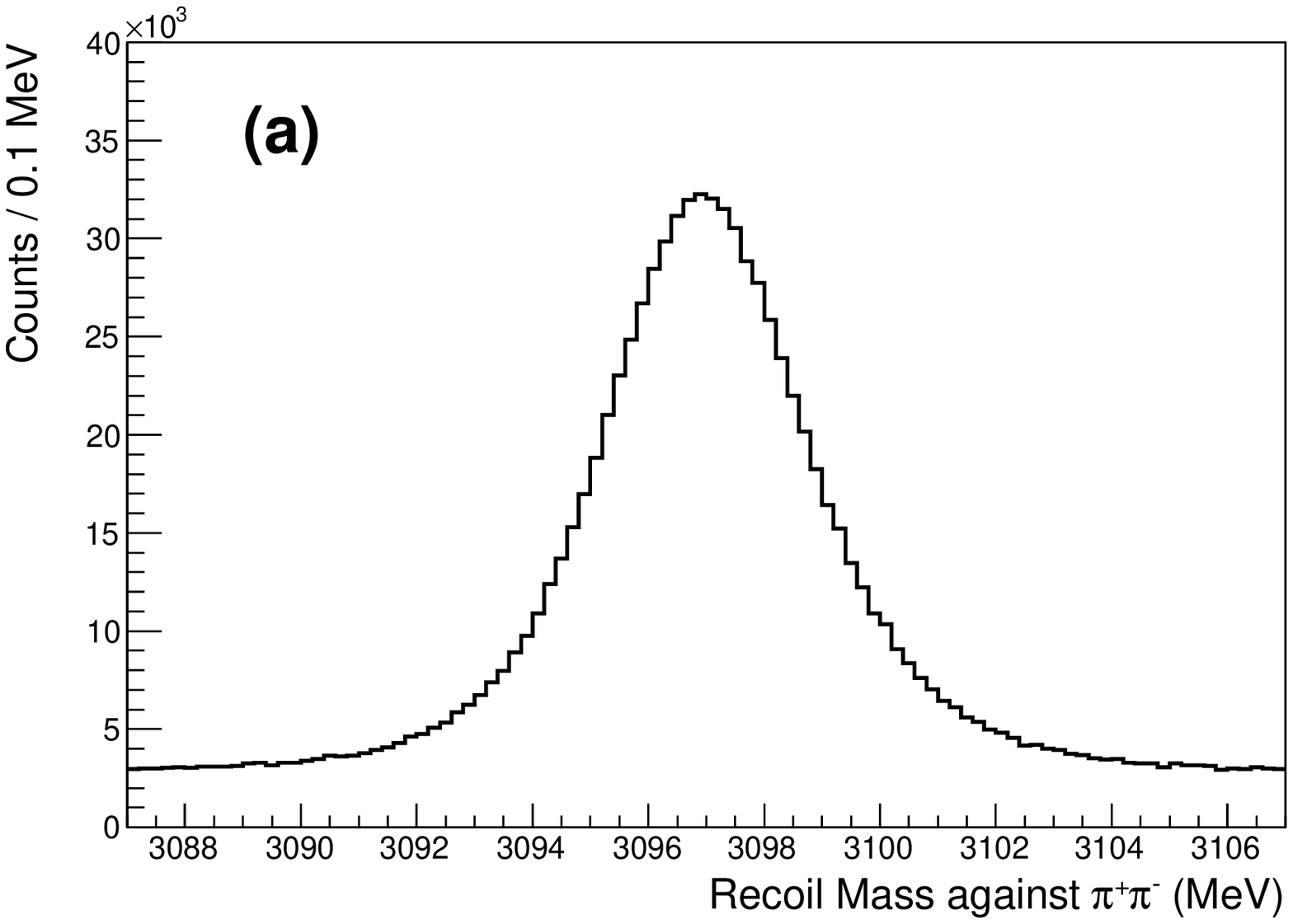}
\includegraphics[width=2.8in]{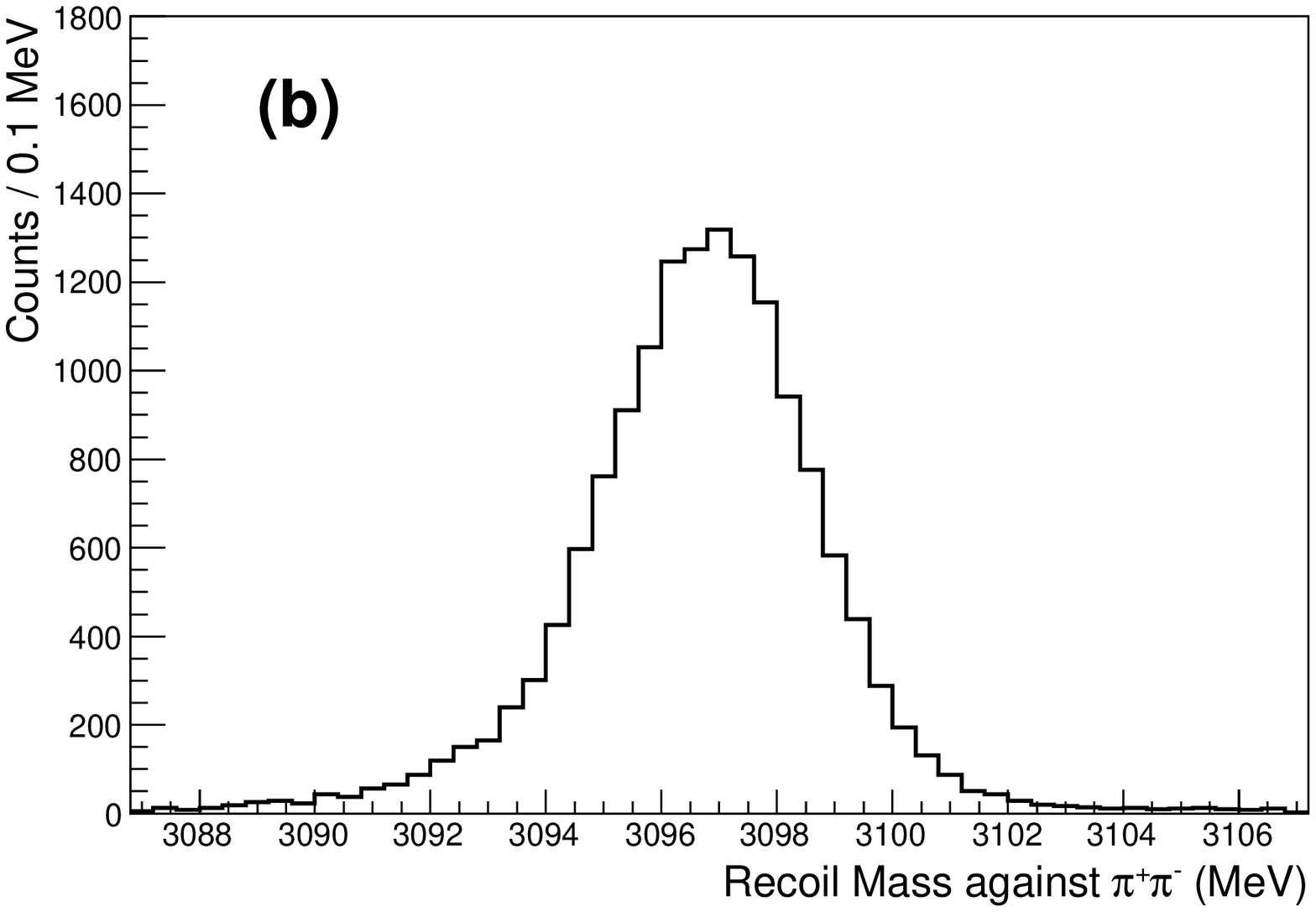}
\end{center}
\caption{Mass distribution of events recoiling against $\pi^+\pi^-$ in $\psi(2S)\to\pi^+\pi^-J/\psi$ decay (a) before and (b) after 4C kinematic fit. The fitted mass of the $J/\psi$ peak is found to be $M(J/\psi)=3097.13\pm0.01(\text{stat})$~MeV.}
\label{fig:jpsirecmass}
\end{figure}

\begin{figure}
\begin{center}
\includegraphics[width=2.8in]{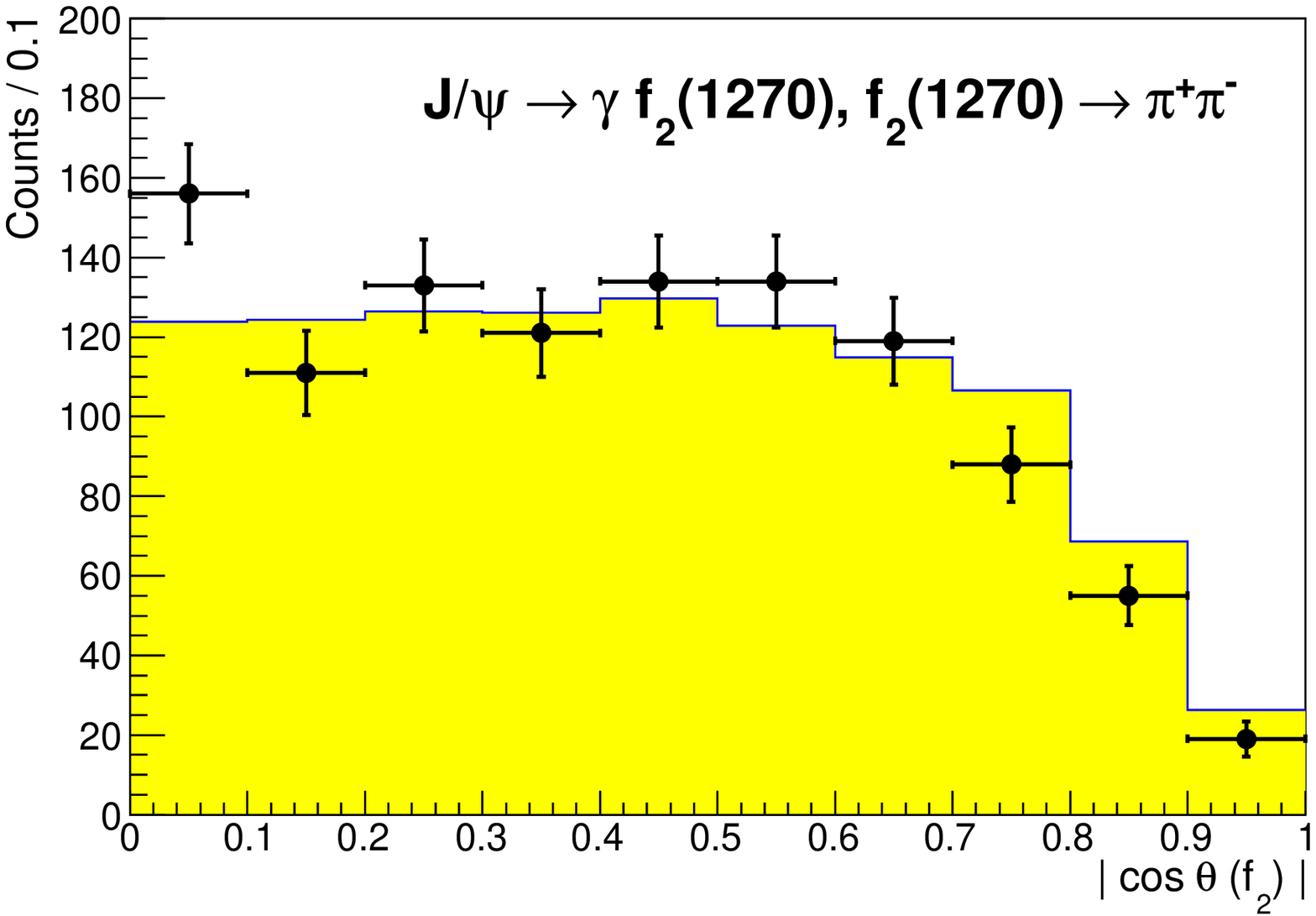}
\includegraphics[width=2.8in]{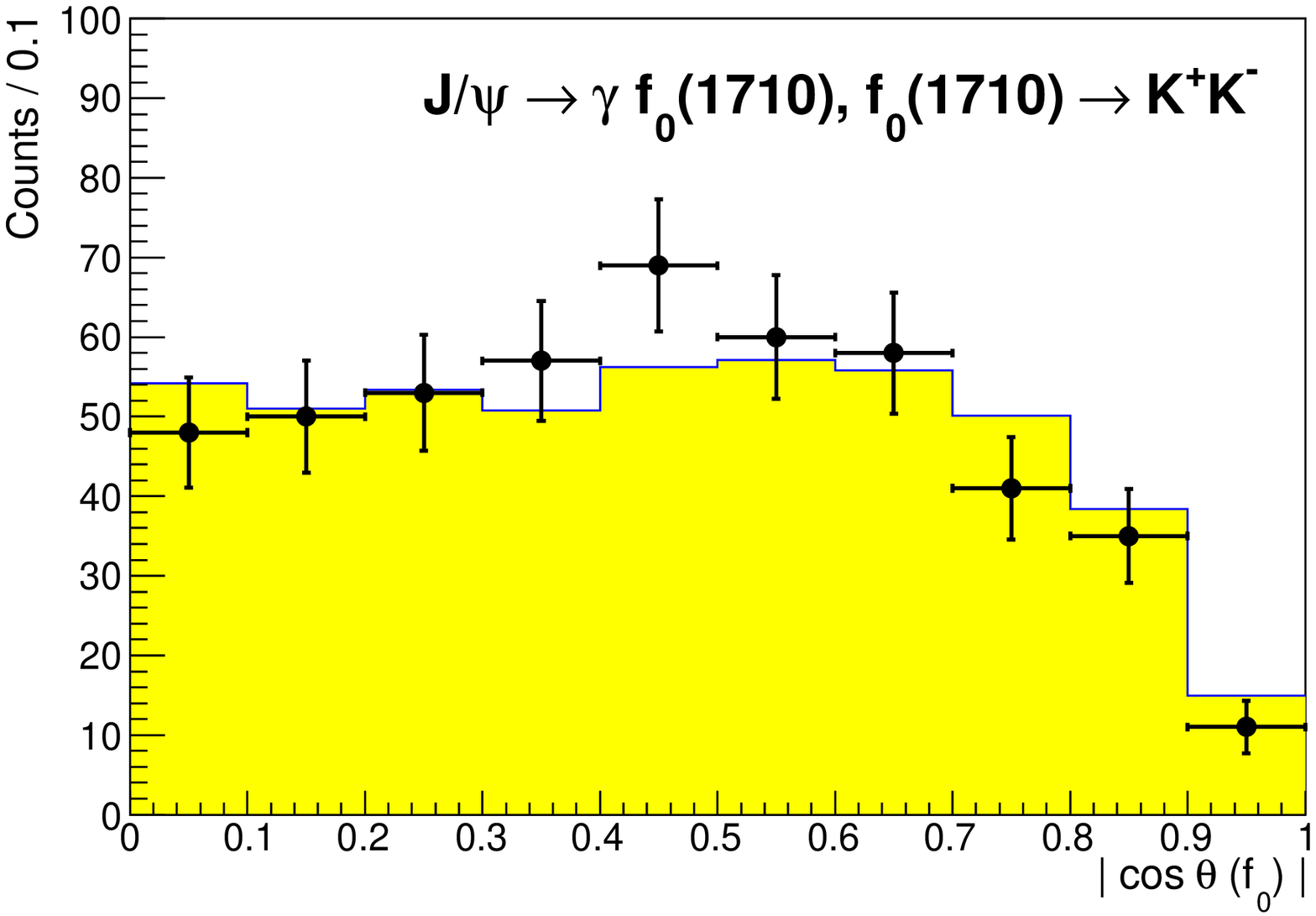}
\end{center}
\caption{Distributions of $|\cos\theta(f_{0,2})|$ for $f_2(1270)\to\pi^+\pi^-$ and $f_0(1710)\to K^+K^-$ decays.  The point show the distributions from data, and the shaded histograms show the results of MC simulations.}
\label{fig:costh}
\end{figure}

\begin{table}[!tb]

\caption{MC--determined efficiencies for the reactions $\psi(2S)\to \gamma R,~R\to \mathrm{PS,PS}$, and $\psi(2S)\to\pi^+\pi^-J/\psi,~J/\psi\to \gamma R,~R\to \mathrm{PS,PS}$, for selected intermediate resonances $R$.}

\begin{ruledtabular}
\setlength{\tabcolsep}{5pt}
\begin{tabular}{lcccc}
\textbf{Resonance} & \multicolumn{2}{c}{\textbf{Efficiency, (\%)}}  \\
$\bm{X}$  & $\bm{J/\psi\to\gamma X}$ &  $\bm{\psi(2S)\to\gamma X}$  \\
\hline
\multicolumn{3}{l}{$\mathrm{X} \to \pi^+\pi^-$} \\
\hline
$f_2(1270)$ & 19.8 & 36.4 \\
$f_0(1500)$ & 18.5 & 35.9 \\
$f_0(1710)$ & 17.1 & 34.5 \\
$f_0(2100)$ & 15.0 & 34.1 \\
$2500-3300$ & ---  & $37.5-29.4$ \\
\hline
\multicolumn{3}{l}{$\mathrm{X} \to \pi^0\pi^0,~~\pi^0\to\gamma\gamma$} \\
\hline
$f_2(1270)$ & 13.1 & 24.5 \\
$f_0(1500)$ & 13.2 & 25.3 \\
$f_0(1710)$ & 13.8 & 25.7 \\
$f_0(2100)$ & 14.8 & 26.9 \\
$2500-3300$ & ---  & $27.4-30.0$ \\ 
\hline
\multicolumn{3}{l}{$\mathrm{X} \to K^+K^-$} \\
\hline
$f_0(1370)$  & 23.3 & 36.1 \\
$f_2'(1525)$ & 23.2 & 35.2 \\
$f_0(1710)$  & 22.4 & 35.6 \\
$f_0(2200)$  & 18.6 & 36.5 \\
$2500-3300$  & ---  & $40.6-40.8$ \\
\hline
\multicolumn{3}{l}{$\mathrm{X} \to K_S^0K_S^0,~~K_S^0\to\pi^+\pi^-$} \\
\hline
$f_0(1370)$  & 14.7 & 23.1 \\
$f_2'(1525)$ & 15.0 & 23.0 \\
$f_0(1710)$  & 15.0 & 23.6 \\
$f_0(2200)$  & 15.9 & 24.1 \\
$2500-3300$  & ---  & $27.3-27.5$ \\
\hline
\multicolumn{3}{l}{$\mathrm{X} \to\eta\eta,~~\eta\to\gamma\gamma$} \\
\hline
$f_0(1710)$ & 11.3 & 18.2 \\
$2500-3300$ & ---  & $17.6-16.0$ \\  
\end{tabular}
\end{ruledtabular}

\label{tbl:effs}
\end{table}

Charged pions and kaons were identified using information from both $dE/dx$ and RICH measurements.  For the pion and kaon tracks with momenta~$>600$~MeV, the variable $\Delta L_{i,j}$ was constructed, combining information from $dE/dx$ measurements, $\sigma_{i,j}^{dE/dx}$, and the RICH  log-likelihood, $L_{i,j}^{RICH}$, for the particle hypotheses $i$ and $j$
$$\Delta L_{i,j} = (\sigma_i^{dE/dx})^2 - (\sigma_j^{dE/dx})^2 + L_i^{RICH} - L_j^{RICH}$$
To identify pions, it was required that $\Delta L_{\pi,K}<0$.  For kaons, it was required that $\Delta L_{\pi,K}>0$.  For pion and kaon tracks with momenta~$<600$~MeV, for which RICH information was not available, it was required that the $dE/dx$ of the track be consistent within $3\sigma$ of the respective mass hypothesis, and that it be more consistent with its mass hypothesis than with the alternate mass hypothesis.

For the $\gamma\pi^+\pi^-$ final state, electron contamination was removed by rejecting events with $E_{CC}/p>0.85$, and muon contamination was removed by rejecting events which have tracks matched to any hits in the muon chambers.

Since we study resonances with masses $<2.8$~GeV for $J/\psi$ decays and $<3.3$~GeV for $\psi(2S)$ decays, we safely require $E_\gamma > 80$~MeV for the ``radiative'' photon candidate.  

We obtain our $J/\psi$ sample from $\psi(2S)$ by tagging events with two oppositely charged tracks, which were assumed to be pions, to have a recoil mass in the range $M(J/\psi)\pm10$~MeV.  The mass distribution of events recoiling against $\pi^+\pi^-$ in the decay $\psi(2S)\to\pi^+\pi^-J/\psi$ is shown in Fig.~\ref{fig:jpsirecmass}(a) before 4C kinematic fit, and Fig.~4(b) after 4C fit.  The strong peak corresponds to $J/\psi$ events.

To reconstruct the full $\psi(2S)$ and $J/\psi$ events, we make a 4C energy/momentum kinematic fit, constraining the sum of the 4-vectors of all final state particles to be equal to the 4-vector of the $e^+e^-$ interaction.  For $J/\psi$ decays, the $J/\psi$ decay products were additionally constrained to have $M(J/\psi)$.    In all cases the final fit was required to have $\chi^2_{4C}<25$.  When constructing the invariant masses from the radiative decays, the 4--vectors resulting from these fits were used.

\begin{figure*}[!tb]

\begin{center}

\includegraphics[width=2.5in]{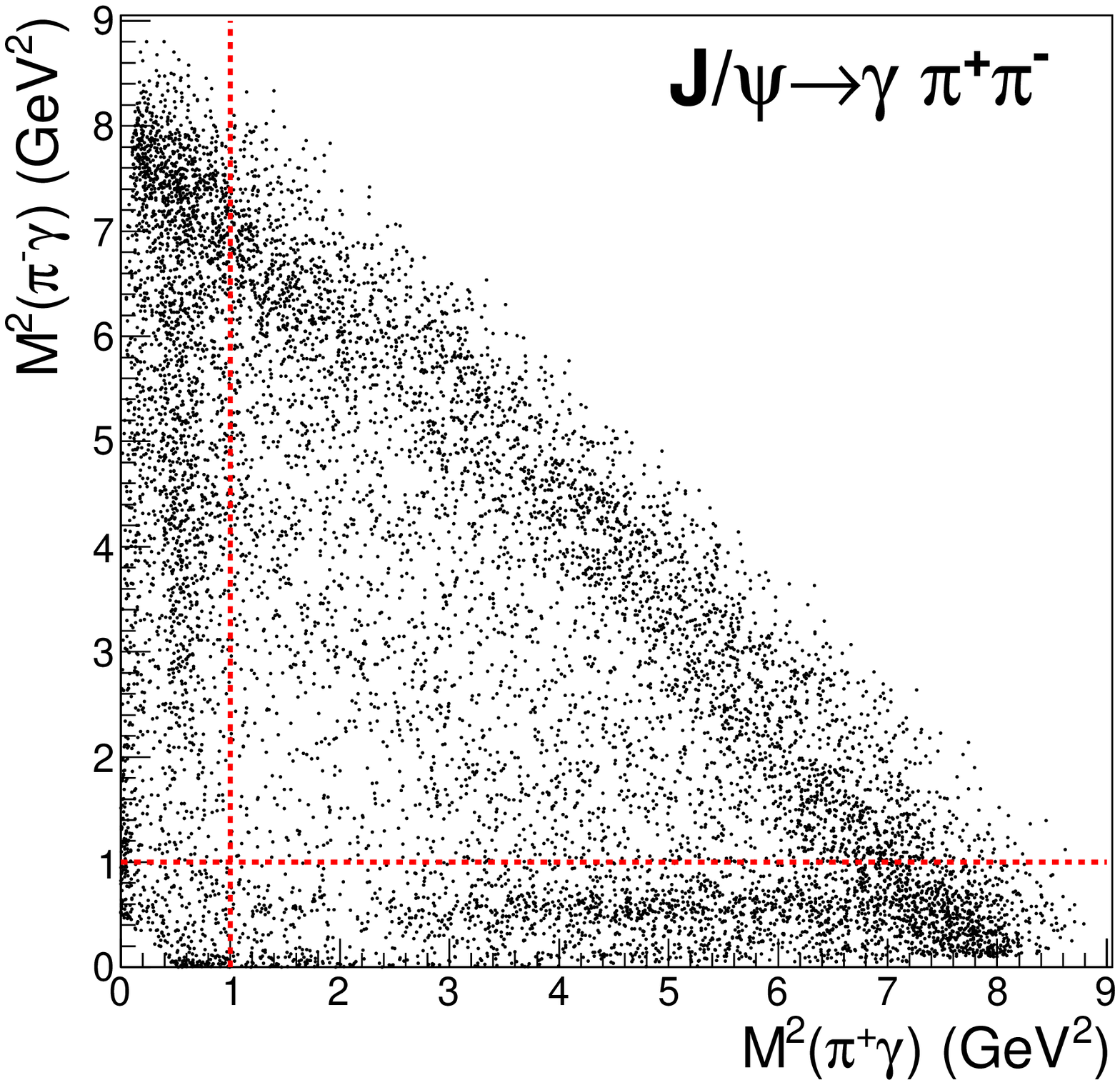}
\includegraphics[width=2.5in]{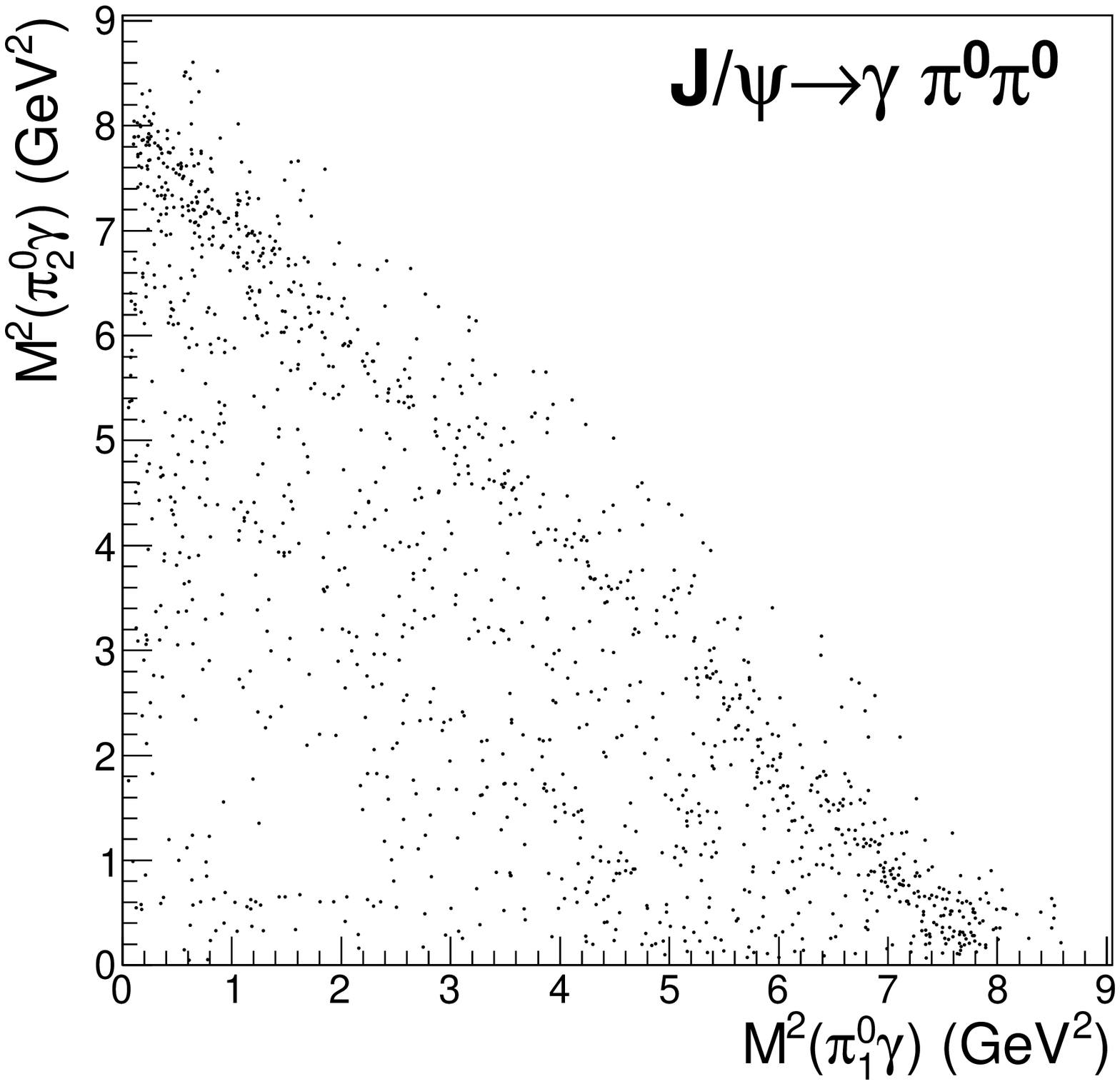}

\includegraphics[width=2.5in]{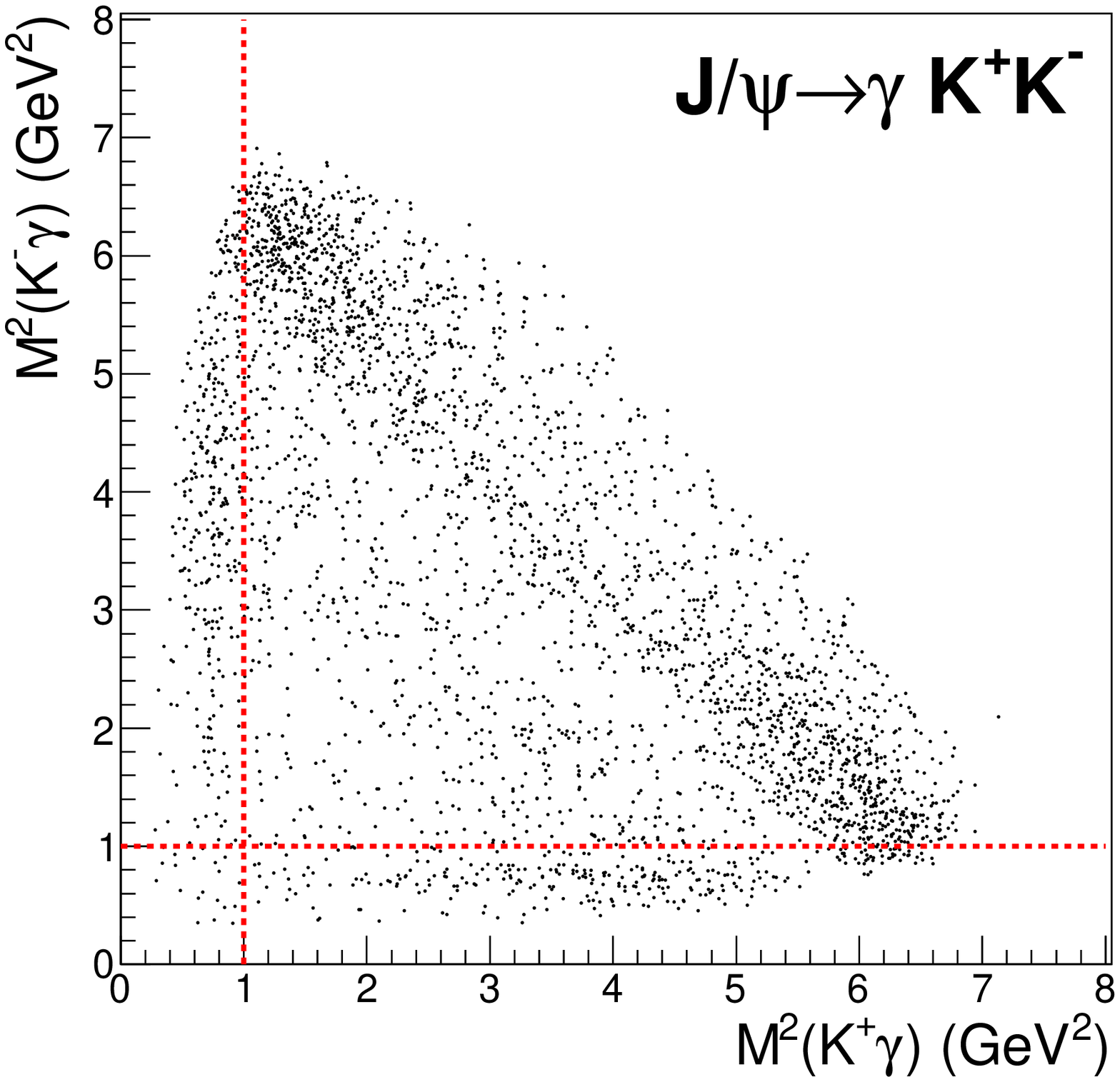}
\includegraphics[width=2.5in]{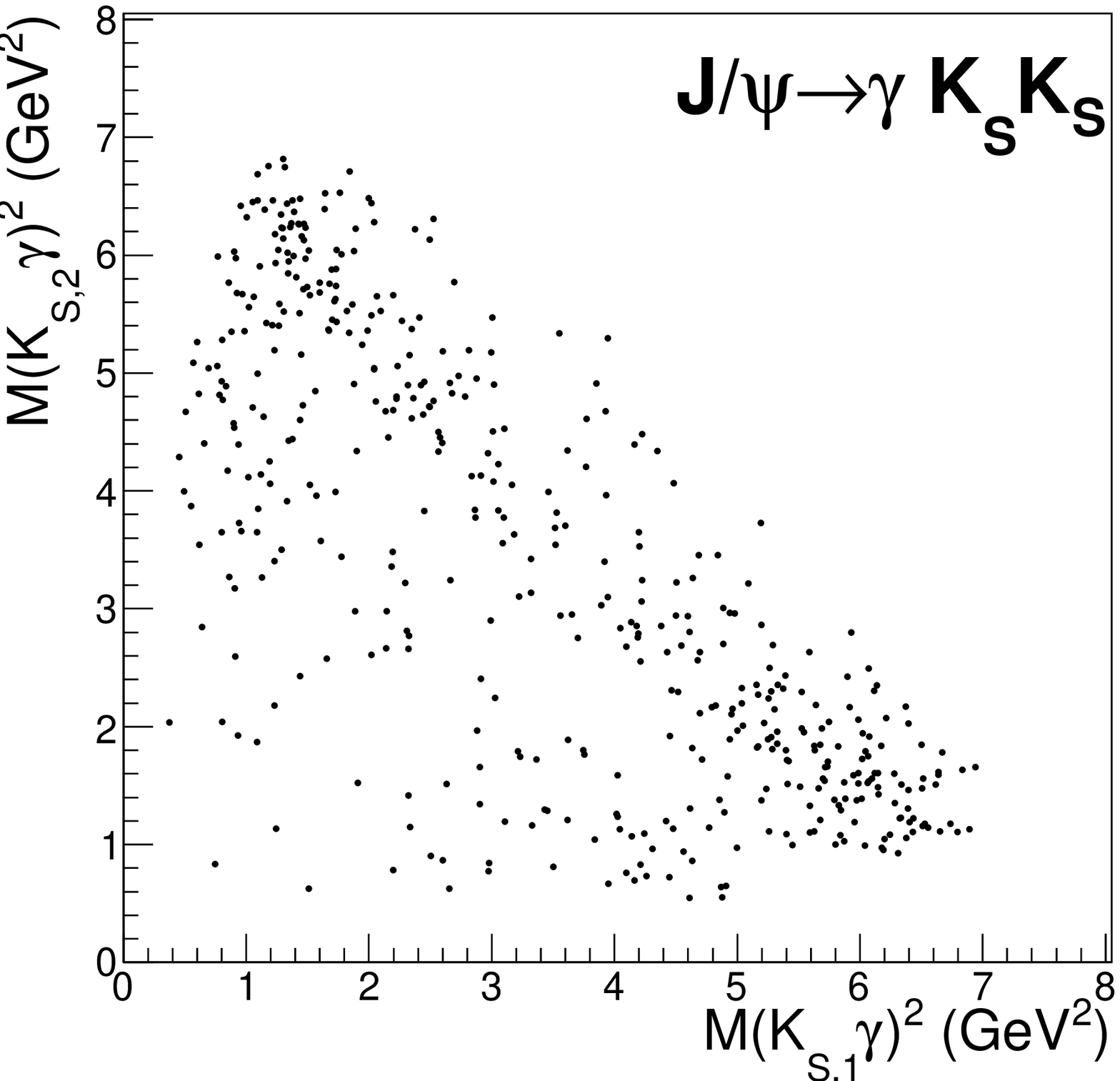}

\end{center}

\caption{Dalitz plots for (top two panels) $J/\psi\to \gamma\pi^+\pi^-$ and $\gamma\pi^0\pi^0$, and (bottom two panels) $J/\psi\to\gamma K^+K^-$ and $\gamma K_S^0K_S^0$.  The dashed lines indicate the regions used to reject $J/\psi\to \rho\pi$ and $ K^*K$ decays in the $\gamma\pi^+\pi^-$ and $\gamma K^+K^-$ decays, respectively.}
\label{fig:dalitz}
\end{figure*}

\begin{figure}
\begin{center}
\includegraphics[width=3.in]{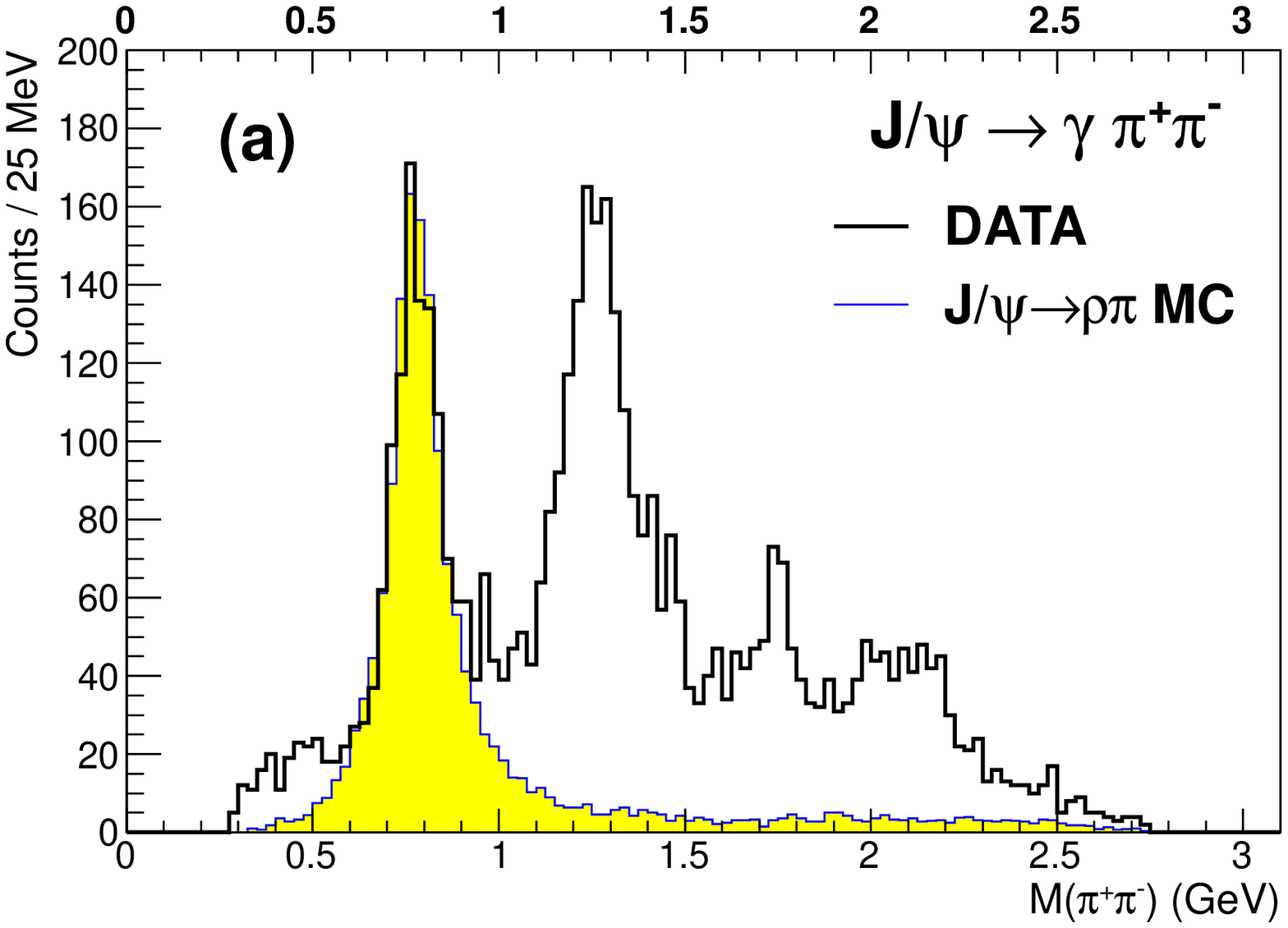}
\includegraphics[width=3.in]{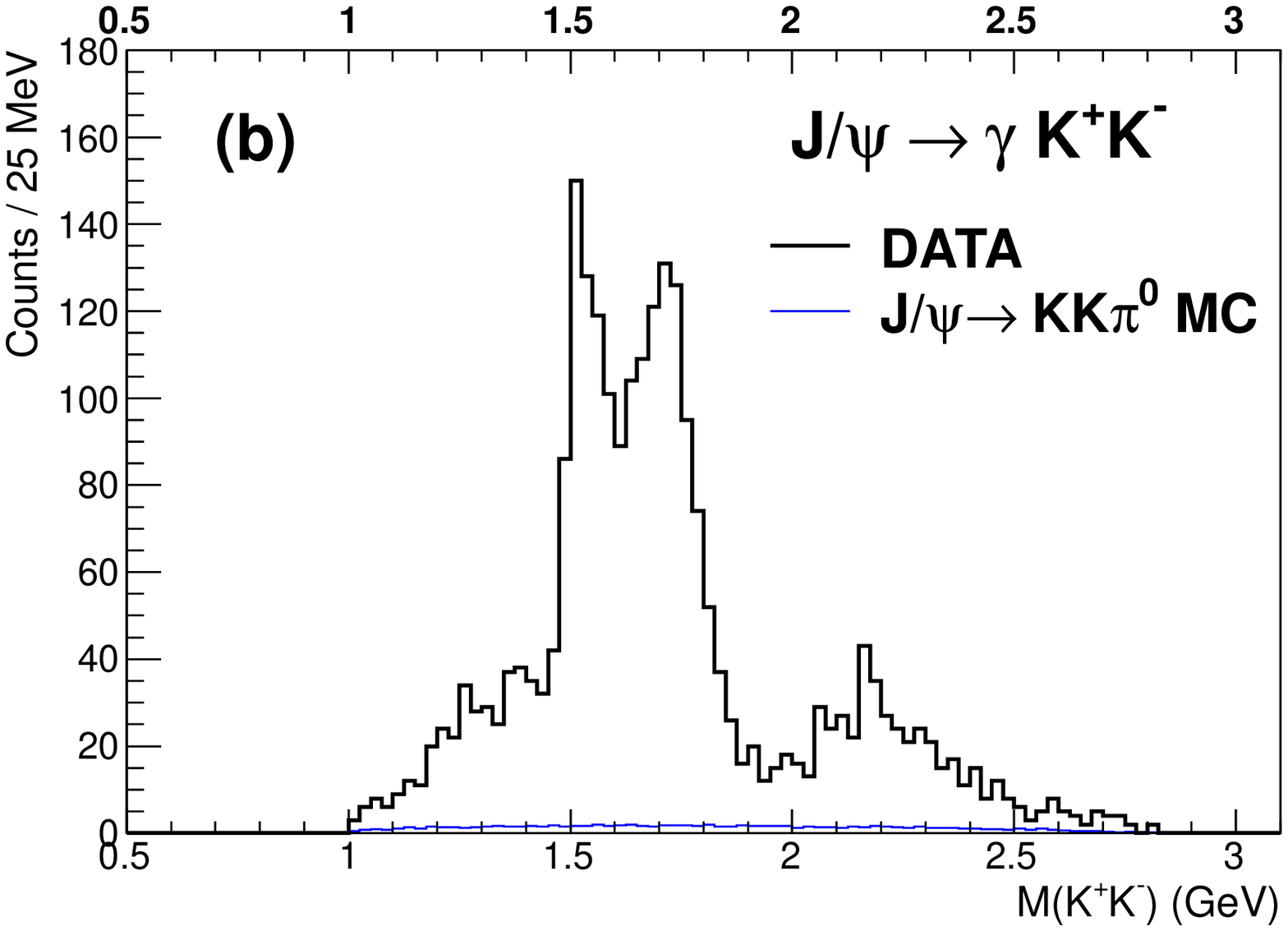}
\end{center}
\caption{(a) Distributions of $\pi^+\pi^-$ invariant mass for $J/\psi\to\gamma\pi^+\pi^-$ decays for data (solid histogram) and $J/\psi\to\rho\pi$ MC decays (shaded histogram), illustrating the small residual non-$\rho$ background from $\rho\pi$ decay.
(b) Distributions of $K^+K^-$ invariant mass for $J/\psi\to\gamma K^+K^-$ decays for data (solid histogram) and $J/\psi\to KK\pi^0$ MC decays (shaded histogram).}
\label{fig:mcbkgd}
\end{figure}

\begin{figure*}[!tb]

\begin{center}

\includegraphics[width=2.5in]{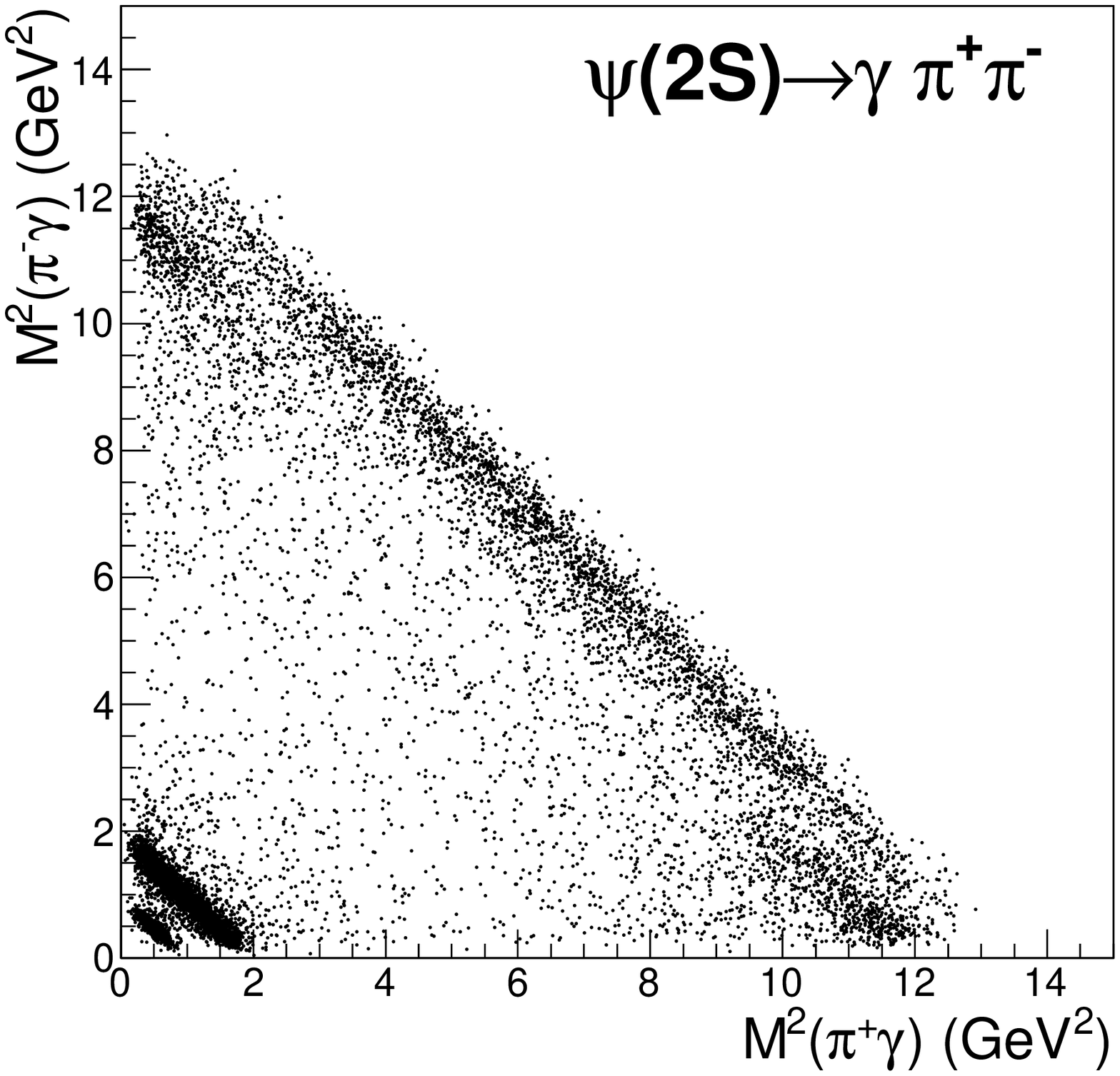}
\includegraphics[width=2.5in]{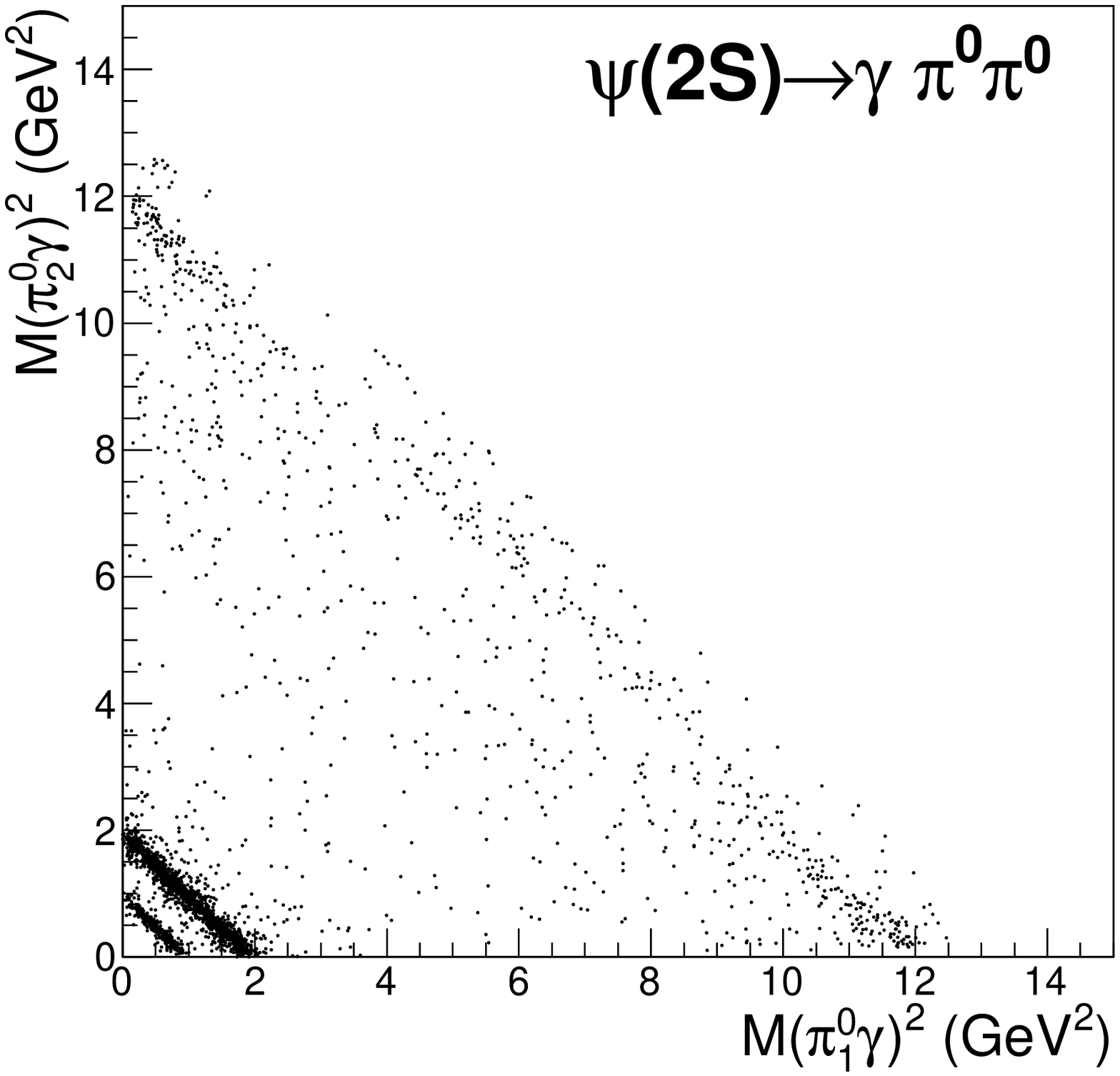}

\includegraphics[width=2.5in]{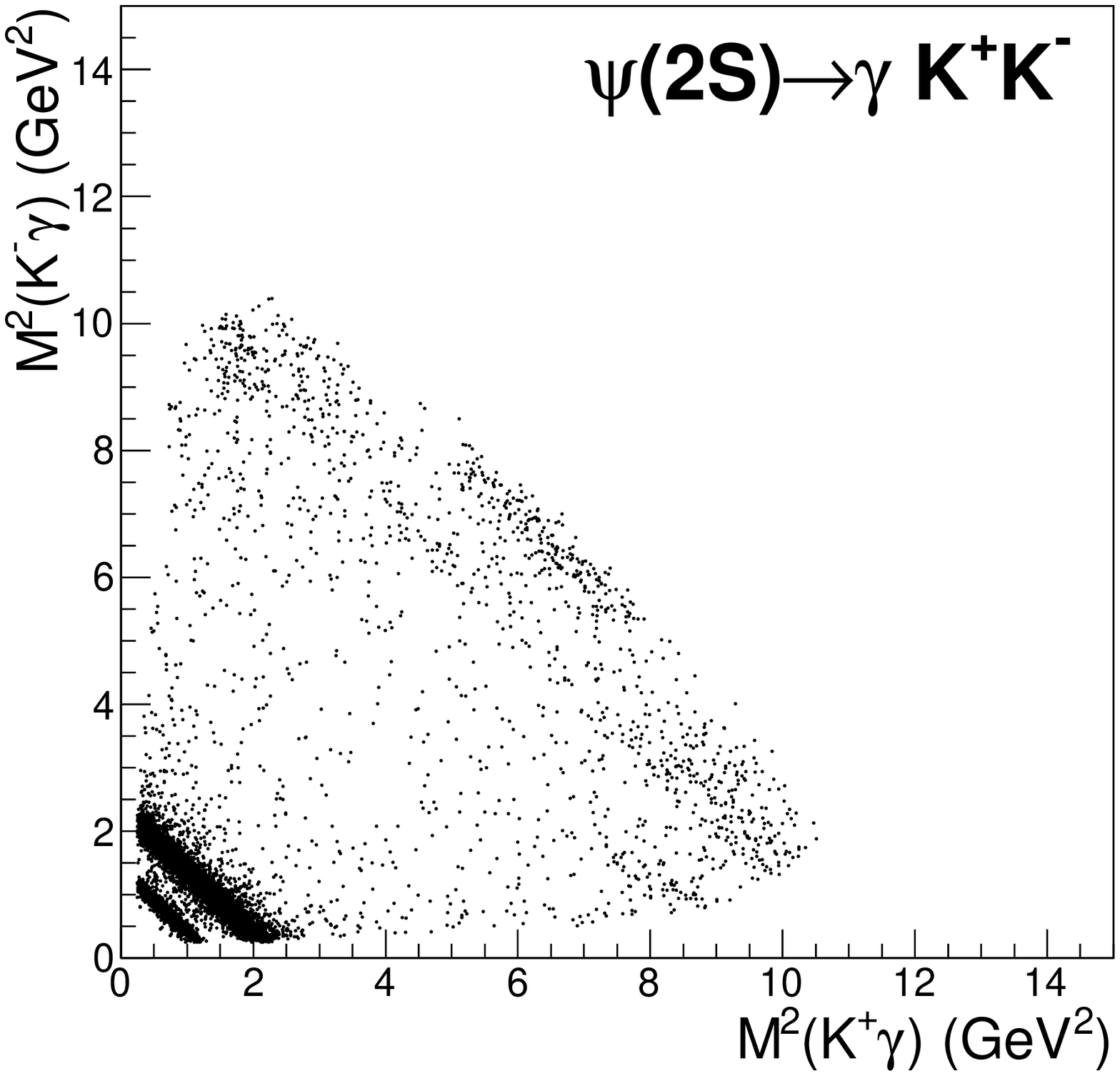}
\includegraphics[width=2.5in]{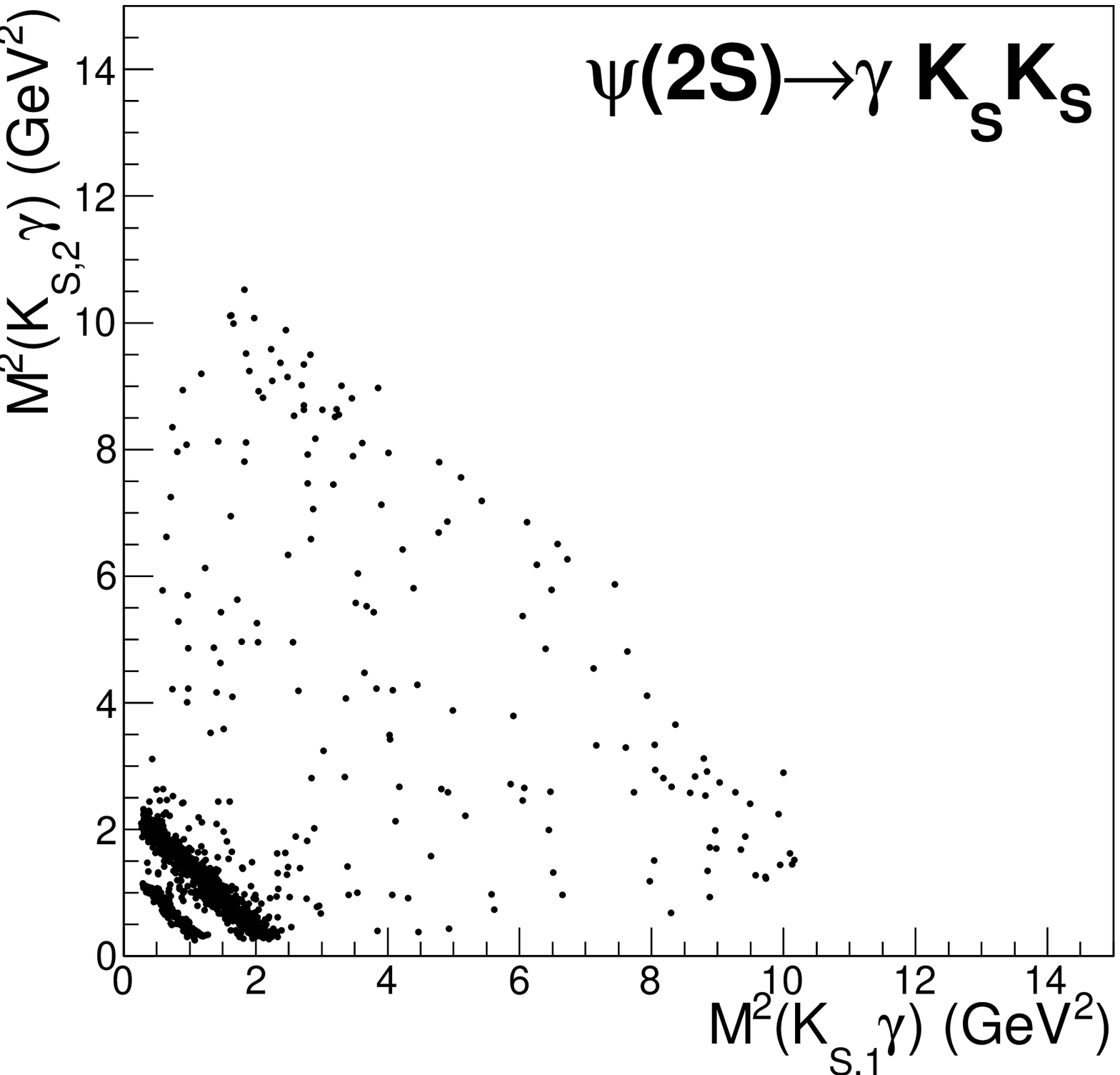}

\end{center}

\caption{Dalitz plots for (top two panels) $\psi(2S)\to \gamma\pi^+\pi^-$ and $\gamma\pi^0\pi^0$, and (bottom two panels)  $\psi(2S)\to \gamma K^+K^-$ and $\gamma K_S^0K_S^0$.}
\label{fig:dalitz2}
\end{figure*}

\section{Monte Carlo Samples and Efficiencies}

Monte Carlo samples of simulated events were generated using EvtGen~\cite{evtgen}, and then passed through the standard detector simulation and reconstruction routines, to determine mass resolutions and experimental efficiencies.  Monte Carlo samples of 200k events were generated for each pseudoscalar pair decay chain for the light quark resonances seen in the data and the presumed glueball candidates in the mass region $2500-3500$~MeV.   To investigate background contributions from other decays, such as $J/\psi\to\rho\pi,K^*K$, Monte Carlo samples of 5~million events were generated.  When considering radiative decays to a particular resonance (e.g. $\psi\to\gamma f_{0,2}$), the expected $1+\alpha\cos^2\theta$ angular distributions for the corresponding E1 transitions were used, with $\alpha=1$ for decays to a $0^{++}$ state and $\alpha=1/13$ for decays to a $2^{++}$ state.  The angular distributions  for scalar and tensor mesons decaying into two pseudoscalars were simulated using the helicity amplitude formalism of EvtGen.
Fig.~\ref{fig:costh} illustrates excellent agreement between the data and simulated angular distribution for the well-separated resonances $f_2(1270)$ and $f_0(1710)$.

The efficiencies obtained for decays to individual resonances are shown in Table~\ref{tbl:effs}.  For $J/\psi$ decays to charged final states the efficiencies range from 15\% to 20\%, and for neutral final states they range form $\sim13\%$ to 16\%.  Corresponding efficiencies for $\psi(2S)$ decays are approximately a factor two larger.

Monte Carlo simulations give instrumental Gaussian resolution widths for masses $<2500$~MeV as $\sigma=3-4$~MeV, for $\pi^+\pi^-$, $K^+K^-$, and $K_S^0K_S^0$, $9-11$~MeV for $\pi^0\pi^0$,  and $7-8$~MeV for $\eta\eta$.

\section{Dalitz Plots for $J/\psi$ and $\psi(2S)$ Radiative Decays}

\noindent
$\bm{J/\psi\to\gamma\pi\pi}$:\\[5pt]
\noindent Dalitz plots for the radiative decays of $J/\psi \to \gamma\pi^+\pi^-$ and $\pi^0\pi^0$ are shown in the top two panels of Fig.~\ref{fig:dalitz}. In the $\gamma\pi^+\pi^-$ plot there are horizontal and vertical bands corresponding to $\rho(770)$. As confirmed by Monte Carlo simulations, these arise from $\rho^0 \pi^0$  decay of $J/\psi$ with one photon from $\pi^0$ missing. In subsequent analysis we remove these by rejecting events with $M^2(\pi^\pm \gamma) < 1.0$~GeV$^2$  . We do not remove $\pi^+\pi^-$ events from $\rho^0$ decays which populate the diagonal band in the plot in order to be able to take the effect of their tails in subsequent fits to the $\pi^+\pi^-$ mass spectrum.
Fig.~\ref{fig:mcbkgd}(a) shows that the tail of the residual $\rho$ peak does not contribute significantly to the $\pi^+\pi^-$ mass spectrum in the $M(\pi^+\pi^-)>1.2$~GeV.

As expected, the $\rho(770)$ contributions are absent in the $\pi^0\pi^0$ Dalitz plot.

The only other feature which is identifiable in these Dalitz plots is the excitation of $f_2(1270)$ as a diagonal band with $J=2$ characteristic angular distribution which produces enhancements near the maxima of $M^2(\pi\gamma)$.
\vspace*{5pt}

\noindent
$\bm{J/\psi\to\gamma K\overline{K}$}:\\[5pt]
\noindent Dalitz plots for the radiative decays of $J/\psi \to \gamma K^+K^-$ and $K_S^0K_S^0$ are shown in the middle two panels of Fig.~\ref{fig:dalitz}. The bands due to $\rho(770)$ excitation seen in the $\gamma\pi^+\pi^-$ plot  are , of course, absent in these plots. Instead, there is evidence for contributions from the decay  $J/\psi \to K^+K^{-*}(892) \to K^+K^-\pi^0 (+c.c.)$, with one photon from $\pi^0$ missing. We remove these events in the horizontal and vertical bands by rejecting events with $M^2(K^\pm\gamma) < 1.0$~GeV$^2$. 
Monte Carlo simulation, shown in Fig.~\ref{fig:mcbkgd}(b), confirms that the residual background, which is too small to be discernible, is completely negligable.
Excitation of $f_2'(1525)$ with its characteristic $J=2$ angular distribution is identifiable as a diagonal band, followed by the nearly uniform population of $f_0(1710)$. 
A weakly populated diagonal band corresponding to $f_0(2200)$ is also discernible.

The $J/\psi \to \gamma K_S^0K_S^0$ Dalitz plot is similar to the of $J/\psi \to \gamma K^+K^-$ plot but has smaller yield.
\vspace*{5pt}

\noindent $\bm{\psi(2S)\to\gamma\pi\pi,\gamma K\overline{K}}$:\\[5pt]
\noindent The four Dalitz plots for the $\pi\pi$ and $KK$ decays of $\psi(2S)$ are shown in Fig.~\ref{fig:dalitz2}. As is well known, all hadronic and radiative decays of $\psi(2S)$ are suppressed by a factor $\sim7$ (the so-called 13\% rule) compared to those of $J/\psi$. As a result all plots in Fig.~\ref{fig:dalitz2} have smaller yields than in the corresponding plots in Fig.~\ref{fig:dalitz} . Because of the well known anomalously small branching fractions for the decays $\psi(2S) \to \rho\pi$ and $K^*K$, the $\rho$ bands and the $K^*$ bands seen in $J/\psi$ decays are not seen in the corresponding $\psi(2S)$ Dalitz plots. However, there is a strong diagonal band for $\rho(770)$ in the $\psi(2S) \to \gamma\pi^+\pi^-$ plot and a narrow band corresponding to $\phi(1020)$ in $\gamma K^+K^-$. 
These arise from ISR contributions as we show later.

The strong diagonal bands due to the population of $\chi_0$ and $\chi_2$ states of charmonium are visible in all four Dalitz plots for $\psi(2S)$ radiative decays.

\begin{figure*}[!tb]
\begin{center}
\includegraphics[width=2.2in]{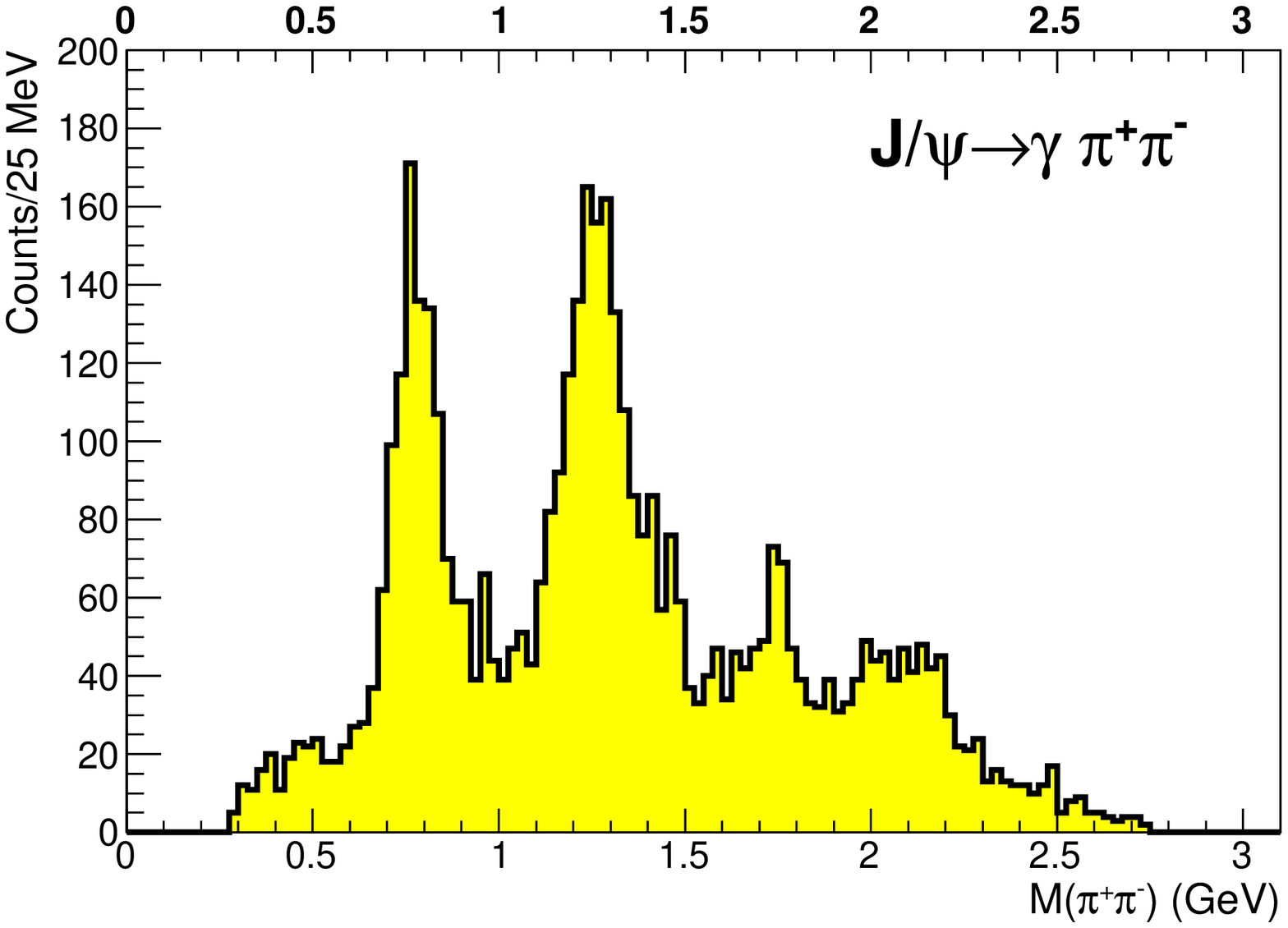}
\includegraphics[width=2.2in]{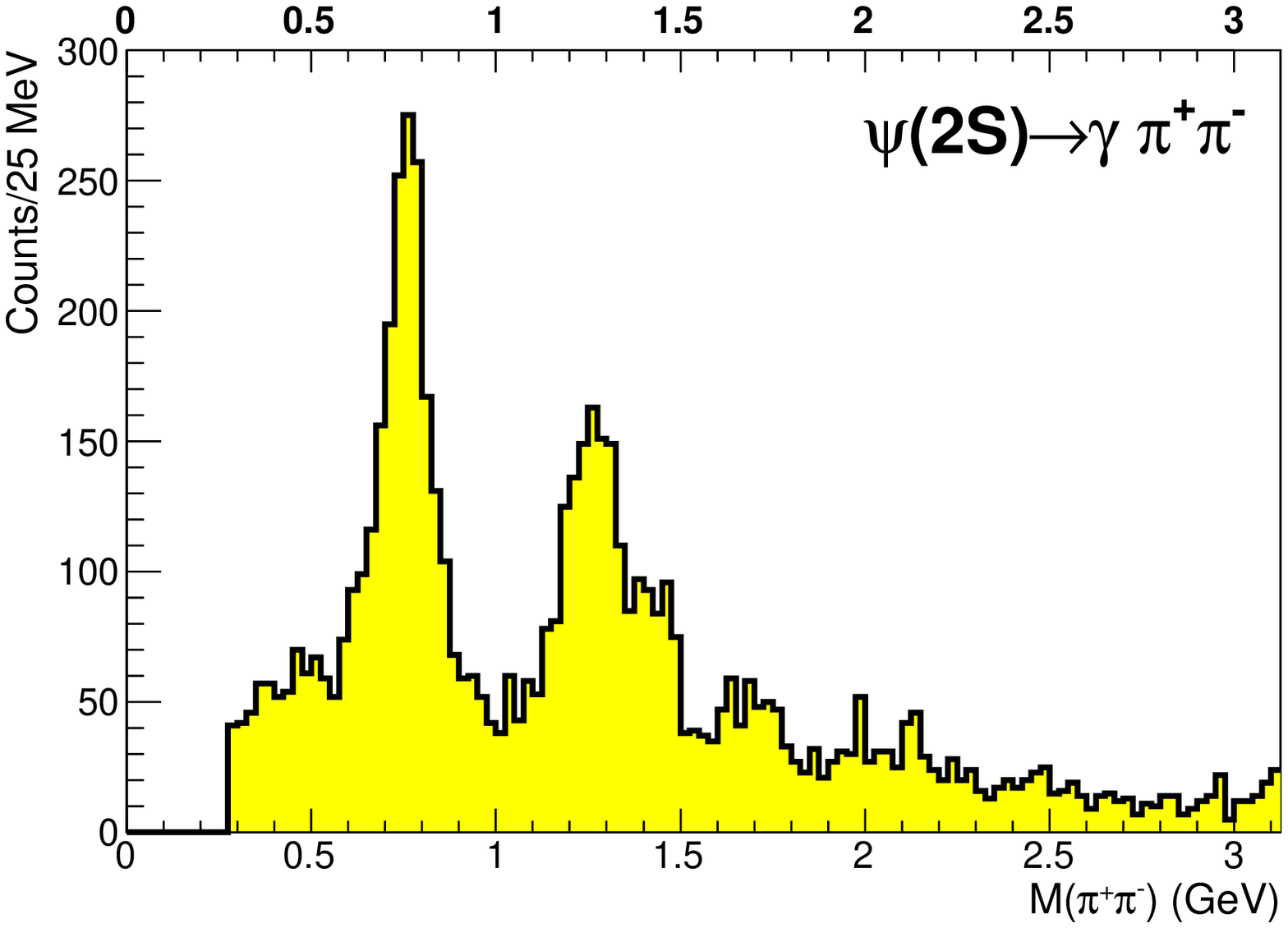}

\includegraphics[width=2.2in]{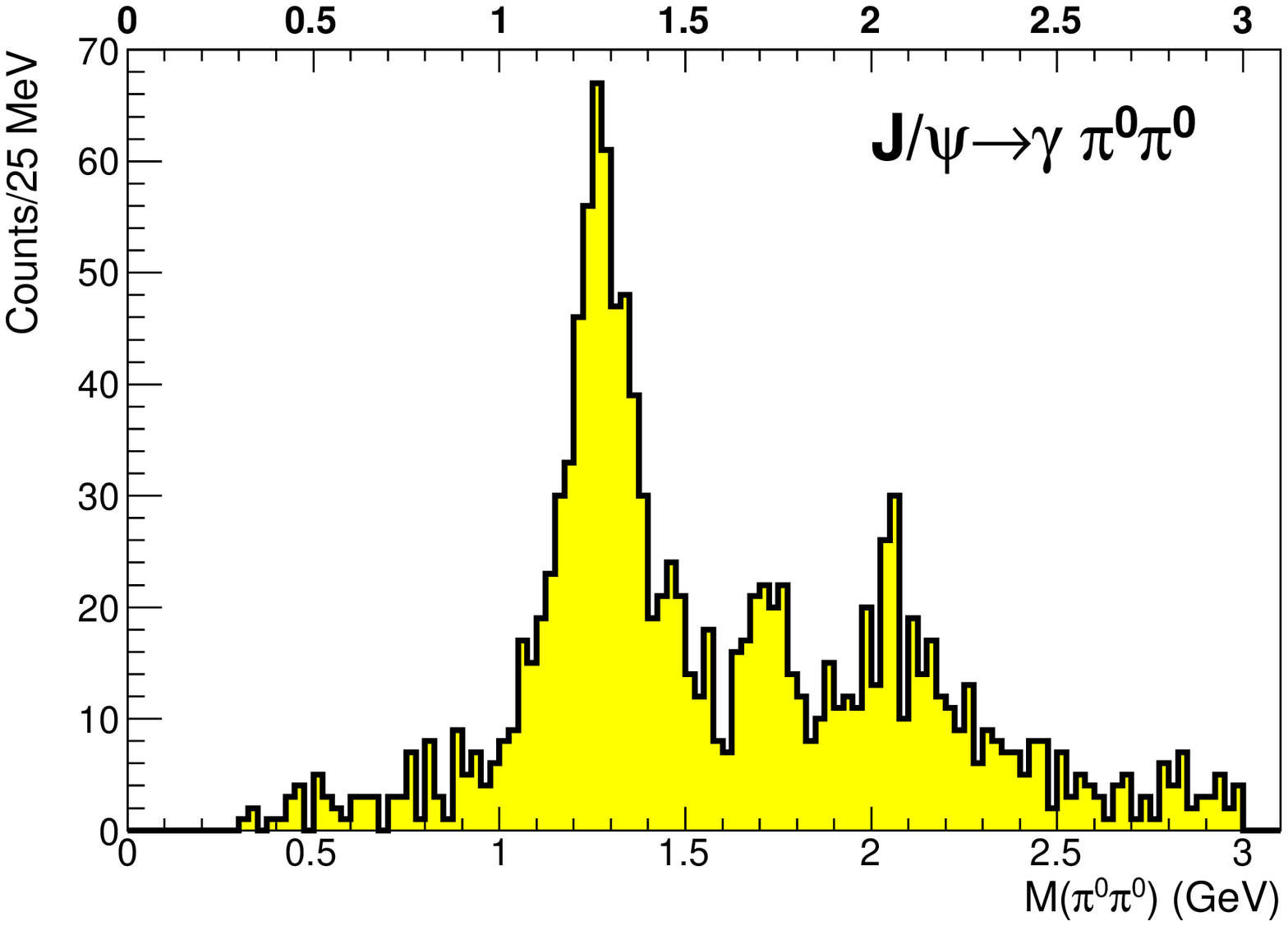}
\includegraphics[width=2.2in]{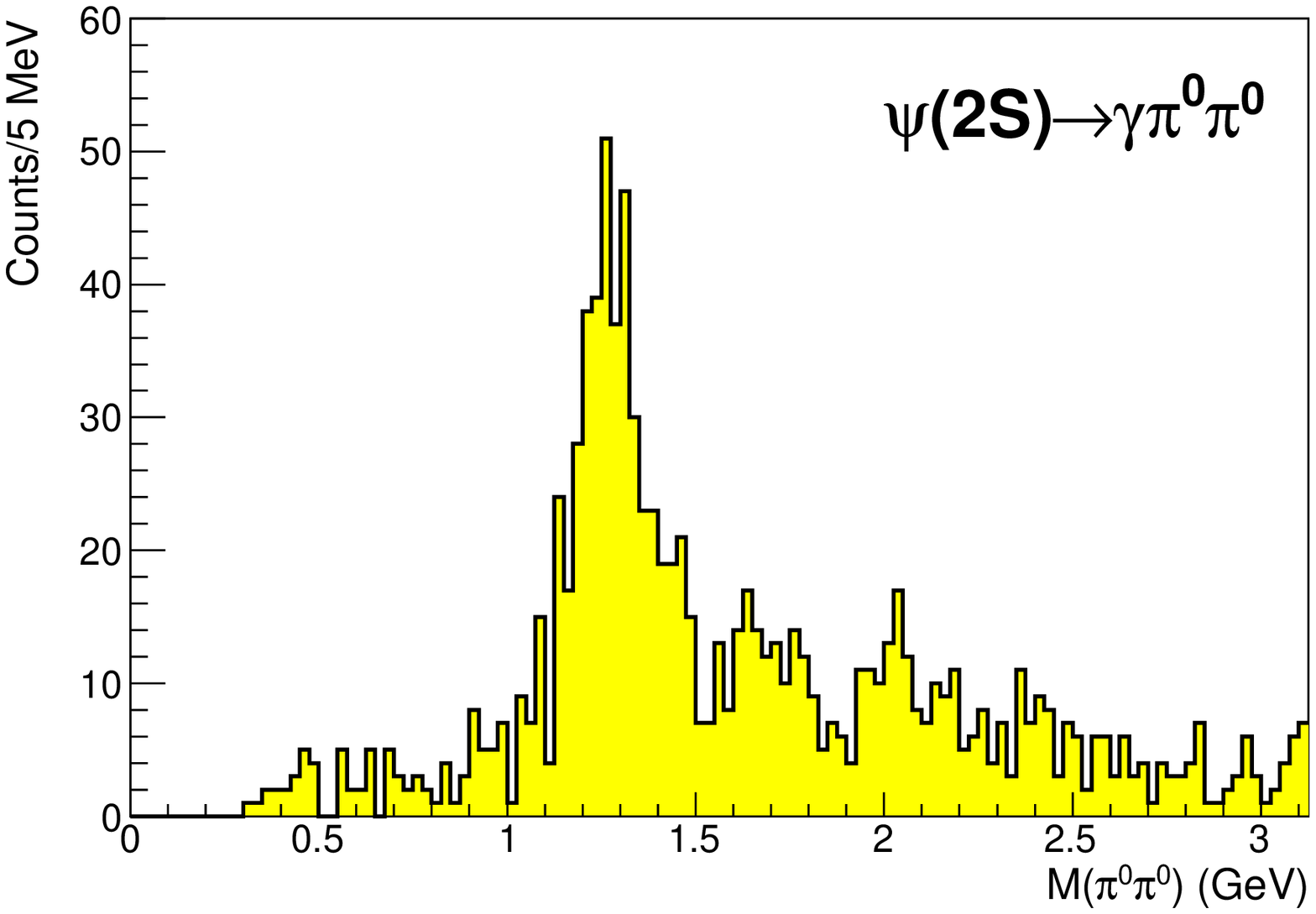}

\includegraphics[width=2.2in]{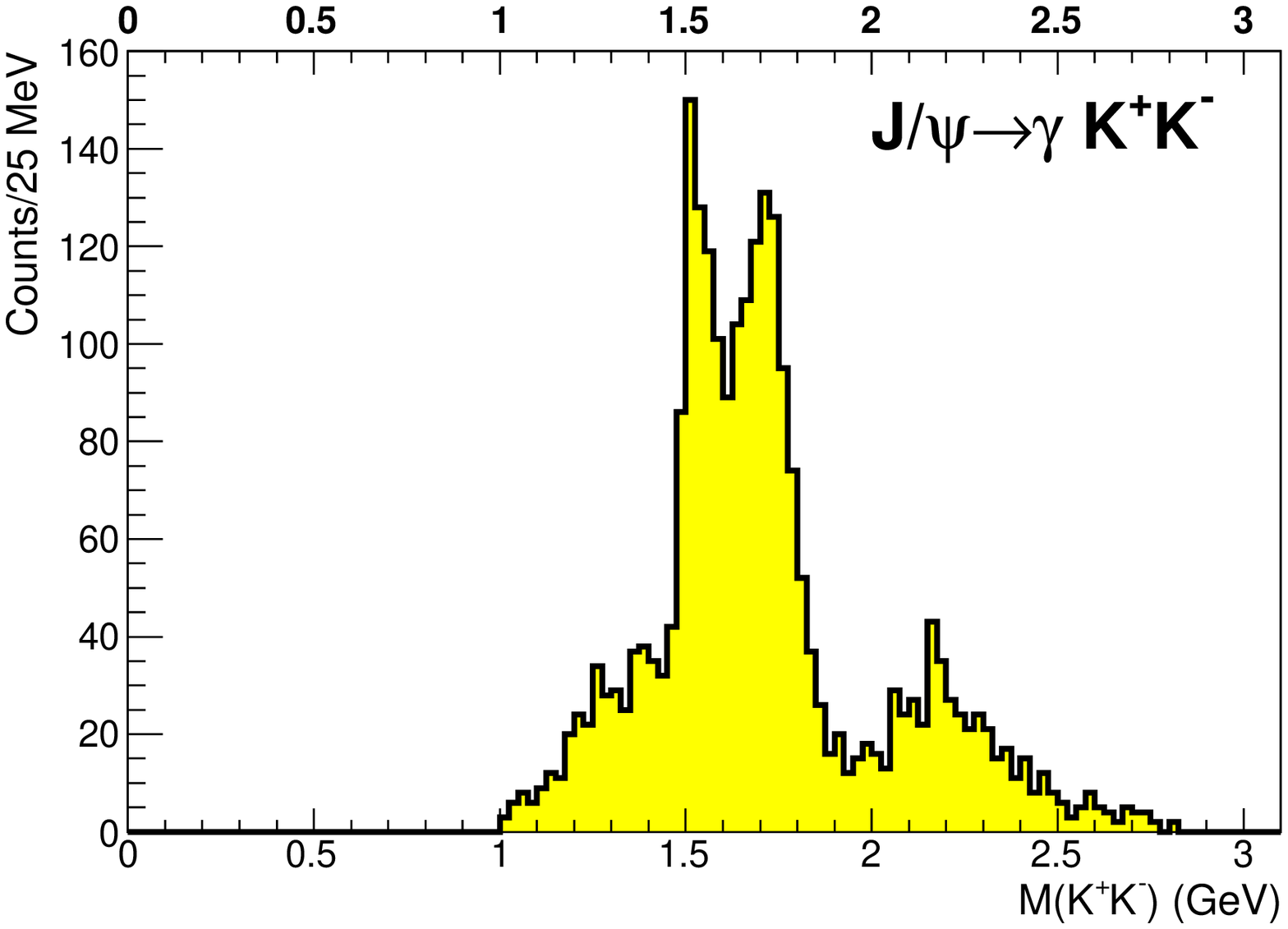}
\includegraphics[width=2.2in]{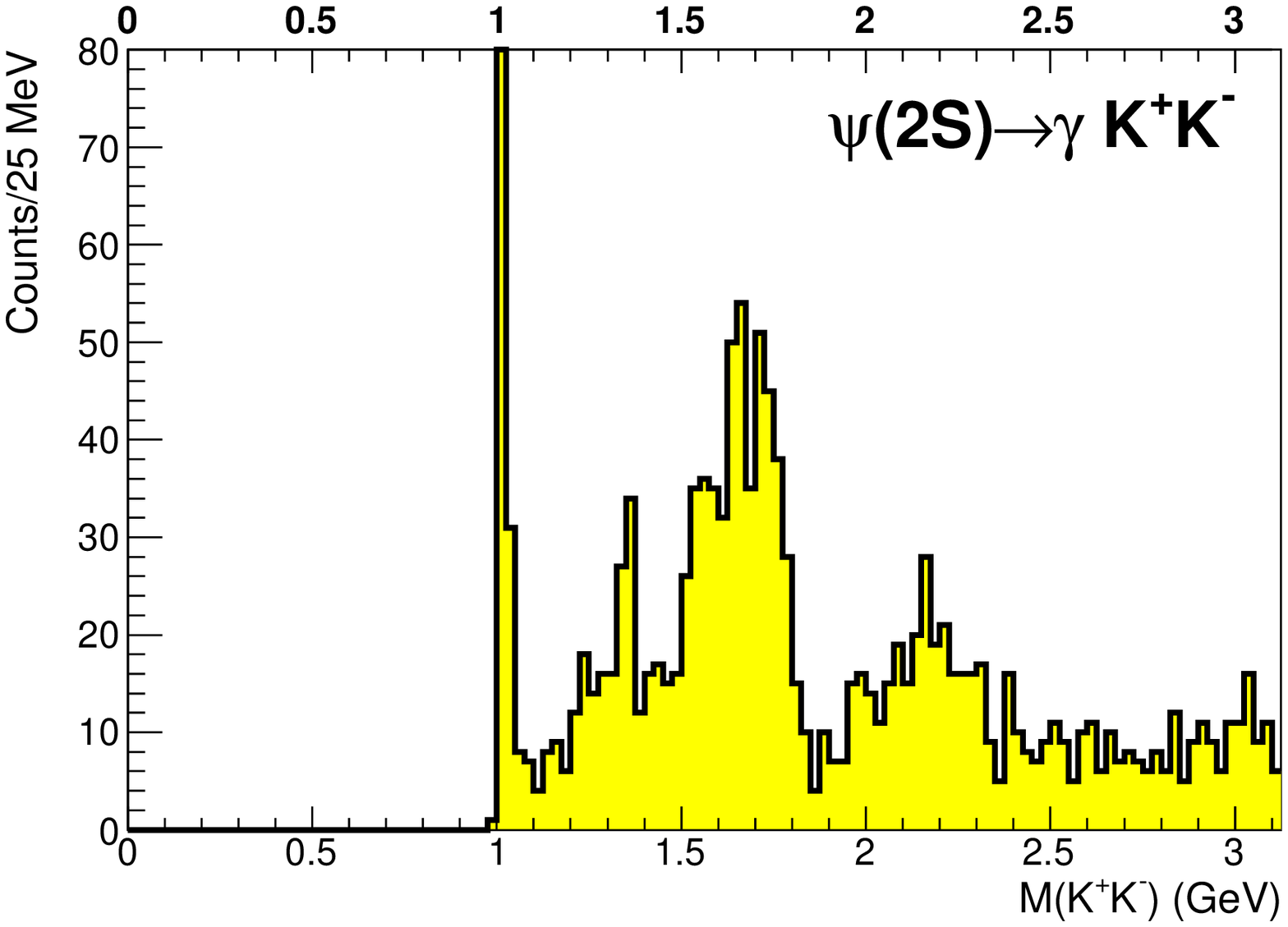}

\includegraphics[width=2.2in]{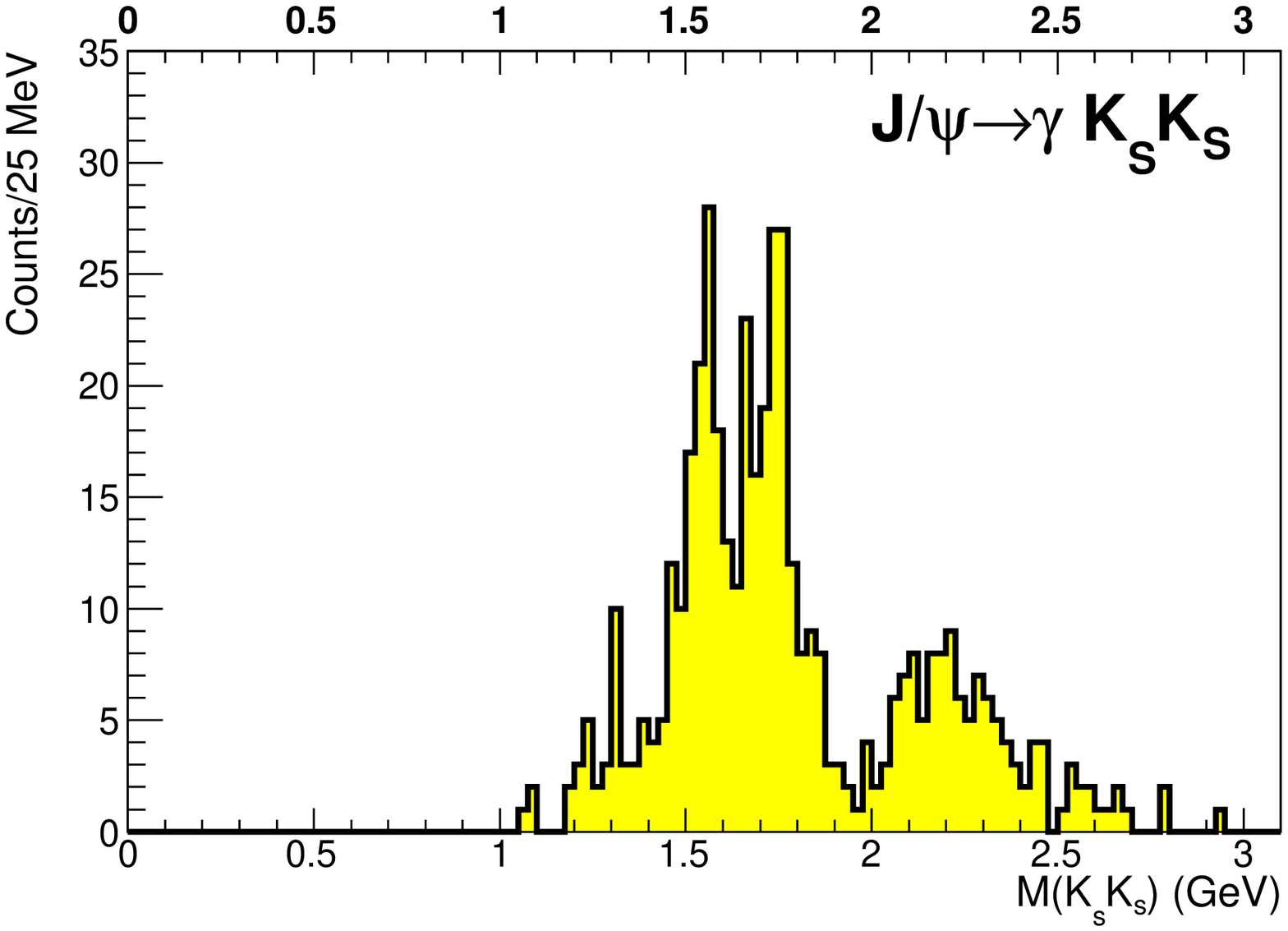}
\includegraphics[width=2.2in]{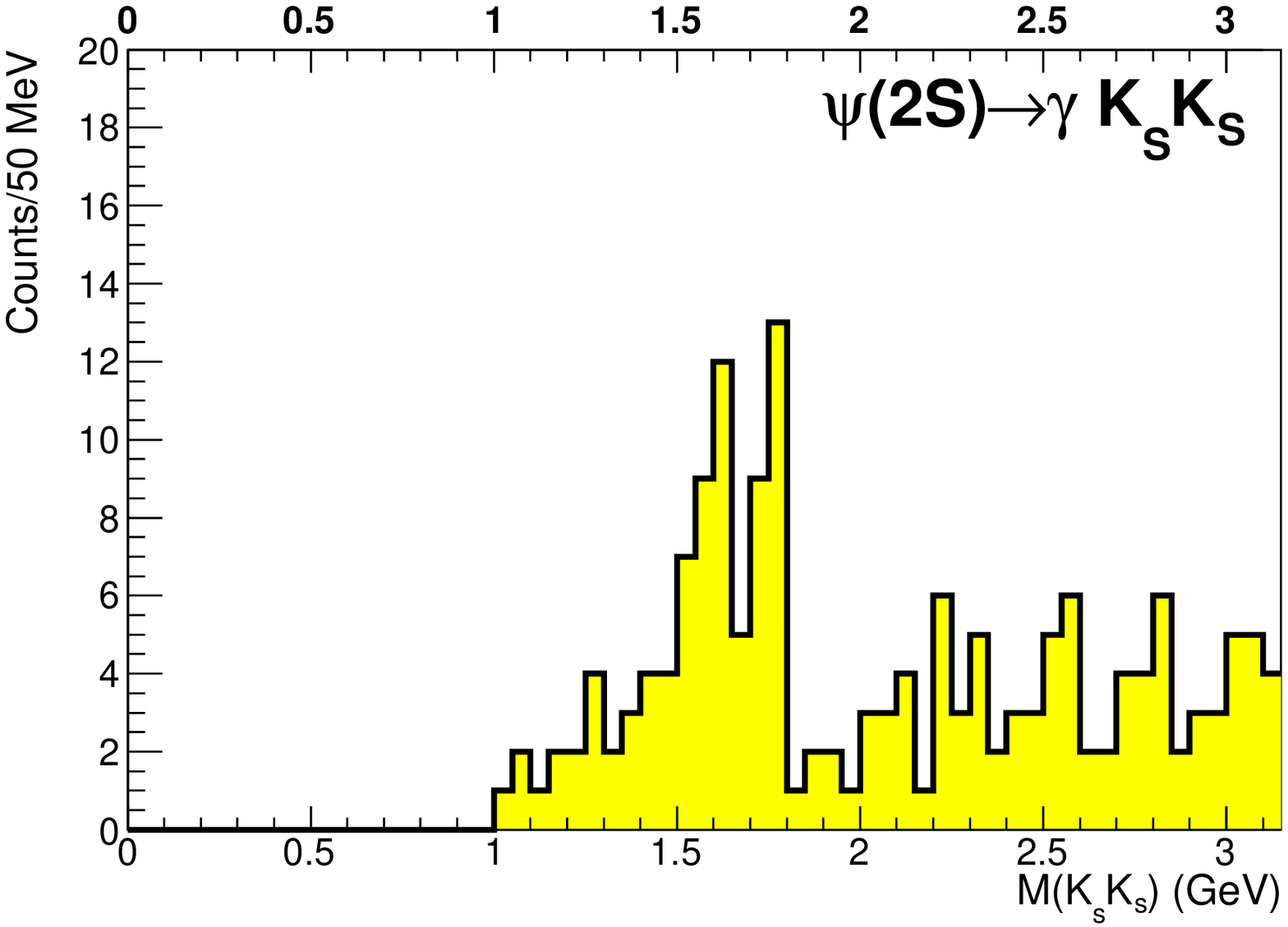}

\includegraphics[width=2.2in]{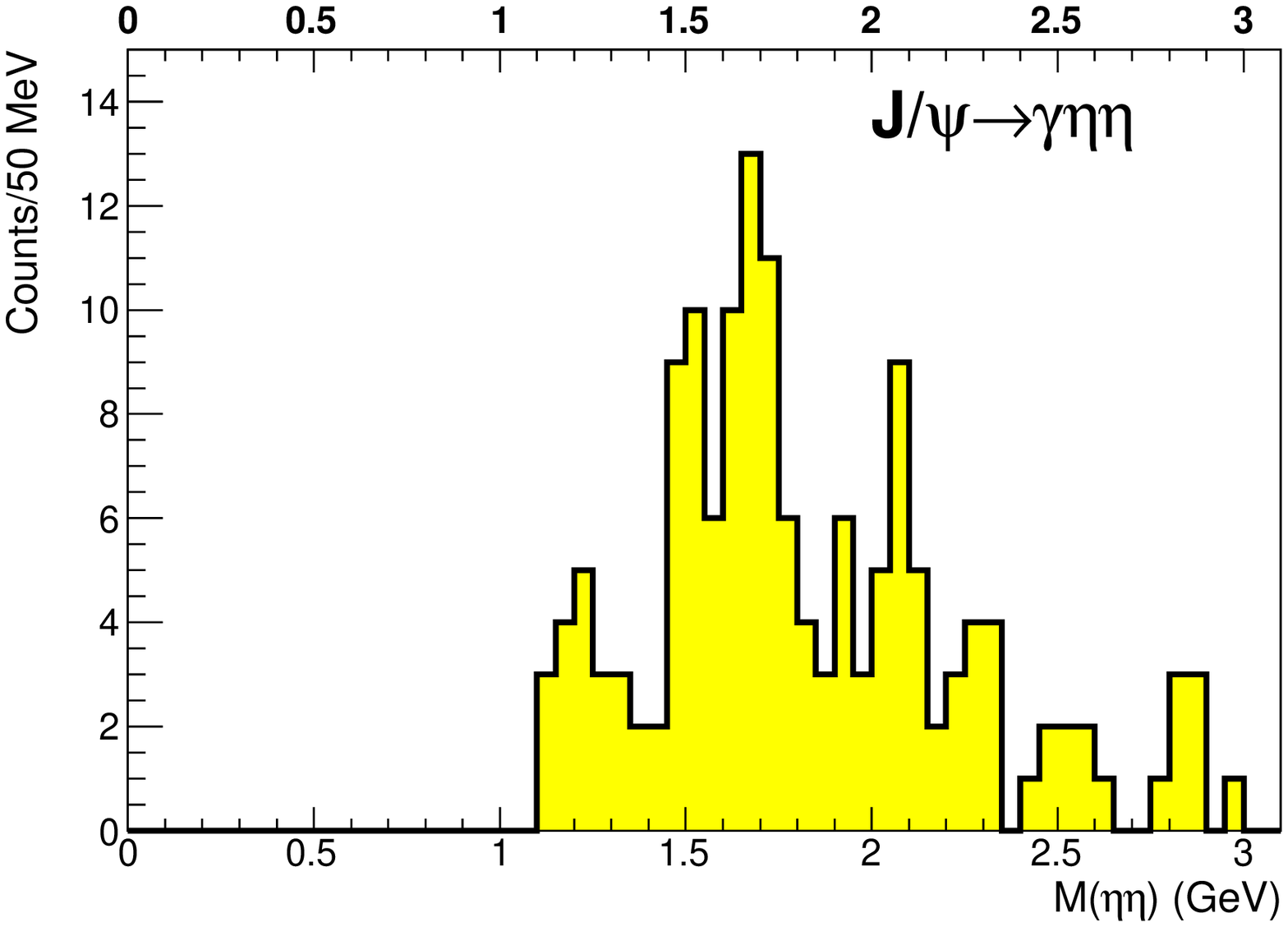}
\includegraphics[width=2.2in]{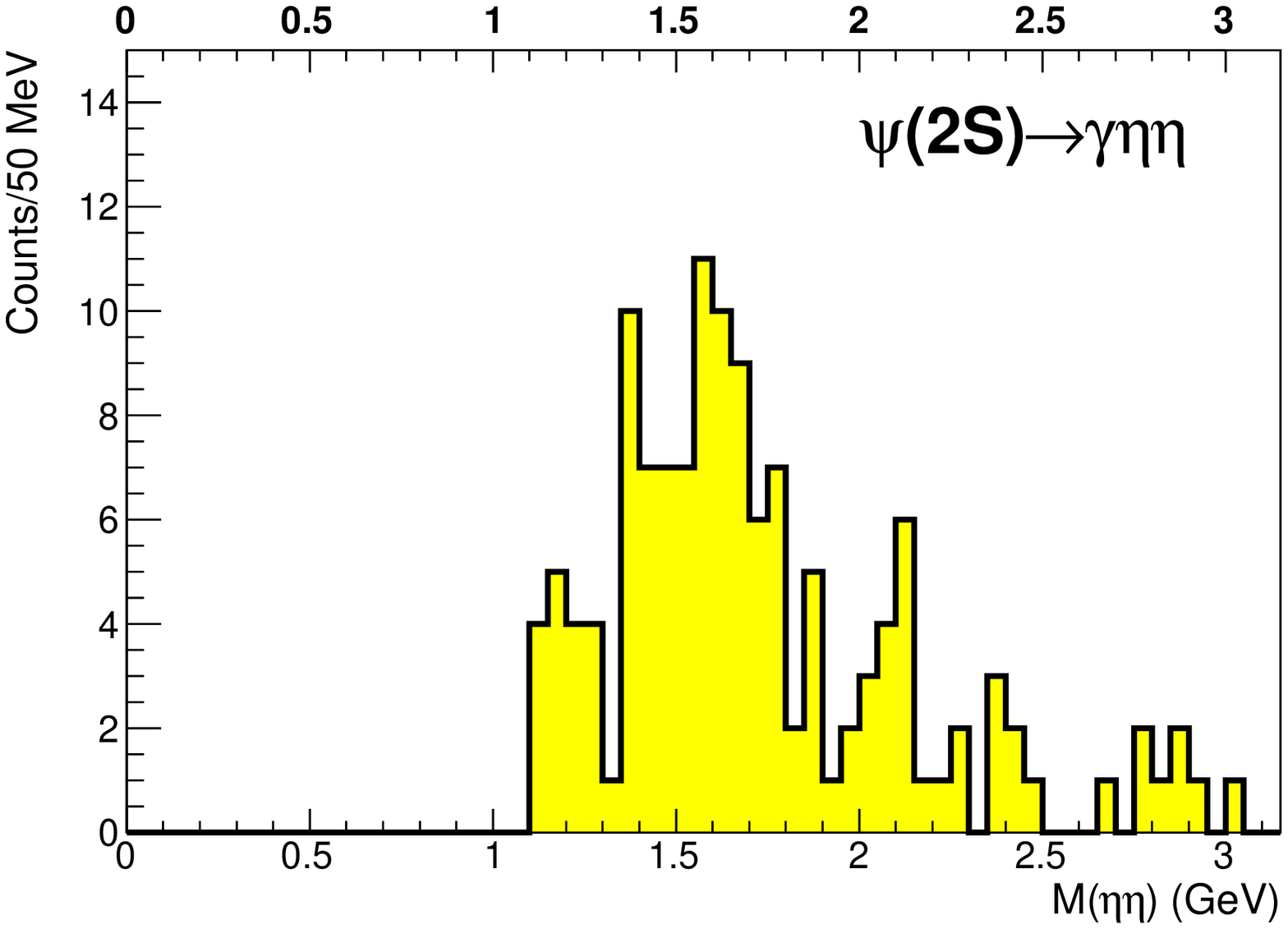}
\end{center}
\caption{Two particle invariant mass distributions for $J/\psi,\psi(2S)\to\gamma \mathrm{PS,PS}$ $(\mathrm{PS}\equiv \pi^\pm,\pi^0,K^\pm,K_S^0,\eta)$ decays, for $M(\mathrm{PS,PS})<3.0$~GeV.}
\label{fig:invmass}
\end{figure*}

\section{Invariant Mass Distributions for $J/\psi$ and $\psi(2S)$ Radiative Decays}

In order to draw quantitative conclusions from our measurements, we make projections from the Dalitz plots for the invariant masses for the $\pi\pi$, $KK$, and $\eta\eta$ final states.  An overall presentation of these is made in Fig.~\ref{fig:invmass}.  Several gross features are apparent in this figure.  The first feature is the general similarity in the observed spectra for $J/\psi$ radiative decays (left five panels) and $\psi(2S)$ radiative decays (right five panels).  The yields for $\psi(2S)$ decays are lower.  As mentioned earlier, this is mainly the consequence of the well-known ``13\% rule'' of pQCD which stipulates that the hadronic and radiative decays of $\psi(2S)$ have nearly a factor seven smaller widths compared to those of $J/\psi$, a ratio which follows that for the leptonic decays.  The second feature is the expected similarity between the corresponding charged and neutral final states, $\pi^+\pi^-$ and $\pi^0\pi^0$, and $K^+K^-$ and $K_S^0K_S^0$. We find that resonance masses determined from charged and neutral spectra are always in agreement within their statistical errors.  However, the mass resolution and statistics is better for the charged final states, as expected.  This leads to our decision to make resonance mass determinations from the charged decay spectra, and to fix them for the final analysis of the neutral decay spectra.  
The limited event statistics also does not allow us to make partial wave analysis of the data.
One additional observation is that the yields for radiative decays to the $\eta\eta$ final states, though suggestive of the population of $f_2'(1525)$ and $f_0(1710)$, are too small to lead to any meaningful results for the branching fractions, and we do not attempt to analyze the $\eta\eta$ spectra. 

We analyze the spectra of invariant mass of pseudoscalar pairs for the radiative decays of $J/\psi$ and $\psi(2S)$ in terms of non-interfering relativistic Breit-Wigner resonances. The resonance masses, $M$, the counts, $N$, and the background are kept free. It was not found possible to keep resonance widths also free, and they were fixed at their PDG values~\cite{pdg}.  The instrumental resolution widths, as given in Table~\ref{tbl:effs}, were convolved with the PDG widths to obtain the peak shapes used in the fits.

\begin{figure*}[!tb]

\begin{center}

\includegraphics[width=2.7in]{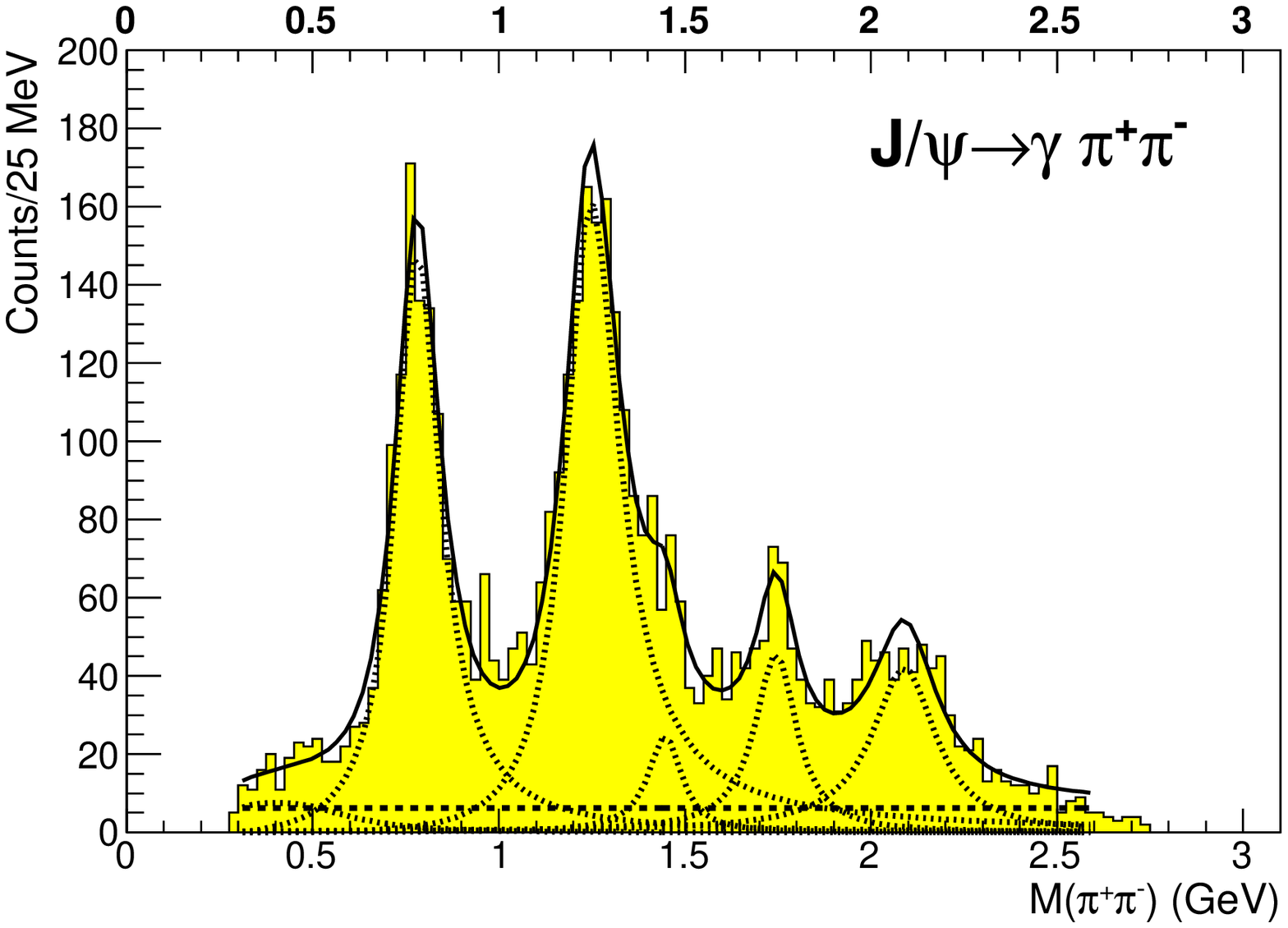}
\includegraphics[width=2.7in]{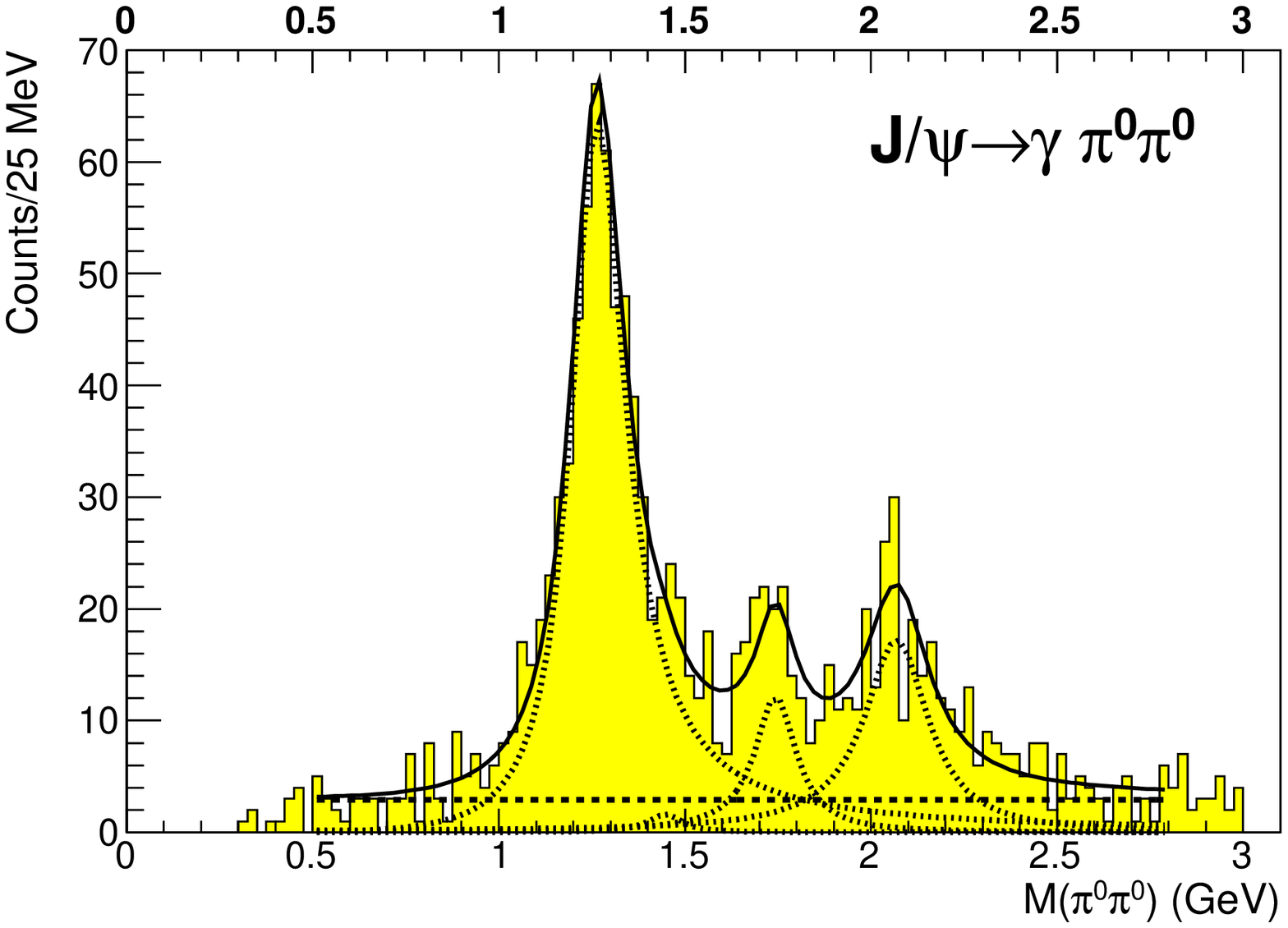}

\includegraphics[width=2.7in]{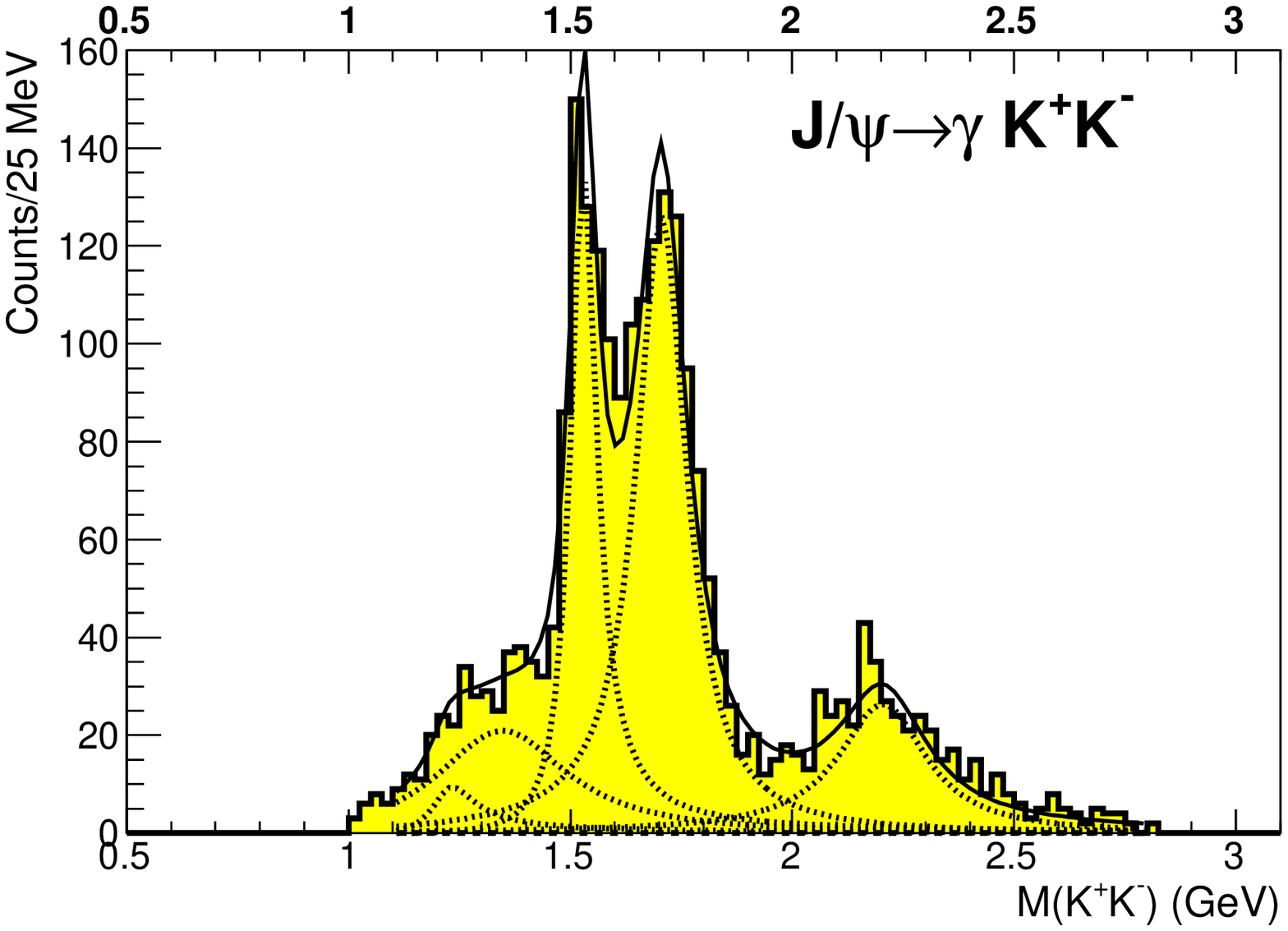}
\includegraphics[width=2.7in]{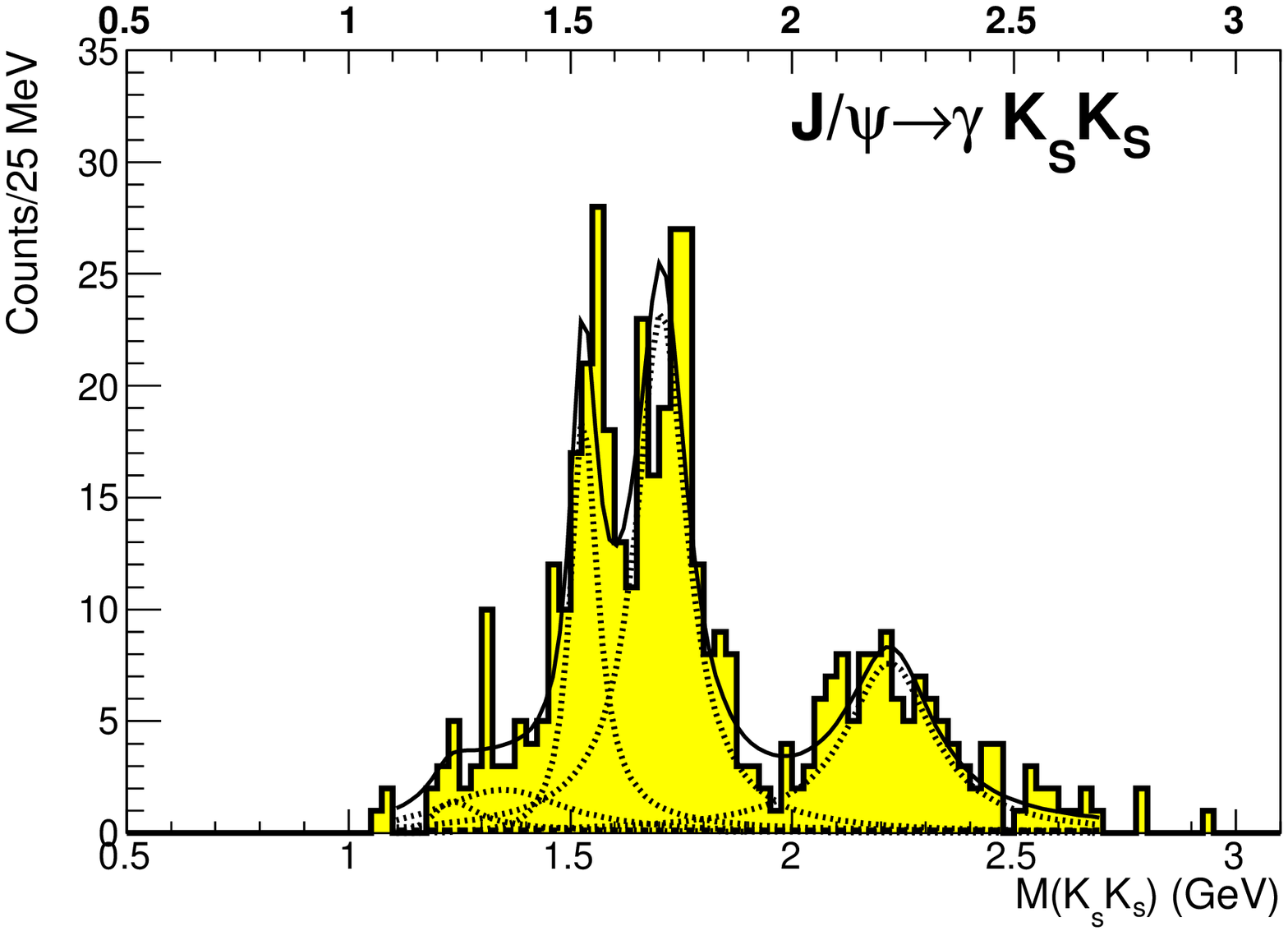}

\end{center}

\caption{Illustrating fits to the invariant mass distributions for $J/\psi \to \gamma \pi\pi$ and $\gamma K\overline{K}$.}

\label{fig:jpsifit}
\end{figure*}

We use the following relativistic Breit-Wigner lineshape
\begin{equation}
\textbf{BW}(M) \propto \frac{ [M \Gamma(M)]^2 }{ [M_0^2 - M^2]^2 + [M\Gamma(M)]^2 }
\end{equation}
where $M_0$ and $\Gamma_0$ are the non-relativistic values for the mass and width of the resonance, respectively.  The mass-dependent resonance width $\Gamma(M)$ is given by
\begin{equation}
\Gamma(M) = \Gamma_0 \left(\frac{M_0}{M}\right) \left(\frac{p}{p_0}\right)^{2l+1} \left[ \frac{B(p)}{B(p_0)} \right]
\end{equation}
where $\Gamma_0$ is the non-relativistic width, $p$ and $p_0$ are the momenta of the resonance's decay products in its center-of-mass frame for masses $M$ and $M_0$, respectively.  $B(p)$ is the $L-\text{dependent}$ Blatt-Weisskopf barrier factor, defined as $B(p)=1$, $1/\sqrt{1+(Rp)^2}$, and $1/\sqrt{1+(Rp)^2/3+(Rp)^4/9}$ for resonances with $L=0,1$, and 2, respectively, and $R=3$~GeV$^{-1}$.

The product branching fractions $\mathcal{B}_1 (J/\psi,\psi(2S) \to \gamma R)\times\mathcal{B}_2(R \to \text{PS,PS})$ are obtained as
\begin{equation}
\mathcal{B}_1 (J/\psi,\psi(2S) \to \gamma R)\times\mathcal{B}_2(R \to \text{PS,PS}) = \frac{ N_R / \epsilon_R }{N(J/\psi,\psi(2S))}
\end{equation}
where $N_R$ is the number of events in the resonance $R$, for which the efficiency is $\epsilon_R$ as given in Table~\ref{tbl:effs}, and $N(J/\psi,\psi(2S))$ are the number of $J/\psi$ and $\psi(2S)$ in the data.

\subsection{Resonances in $J/\psi$ Radiative Decays to Pseudoscalar Pairs}

The invariant mass spectra for $J/\psi$ radiative decays and fits to the observed resonances are shown in Fig.~\ref{fig:jpsifit}.  Table~\ref{tbl:jpsibrs} lists the fit results for the resonances.
We note that in general, the results shown in Table~\ref{tbl:jpsibrs} show that for both $\pi\pi$ and $KK$ the masses determined for charged and neutral decays are consistent within their respective statistical errors, and the branching fractions are consistent with the isospin expectation of factor two between charged and neutral decays.

We now discuss the results for individual resonances separately.
\vspace*{5pt}

\noindent
\textbf{The scalar $\bm{f_0(500)}$ in $\bm{J/\psi}$ decay}\\[5pt]
\noindent \textit{The $\pi\pi$ spectrum:} In the $\pi^+\pi^-$ spectrum there is a non-trivial yield in a broad bump in the low energy tail region of the strong peak of $\rho(770)$.
It is tempting to ascribe it to $f_0(500)$. However, such an assignment is unsupportable because we note that there is no evidence for a corresponding enhancement in the $\pi^0\pi^0$ spectrum which is cleaner in that mass region because of the absence of the $\rho^0$ peak. 
\vspace*{5pt}

\noindent
\textbf{The vector $\bm{\rho(770)}$ in $\bm{J/\psi}$ decay}\\[5pt]
\noindent As mentioned in Sec.~IV the prominent peak for $\rho(770)$ arises in the $\pi^+\pi^-$ spectrum because of the decay with large branching fraction $\mathcal{B}(J/\psi \to \rho^0 \pi^0) = 5.6(7) \times 10^{-3}$, $\rho^0 \to \pi^+\pi^-$, with one photon missing from the $\pi^0 \to \gamma\gamma$ decay.  
As expected, the peak is missing in the $\pi^0\pi^0$ spectrum.
\vspace*{5pt}

\begin{table*}[!tb]
\caption{Fit results for $J/\psi$ decays to $\gamma \mathrm{R}$, $\mathrm{R}\to\mathrm{PS,PS}$.  For each observed resonance $\mathrm{R}$, results are presented for decays to $\pi^+\pi^-$ ($\chi^2/d.o.f=1.36$), decays to $\pi^0\pi^0$ ($\chi^2/d.o.f.=1.00$), decays to $K^+K^-$ ($\chi^2/d.o.f.=1.38$), and decays to $K_S^0K_S^0$ ($\chi^2/d.o.f.=1.00$).  For fits to $\pi^+\pi^-$ and $K^+K^-$, the resonance masses were kept free.  For all fits their widths were fixed at their PDG2014 values.  The summed branching fractions are $\mathcal{B}_1\times\mathcal{B}_2(\pi\pi) = \mathcal{B}_1\times\mathcal{B}_2(\pi^+\pi^-) + \mathcal{B}_1\times\mathcal{B}_2(\pi^0\pi^0)$, and $\mathcal{B}_1\times\mathcal{B}_2(K\overline{K}) = [\mathcal{B}_1\times\mathcal{B}_2(K^+K^-) + \mathcal{B}_1\times\mathcal{B}_2(K_S^0K_S^0)] \times (4/3)$.}
\begin{center}
\setlength{\tabcolsep}{10pt}
\begin{tabular}{l|c|ccc}
\hline\hline
$\bm{J/\psi\to\gamma\mathrm{R}}$ & & \multicolumn{3}{c}{$\bm{\mathrm{R}\to\pi\pi}$}  \\
R, & & & & \\
$M(\mathrm{R}),~\Gamma(\mathrm{R})$, MeV &  & $M(\mathrm{MeV})$ & N & $\mathcal{B}_1\!\times\!\mathcal{B}_2\!\times\!10^5$  \\
\hline
$f_2(1270)$,  & $\pi^+\pi^-$ & $1259(4)$ &  $1722(61)$ &  $108.8(39)$  \\
$1275(1),~185.0(^{2.9}_{2.4})$  & $\pi^0\pi^0$ &  &  $688(36)$ &  $65.6(34)$  \\
  & $\pi\pi$ &  & $2410(71)$ & $174.4(52)$ \\[5pt]

$f_0(1500)$, & $\pi^+\pi^-$ & $1447(16)$ &  $163(36)$ &  $11.0(24)$    \\
$1505(6),~109(7)$  & $\pi^0\pi^0$ &  &  $11(18)$ &  $1.1(17)$ \\
  & $\pi\pi$ &  & $174(40)$ & $12.1(29)$ \\[5pt]

$f_0(1710)$, & $\pi^+\pi^-$ & $1744(7)$ &  $381(34)$ &  $27.9(25)$  \\
$1720(6),~135(8)$  & $\pi^0\pi^0$ &  &  $102(18)$ &  $9.3(16)$   \\
  & $\pi\pi$ &  & $483(38)$ & $37.2(30)$ \\[5pt]

$f_0(2100)$, & $\pi^+\pi^-$ & $2090(10)$ &  $529(39)$ &  $44.3(33)$  \\
$2103(8),~209(19)$  & $\pi^0\pi^0$ &  &  $215(23)$ &  $18.1(19)$   \\
  & $\pi\pi$ &  & $744(45)$ & $62.4(48)$ \\[5pt]

\hline
$\bm{J/\psi\to\gamma\mathrm{R}}$ &  & \multicolumn{3}{c}{$\bm{\mathrm{R}\to K\overline{K}}$}  \\
R, & & & &  \\
$M(\mathrm{R}),~\Gamma(\mathrm{R})$, MeV &  & $M(\mathrm{MeV})$ & N & $\mathcal{B}_1\!\times\!\mathcal{B}_2\!\times\!10^5$  \\
\hline
$f_0(1370)$, & $K^+K^-$ & $1360(31)$ &  $430(90)$ &  $23.1(48)$   \\
$1350(150),~346(77)$  & $K_S^0K_S^0$ &  &  $48(15)$ &  $8.3(26)$   \\
  & $K\overline{K}$ &  & $478(91)$ & $41.9(73)$ \\[5pt]

$f_2(1525)$, & $K^+K^-$ & $1532(3)$ &  $644(47)$ &  $34.8(25)$   \\
$1525(5),~73(^{6}_{5})$ & $K_S^0K_S^0$ &  &  $106(15)$ &  $18.4(24)$  \\
  & $K\overline{K}$ &  & $750(49)$ & $70.9(46)$ \\[5pt]

$f_0(1710)$, & $K^+K^-$ & $1706(4)$ &  $1004(47)$ &  $56.2(26)$  \\
$1720(6),~135(8)$ & $K_S^0K_S^0$ & &  $185(18)$ &  $32.0(31)$  \\
  & $K\overline{K}$ &  & $1189(50)$ & $117.6(54)$ \\[5pt]

$f_0(2200)$,  & $K^+K^-$ & $2206(12)$ &  $381(35)$ &  $26.1(19)$ \\
$2189(13),~238(50)$  & $K_S^0K_S^0$ & &  $109(19)$ &  $17.8(31)$   \\
  & $K\overline{K}$ &  & $490(40)$ & $58.6(49)$ \\

\hline \hline
\end{tabular}
\end{center}
\label{tbl:jpsibrs}
\end{table*}

\noindent
\textbf{The tensor resonance $\bm{f_2(1270)}$ in $\bm{J/\psi}$ decay}\\[5pt]
\noindent
\textit{The $\pi\pi$ Spectra:} This is the strongest excited peak in both the $\pi^+\pi^-$ and $\pi^0\pi^0$ spectra. Fitted separately we obtain for $\pi^+\pi^-$ decay the mass $M(\pi^+\pi^-) = 1259(4)$~MeV, and $\mathcal{B}_1\times\mathcal{B}_2 (\pi^+\pi^-) \times 10^5 = 108.8(39)$ and for $\pi^0\pi^0$ decay and $\mathcal{B}_1\times\mathcal{B}_2 (\pi^0\pi^0) \times 10^5 = 65.6(34)$.  The total branching fraction for $\pi\pi$ decay is therefore  $\mathcal{B}_1\times\mathcal{B}_2 (\pi\pi) \times  10^5 = 174.4(52)$.
These results compare with $M = 1262(8)$~MeV, $\mathcal{B}_1\times\mathcal{B}_2 (\pi^+\pi^-) \times 10^5 = 91.4(7)(148))$,  $\mathcal{B}_1\times\mathcal{B}_2 (\pi^0\pi^0) \times 10^5 = 40.0(9)(58)$, and  $\mathcal{B}_1\times\mathcal{B}_2 (\pi\pi) \times 10^5 = 131.4(11)(159)$, obtained by BES~II~\cite{besii2} for directly produced $J/\psi$. 
We note that the BES spectra have much larger backgrounds which may account for the substantial difference from our results.

\noindent
\textit{The $K\overline{K}$ Spectra:} The $f_2(1270)$ is not strongly excited in $K\overline{K}$ decays. In the $K^+K^-$ spectrum there is a prominent bump in the low energy tail of the strong peak due to the excitation of $f_2'(1525)$ resonance. It can not be fitted with a single peak corresponding to $f_2(1270)$, and requires including another larger mass resonance, for which the best candidate is the controversial $f_0(1370)$, which we discuss below. Because of the isospin difference of factor two, and the nearly factor two smaller efficiency, the corresponding enhancement the $K_S^0K_S^0$ spectrum has too small statistics to make independent analysis.
We do not fit for $f_2(1270)\to K\overline{K}$, but include its small contribution by appropriately scaling down $N_{\pi\pi}(f_2(1270))$ by the PDG~\cite{pdg} ratio $\mathcal{B}(f_2(1270)\to KK)/\mathcal{B}(f_2(1270)\to \pi\pi) = 0.054(6)$, to obtain $N_{K^+K^-}(f_2(1270))=82$ and $N_{K_S^0K_S^0}(f_2(1270))=13$.  We do not, however, report $\mathcal{B}_1\times\mathcal{B}_2(K\overline{K})$ for $f_2(1270)$.
BES~II reports that there is evidence for $f_0(1270)$ decays into $K\overline{K}$, but also does not report any numerical results~\cite{besii3}.
\vspace*{5pt}

\noindent
\textbf{The scalars $\bm{f_0(1370)}$ and $\bm{f_0(1500)}$ in $\bm{J/\psi}$ decay}\\[5pt]
\noindent
\textit{The $\pi\pi$ Spectra:} The resonance $f_0(1370)$ is considered to be the crucial ingredient in the search for scalar glueballs~[4,5]. However, its mere existence is considered questionable. As a result, PDG~(2014) estimates the mass of $f_0(1370)$ to be in the range of $1200-1500$~MeV and its width to be between $300-500$~MeV. In both the $\pi^+\pi^-$ and $\pi^0\pi^0$ spectra a small, relatively narrow, ``bump'' is visible in the high energy tail of the large $f_2(1270)$ peak.  Since the well-known resonance $f_0(1500)$ is expected in the same tail region, it is difficult to assign the observed bump uniquely to either $f_0(1370)$ or $f_0(1500)$,  This problem was also noted by Belle~\cite{belle1370}. Our attempt to make the best possible choice is based on the following exercise. 

We analyze our $\pi^+\pi^-$ spectrum assuming that a relatively narrow resonance with width between 100 and 200~MeV is responsible for the ``bump'' we observe in the high energy tail region of $f_2(1270)$. As noted before, we do not have enough statistics to fit the observed structures keeping both the mass and width of this resonance free. We therefore assume five different values of the width, $\Gamma = 100,125, 150,175$, and 200~MeV, and fit the $\pi^+\pi^-$ spectrum for the best value of the mass corresponding to each assumed value of the width. The corresponding best fit values are found to be respectively: $M\,(\text{MeV}) = 1452(13)$, $1445(15)$, $1439(17)$, $1432(19)$, and $1424(23)$.
These results indicate that the observed narrow ``bump'' can be only assigned to $f_0(1500)$ (for which many recent measurements report $M\approx1470$~MeV~\cite{pdg}), and not to $f_0(1370)$. This does not mean that a broad $f_0(1370)$ does not exist, but only that we have no evidence for its existence in our $\pi\pi$ spectra which are dominated by the $f_2(1270)$ resonance. 

Our attribution of the ``bump'' to the excitation of $f_0(1500)$, assuming the PDG~\cite{pdg} values of the width $\Gamma=109$~MeV~\cite{pdg}, leads to $M(f_0(1500)) = 1447(16)$~MeV, $\mathcal{B}_1\times\mathcal{B}_2 (\pi^+\pi^-)  \times 10^5= 11.0(24)$ and $\mathcal{B}_1\times\mathcal{B}_2 (\pi^0\pi^0)  \times 10^5= 1.1(17)$.  The summed branching fraction is $\mathcal{B}_1\times\mathcal{B}_2 (\pi\pi) \times  10^5 = 12.1(29)$.
BES~II has analyzed their data for radiative decays of directly produced $J/\psi$ into pion pairs, and reported $M(1500) = 1466(21)$~MeV, $\mathcal{B}_1\times\mathcal{B}_2 (\pi^+\pi^-)\times  10^5 = 6.7(2)(30)$, $\mathcal{B}_1\times\mathcal{B}_2 (\pi^0\pi^0)\times  10^5 = 3.4(3)(15)$, and $\mathcal{B}_1\times\mathcal{B}_2 (\pi\pi)\times  10^5 = 10.1(4)(34)$~\cite{besii3}.
\vspace*{5pt}

\noindent
\textit{The $K\overline{K}$ Spectra:} Both the $K^+K^-$ and $K_S^0K_S^0$ spectra are dominated by the $f_2'(1525)$ and $f_0(1710)$ resonances. However, there is appreciable yield spread over more than 200~MeV in the $K^+K^-$ spectrum in the region of the low energy tail of $f_2'(1525)$. To determine whether this yield should be assigned to the resonance $f_2(1270)$ or $f_0(1370)$, we have done the same exercise as described above for $f_0(1500)$.  We fit the $K^+K^-$ spectrum to determine values of masses in this region with assumed values of width, $\Gamma=200,250,300,350$, and 400~MeV.  The best fit masses are found to vary from 1330 to 1353~MeV.  Allowing the width to be a free parameter results in $M = 1360(31)$~MeV, $\Gamma = 346(77)$~MeV.  We therefore assign this enhancement to $f_0(1370)$. 
The $K^+K^-$ spectrum leads to $\mathcal{B}_1\times\mathcal{B}_2 (K^+K^-) \times10^5 = 23.1(48)$, and the $K_S^0K_S^0$ spectrum leads to $\mathcal{B}_1\times\mathcal{B}_2 (K_S^0K_S^0) \times10^5 = 8.3(26)$.  The summed result is $\mathcal{B}_1\times\mathcal{B}_2 (K\overline{K}) \times10^5 = 41.9(73)$, taking account of a factor $4/3$ for the undetected $K_L^0K_L^0$ decay, as noted for all kaon decays in the Table~\ref{tbl:jpsibrs} caption.
BES~II also observes kaon yields in this general mass region, but does not quote any numerical results~\cite{besii2}.
\vspace*{5pt}

\begin{figure*}[!tb]

\begin{center}

\includegraphics[width=2.7in]{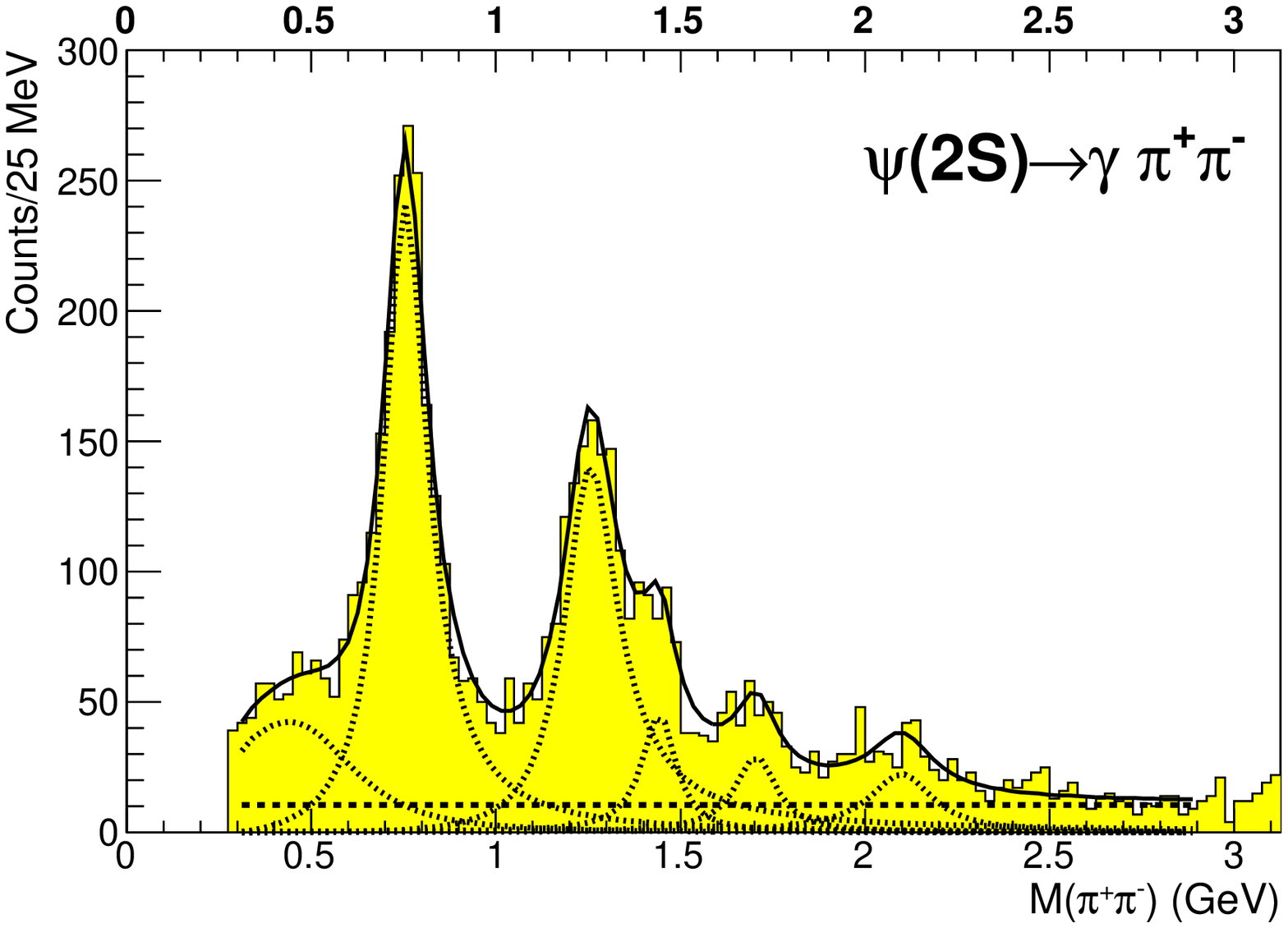}
\includegraphics[width=2.7in]{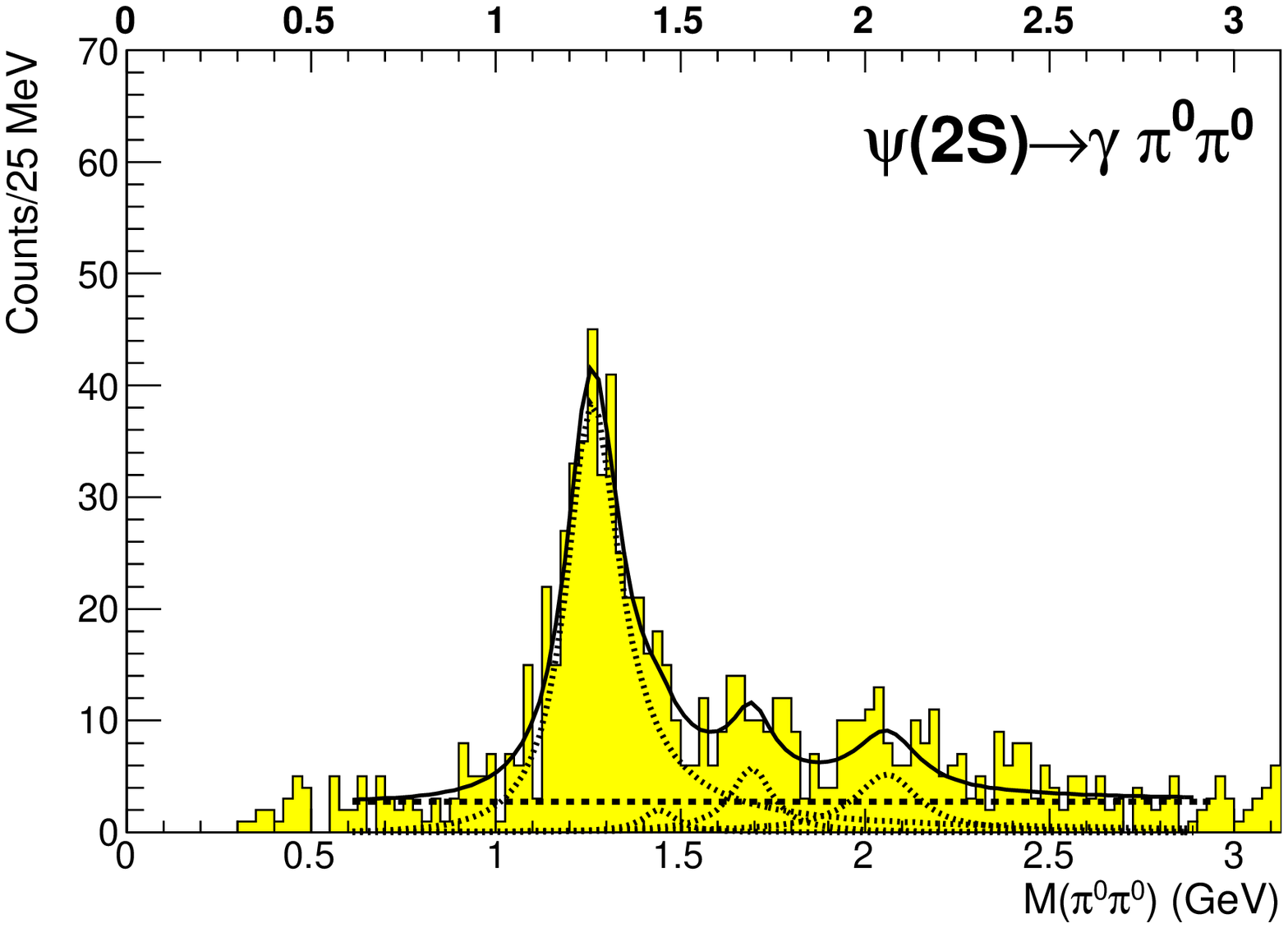}

\includegraphics[width=2.7in]{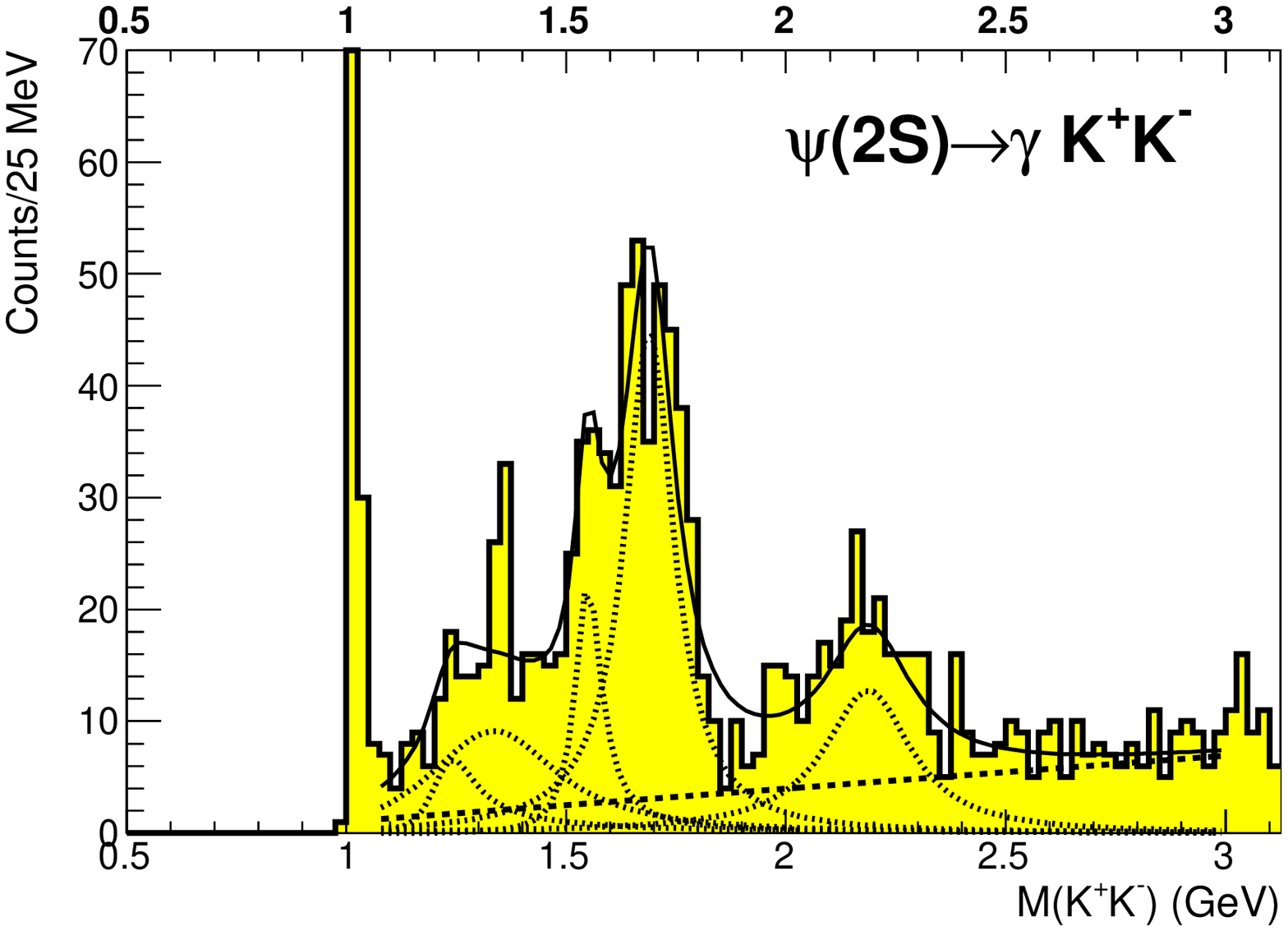}
\includegraphics[width=2.7in]{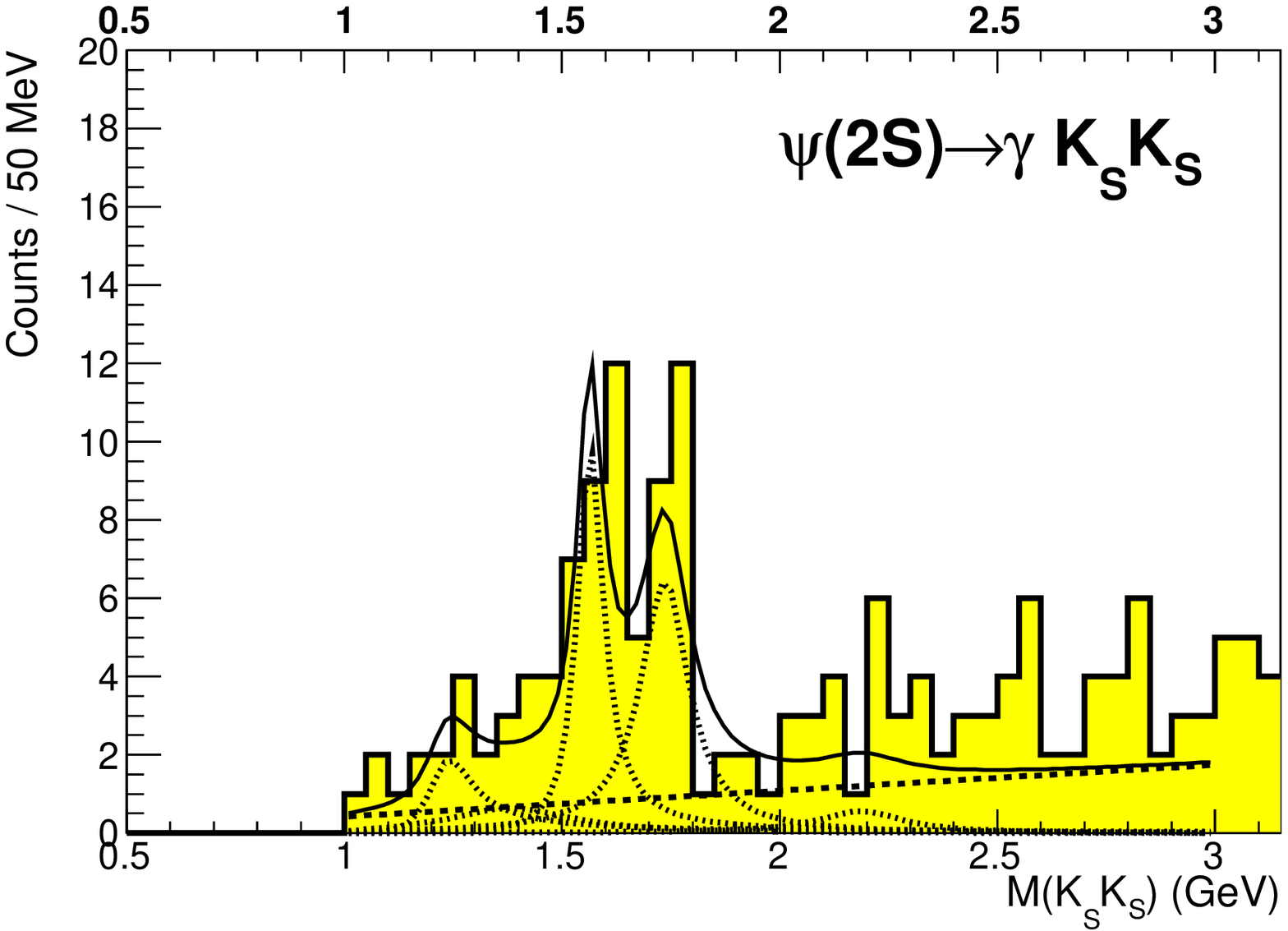}

\end{center}

\caption{Illustrating fits to the invariant mass distributions for $\psi(2S) \to \gamma \pi\pi$ and $\gamma K\overline{K}$.}

\label{fig:psipfit}
\end{figure*}

\begin{figure}[!tb]
\begin{center}
\includegraphics[width=2.7in]{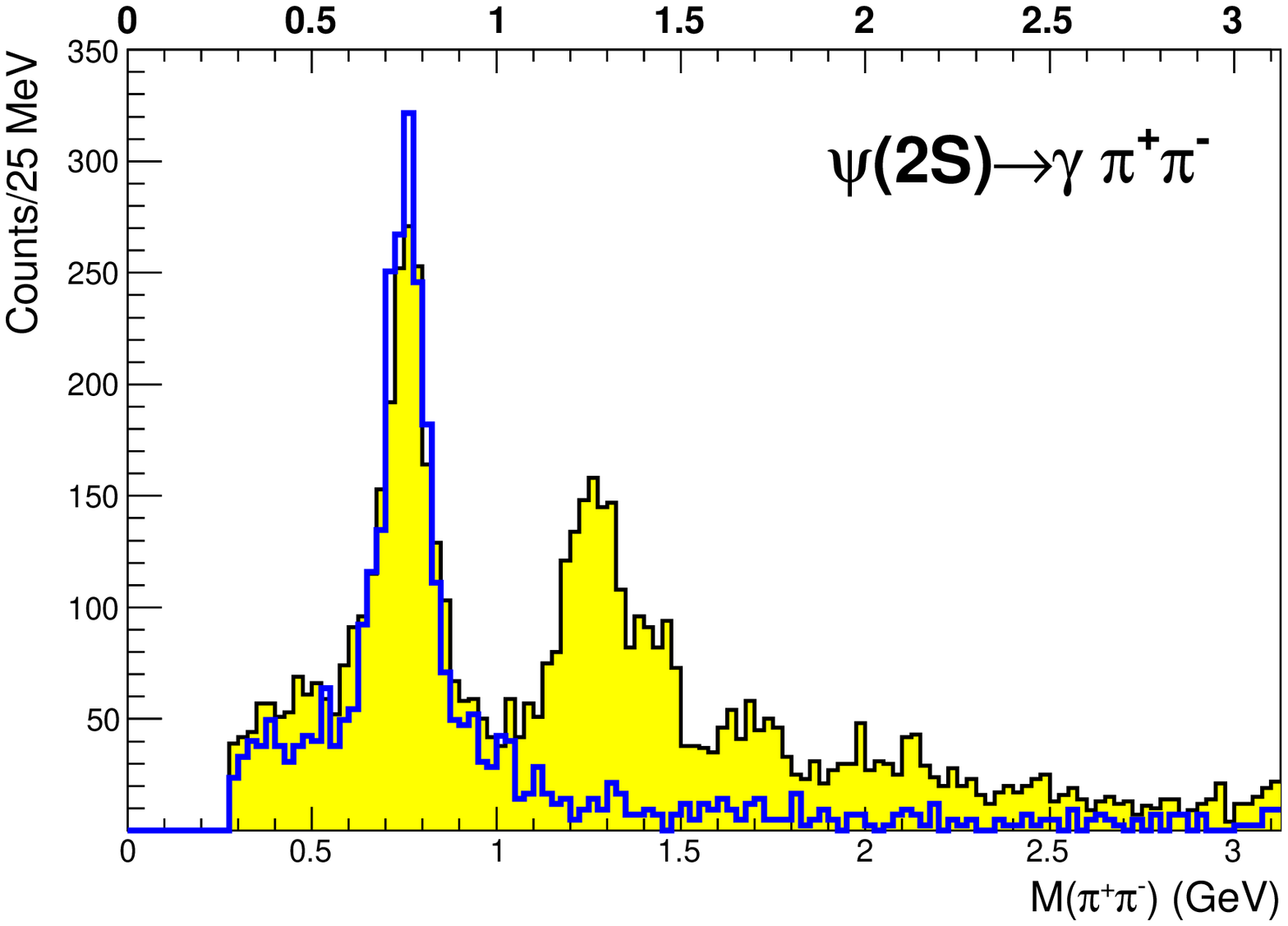}
\includegraphics[width=2.7in]{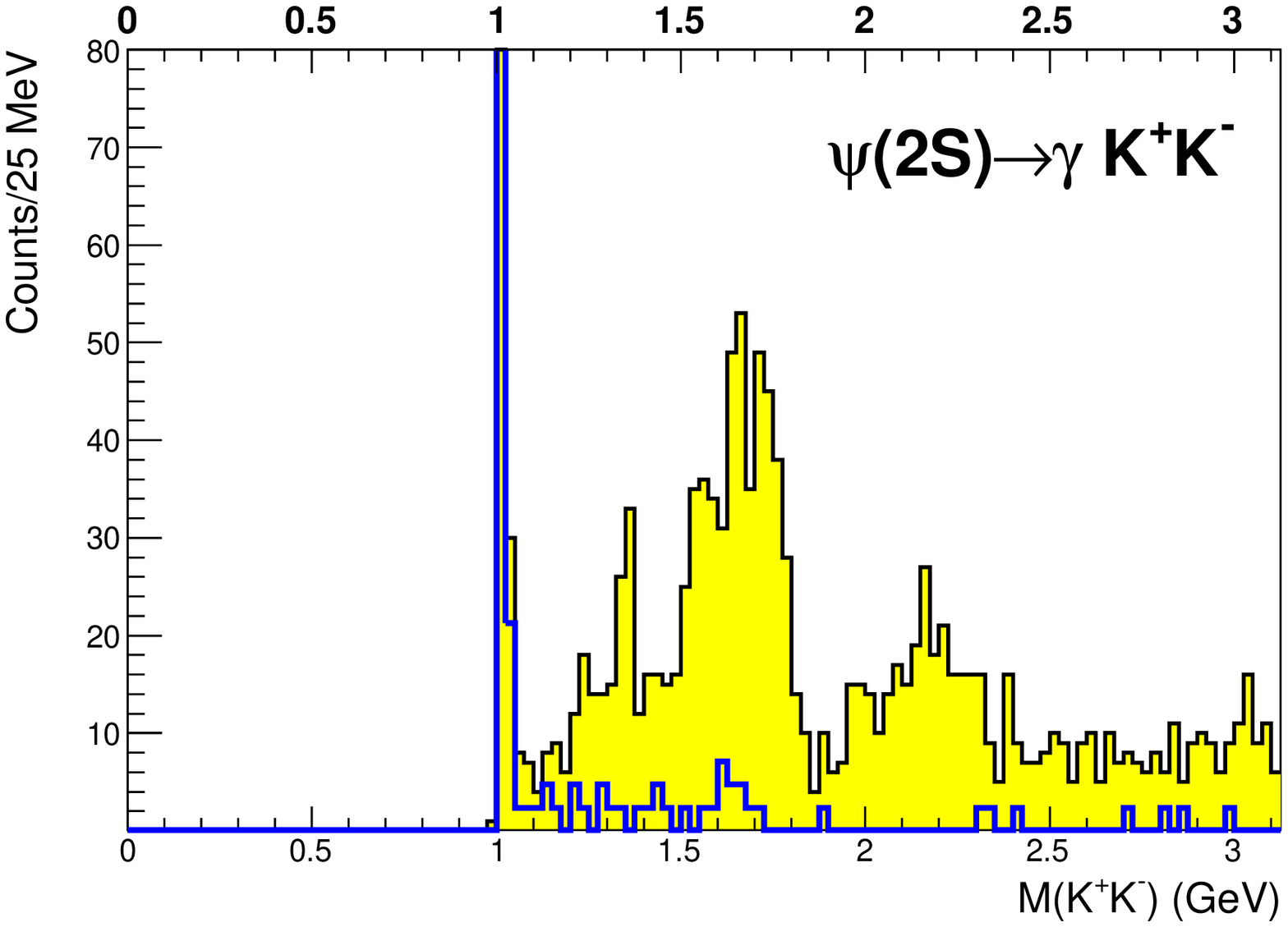}
\end{center}
\caption{Data for $\pi^+\pi^-$ and $K^+K^-$ mass distributions with superposed continuum contributions (in blue) estimated by normalizing continuum yields observed in off-resonance data taken at $\sqrt{s}\approx3871$~MeV.}
\label{fig:contspectra}
\end{figure}

\noindent
\textbf{The tensor $\bm{f_2'(1525)}$ in $\bm{J/\psi}$ decay}\\[5pt]
\noindent There is no evidence for the excitation of $f_2'(1525)$ in the $\pi\pi$ decay spectra. In the $K\overline{K}$ spectra $f_2'(1525)$ is prominently excited, with the $K^+K^-$ spectrum leading to $M(f_2'(1525)) = 1532(3)$~MeV. The $K^+K^-$ spectrum leads to $\mathcal{B}_1\times\mathcal{B}_2 (K^+K^-) \times10^5 = 34.8(25)$, and the $K_S^0K_S^0$ spectrum leads to $\mathcal{B}_1\times\mathcal{B}_2 (K_S^0K_S^0) \times10^5 = 18.4(24)$.
The summed result is $\mathcal{B}_1\times\mathcal{B}_2 (K\overline{K}) \times10^5 = 70.9(46)$.
The result from BES~II with large background in the invariant mass spectra is $\mathcal{B}_1\times\mathcal{B}_2 (K\overline{K}) \times10^5= 34.2(15)(^{69}_{65})(^{155}_{0})$~\cite{besii3}.  
\vspace*{5pt}

\noindent
\textbf{The scalar $\bm{f_0(1710)}$ in $\bm{J/\psi}$ decay}\\[5pt]
\noindent
The scalar $f_0(1710)$ is prominently excited in both $\pi\pi$ decays and $K\overline{K}$ decays (Fig.~\ref{fig:jpsifit}). The $\pi^+\pi^-$ spectrum leads to the mass $M(f_0(1710)) = 1744(7)$~MeV and $\mathcal{B}_1\times\mathcal{B}_2(\pi^+\pi^-)\times10^5  = 27.9(25)$, and the $\pi^0\pi^0$ spectrum leads to $\mathcal{B}_1\times\mathcal{B}_2(\pi^0\pi^0)\times10^5  = 9.3(16)$, with the sum $\mathcal{B}_1\times\mathcal{B}_2 (\pi\pi) \times  10^5 = 37.2(30)$.  BES~II has reported $\mathcal{B}_1\times\mathcal{B}_2(\pi^+\pi^-)\times10^5  = 26.4(4)(75)$ and $\mathcal{B}_1\times\mathcal{B}_2(\pi^0\pi^0)\times10^5  = 13.3(5)(88)$, and therefore  $\mathcal{B}_1\times\mathcal{B}_2(\pi\pi)\times10^5  = 39.7(6)(116)$~\cite{besii2}.
\vspace*{5pt}

\noindent
\textit{The $K\overline{K}$ Spectra:} The $K^+K^-$ spectrum leads to $M(1710) = 1706(4)$~MeV, and $\mathcal{B}_1\times\mathcal{B}_2 (K^+K^-) \times10^5= 56.2(26)$, the $K_S^0K_S^0$ spectrum leads to $\mathcal{B}_1\times\mathcal{B}_2 (K_S^0K_S^0) \times10^5= 32.0(31)$. 
Our summed result is $\mathcal{B}_1\times\mathcal{B}_2 (K\overline{K}) \times10^5 = 117.6(54)$.
The corresponding result of BES~II is $\mathcal{B}_1\times\mathcal{B}_2 (K\overline{K}) \times10^5= 96.2(29)(^{211}_{186})(^{281}_{0})$~\cite{besii3}.  

Of the seven resonances listed in Table~\ref{tbl:jpsibrs}, unambiguous results for both $\pi\pi$ and $K\overline{K}$ decays are observed for only one resonance, $f_0(1710)$. This is largely due to the fact that all the resonances have widths much larger than the experimental resolution widths, and they strongly overlap.  As a result, as described in Sec.~IX, only for $f_0(1710)$ we are able to test the flavor independent character attributed to pure glueballs.

Both $K^+K^-$ and $K_S^0K_S^0$ spectra are very well defined in their high energy tail region, and show no evidence for enhancement in the $M=1790$ region which has been occasionally reported.
\vspace*{5pt}

\noindent
\textbf{The $\bm{M = 1800-2500}$~MeV region in $\bm{J/\psi}$ decay}\\[5pt]
\noindent
The mass region above $f_0(1710)$ is complicated with several $f_0$, and $f_2$ resonances reported~\cite{pdg}. In particular, radiative decays of $J/\psi$ have been reported to $f_0(1800)$, $f_0(2200)$, $f_2(1910)$, $f_2(1950)$, and to the much talked about tensor $f_2(2220)$. In the $\pi\pi$ and $K\overline{K}$ mass spectra in the present investigation a broad enhancement at $M \approx 2100$~MeV is seen in the $\pi\pi$ decays, and another broad enhancement at $M \approx 2200$~MeV is seen in the $K\overline{K}$ decays.  The background is generally much larger than in the lower mass regions.  This makes resonance assignments in this mass region rather speculative. With this caveat we fit the $\pi\pi$ spectra in this region with $f_0(2100)$ with the PDG width of 209~MeV, and the $K\overline{K}$ spectra with $f_0(2200)$ with the PDG width of 238~MeV.
Fits to the $\pi\pi$ spectra lead to $M(f_0(2100)) = 2090(10)$~MeV,  $\mathcal{B}_1\times\mathcal{B}_2(\pi^+\pi^-)\times10^5  = 44.3(33)$,  $\mathcal{B}_1\times\mathcal{B}_2(\pi^0\pi^0)\times10^5  = 18.1(19)$,  and the sum $\mathcal{B}_1\times\mathcal{B}_2 (\pi\pi) \times10^5 = 62.4(48)$. 
Fits to the $K\overline{K}$ spectra lead to $M(f_0(2200)) = 2206(12)$~MeV,  $\mathcal{B}_1\times\mathcal{B}_2 (K^+K^-) \times10^5= 26.1(19)$, and $\mathcal{B}_1\times\mathcal{B}_2 (K_S^0K_S^0) \times10^5= 17.8(51)$, and the total $\mathcal{B}_1\times\mathcal{B}_2 (K\overline{K}) \times10^5 = 58.6(49)$.

\begin{table*}[!tb]

\caption{Fit results for $\psi(2S)$ decays to $\gamma \mathrm{R}$, $\mathrm{R}\to\mathrm{PS,PS}$.  For each observed resonance $\mathrm{R}$, results are presented for decays to $\pi^+\pi^-$ ($\chi^2/d.o.f.=1.24$), decays to $\pi^0\pi^0$ ($\chi^2/d.o.f.=1.19$), decays to $K^+K^-$ ($\chi^2/d.o.f.=1.29$), and decays to $K_S^0K_S^0$ ($\chi^2/d.o.f.=0.94$.  For fits to $\pi^+\pi^-$ and $K^+K^-$, the resonance masses were kept free.  For all fits their widths were fixed at their PDG2014 values. The summed branching fractions are $\mathcal{B}_1\times\mathcal{B}_2(\pi\pi) = \mathcal{B}_1\times\mathcal{B}_2(\pi^+\pi^-) + \mathcal{B}_1\times\mathcal{B}_2(\pi^0\pi^0)$ and $\mathcal{B}_1\times\mathcal{B}_2(K\overline{K}) = [\mathcal{B}_1\times\mathcal{B}_2(K^+K^-) + \mathcal{B}_1\times\mathcal{B}_2(K_S^0K_S^0)] \times (4/3)$. }

\begin{center}
\setlength{\tabcolsep}{10pt}
\begin{tabular}{l|c|ccc}
\hline\hline
$\bm{\psi(2S)\to\gamma\mathrm{R}}$ & & \multicolumn{3}{c}{$\bm{\mathrm{R}\to\pi\pi}$}  \\
R, & & & & \\
$M(\mathrm{R}),~\Gamma(\mathrm{R})$, MeV &  & $M(\mathrm{MeV})$ & N & $\mathcal{B}_1\!\times\!\mathcal{B}_2\!\times\!10^5$ \\
\hline
$f_2(1270)$,  & $\pi^+\pi^-$ & $1267(4)$ &  $1509(63)$ &  $17.0(7)$   \\
$1275(1),~185.0(^{2.9}_{2.4})$  & $\pi^0\pi^0$ &  &  $415(30)$ &  $6.9(5)$   \\
  & $\pi\pi$ &  & $1924(70)$ & $23.9(9)$ \\[5pt]

$f_0(1500)$,  & $\pi^+\pi^-$ & $1442(9)$ &  $261(40)$ &  $3.0(5)$  \\
$1505(6),~109(7)$  & $\pi^0\pi^0$ &  &  $13(17)$ &  $0.2(3)$   \\
  & $\pi\pi$ &  & $274(43)$ & $3.2(6)$ \\[5pt]

$f_0(1710)$,  & $\pi^+\pi^-$ & $1705(11)$ &  $237(30)$ &  $2.8(4)$     \\
$1720(6),~135(8)$  & $\pi^0\pi^0$ &  &  $53(14)$ &  $0.8(2)$   \\
  & $\pi\pi$ &  & $290(33)$ & $3.6(4)$ \\[5pt]

$f_0(2100)$,  & $\pi^+\pi^-$ & $2099(17)$ &  $283(35)$ &  $3.4(4)$  \\
$2103(8),~209(19)$  & $\pi^0\pi^0$ &  &  $90(17)$ &  $1.4(3)$  \\
  & $\pi\pi$ &  & $373(39)$ & $4.8(5)$ \\[5pt]

\hline
$\bm{\psi(2S)\to\gamma\mathrm{R}}$ &  & \multicolumn{3}{c}{$\bm{\mathrm{R}\to K\overline{K}}$}  \\
R, & & & &  \\
$M(\mathrm{R}),~\Gamma(\mathrm{R})$, MeV &  & $M(\mathrm{MeV})$ & N & $\mathcal{B}_1\!\times\!\mathcal{B}_2\!\times\!10^5$ \\
\hline
$f_0(1370)$, & $K^+K^-$ & $1350(48)$ &  $168(50)$ &  $1.9(6)$    \\
$1350(150),~346(77)$   & $K_S^0K_S^0$ & &  $7(12)$ & $0.4(4)$  \\
  & $K\overline{K}$ &  & $175(51)$ & $3.1(10)$ \\[5pt]

$f_2(1525)$, & $K^+K^-$ & $1552(9)$ &  $113(24)$ &  $1.3(3)$   \\
$1525(5),~73(^{6}_{5})$  & $K_S^0K_S^0$ & &  $23(8)$ &  $0.9(3)$ \\
  & $K\overline{K}$ &  & $136(25)$ & $2.9(6)$ \\[5pt]

$f_0(1710)$,  & $K^+K^-$ & $1690(8)$ &  $349(30)$ &  $4.1(3)$   \\
$1720(6),~135(8)$  & $K_S^0K_S^0$ &  & $26(8)$ &  $0.9(3)$   \\
  & $K\overline{K}$ &  & $375(31)$ & $6.7(6)$ \\[5pt]

$f_0(2200)$,  & $K^+K^-$ & $2188(17)$ &  $203(30)$ &  $2.3(3)$    \\
$2189(11),~238(50)$  & $K_S^0K_S^0$ &  & $4(9)$  & $0.1(3)$  \\
  & $K\overline{K}$ &  & $207(31)$ & $3.2(6)$ \\[5pt]

\hline \hline
\end{tabular}
\end{center}
\label{tbl:psipbrs}
\end{table*}

\subsection{Resonances in $\psi(2S)$ Radiative Decays to Pseudoscalar Pairs}

While searches for glueballs in radiative decays of $J/\psi$ have been reported before, similar searches in $\psi(2S)$ radiative decays have not been reported before. The only reports for $\psi (2S)$ radiative decays  which exist are a BES~II publication with a sample of 4.02~million~$\psi(2S)$~\cite{besii2}, and an unpublished report with 14~million~$\psi(2S)$~\cite{besii4}. We present in this paper our results for the radiative decays observed in a sample of 24.5~million~$\psi(2S)$.

As pointed out in Sections~IV and V, and shown in Fig.~7, the results for $\psi(2S)$ decays are very similar to those of $J/\psi$ decays. The event statistics is smaller, due primarily to the ``13\% rule'', and the backgrounds are relatively larger. The fits to the invariant mass spectra, and the determination of the product branching fractions for the $\psi(2S)$ decays was done in the same manner as for $J/\psi$ decays. The results are tabulated in Table~\ref{tbl:psipbrs}, and the fits to the charged and neutral decays for $\pi\pi$ and $K\overline{K}$ spectra are shown in Fig.~\ref{fig:psipfit}.
\vspace*{5pt}

\noindent \textbf{The vectors $\bm{\rho(770)}$ and $\bm{\phi(1020)}$ in $\bm{\psi(2S)}$ decay}\\[5pt]
\noindent
In the $\pi^+\pi^-$ and $K^+K^-$ invariant mass spectra we note the presence of the $\rho(770)$ and $\phi(1020)$ peaks as also indicated by the diagonal enhancements in the corresponding Dalitz plots of Fig.~\ref{fig:dalitz2}.  These arise due to the production of $\rho(770)$  and $\phi(1020)$ vector mesons via initial state radiation.  To confirm this, we have analyzed our off-$\psi(2S)$ resonance data at $\sqrt{s}=3671$~MeV with the integrated luminosity of 20.6~pb$^{-1}$.  To compare the $\pi^+\pi^-$ and $K^+K^-$ yields observed in these data with those observed in $\psi(2S)$ data we multiply the yields by the relative luminosity factor $\mathcal{L}(\psi(2S))/\mathcal{L}(3671) = 53/20.6 = 2.57$, and the factor $(s'/s)^3 = (3686/3671)^3 = 1.06$.  The normalized yields are shown in Fig.~\ref{fig:contspectra} superposed on the $\psi(2S)$ data.  It is seen that the normalized contributions fully account for the yield of $\rho(770)$ in the $\pi^+\pi^-$ spectrum, and for the yield of $\phi(1020)$ in the $K^+K^-$ spectrum.  It is also seen that the background contribution due to this ISR yield is predicted to be negligible for masses $>1$~GeV.
\vspace*{5pt}

\noindent \textbf{The tensor $\bm{f_2(1270)}$ in $\bm{\psi(2S)}$ decay} \\[5pt]
\noindent
As in $J/\psi$ decay, the $f_2(1270)$ is prominently excited in both $\pi^+\pi^-$ and $\pi^0\pi^0$ decays. On the high energy shoulders in both spectra we again see enhancement which is well fitted with the narrow $f_0(1500)$ rather than narrow or wide $f_0(1370)$.
The fit to $f_2(1270)$ results in $M=1267(4)$~MeV, $\mathcal{B}_1\times\mathcal{B}_2(\pi^+\pi^-)\times10^5  = 17.0(7)$, $\mathcal{B}_1\times\mathcal{B}_2(\pi^0\pi^0)\times10^5  = 6.9(5)$, and the sum $\mathcal{B}_1\times\mathcal{B}_2 (\pi\pi) \times  10^5 = 23.9(9)$.

In the $K^+K^-$ spectrum there is a broad enhancement below 1500~MeV which cannot be assigned to $f_2(1270)$. Instead, it is well fitted with a wide resonance with mass around 1350~MeV, as discussed below.
\vspace*{5pt}

\noindent \textbf{The scalar $\bm{f_0(1370)}$ in $\bm{\psi(2S)}$ decay} \\[5pt]
\noindent
In the $J/\psi$ decay to $\pi\pi$ it was not possible to discern the possible excitation of $f_0(1370)$ because of the strong excitation of $f_2(1270)$.  As described in Sec.~V.A, the small shoulder seen on the high energy tail of $f_2(1270)$ could be assigned to $f_0(1500)$ but not to $f_0(1370)$. In the $K\overline{K}$ spectra for $\psi(2S)$ the $f_2(1270)$ excitation is much weaker. This allows us the opportunity to fit the mass region below 1525~MeV. 
The $\psi(2S)$ spectra have much smaller statistics than the $J/\psi$ spectra, which makes it difficult to determine the properties of a wide, weakly excited resonance like the $f_0(1370)$.  To explore its possible excitation, we have fitted the $K^+K^-$ spectrum with the width of $f_0(1370)$ fixed to $\Gamma=250,300,350$~MeV.  The resulting masses were $M=1356(41)$, $1353(45)$, $1349(48)$~MeV, i.e., in essential agreement with the determination of $M=1360(31)$~MeV and $\Gamma=346$~MeV for $J/\psi$ decay.

For our final fit, we therefore assume a state with width of 346~MeV. The fit to the $K^+K^-$ spectrum yields a mass of $M = 1350(48)$~MeV.  We identify it with $f_0(1370)$.  The fit results to $\mathcal{B}_1\times\mathcal{B}_2 (K^+K^-) \times10^5= 1.9(6)$, $\mathcal{B}_1\times\mathcal{B}_2 (K_S^0K_S^0) \times10^5= 0.4(4)$, and the total $\mathcal{B}_1\times\mathcal{B}_2 (K\overline{K}) \times10^5 = 3.1(10)$.
This fit has been made excluding the two bins near $M\sim1350$~MeV, which would correspond to a very narrow state of $\Gamma<50$~MeV, for which no plausible candidate exists.
\vspace*{5pt}

\noindent \textbf{The scalar $\bm{f_0(1500)}$ in $\bm{\psi(2S)}$ decay} \\[5pt]
\noindent
The absence of $f_2'(1525)$ in $\pi\pi$ decay allows us to fit for $f_0(1500)$ with the results: $M(f_0(1500)) = 1442(9)$~MeV,  $\mathcal{B}_1\times\mathcal{B}_2(\pi^+\pi^-)\times10^5  = 3.0(5)$, $\mathcal{B}_1\times\mathcal{B}_2(\pi^0\pi^0)\times10^5  = 0.2(3)$,  and the sum $\mathcal{B}_1\times\mathcal{B}_2 (\pi\pi) \times10^5 = 3.2(6)$.

The strong excitation of $f_2'(1525)$ in $K\overline{K}$ decays makes it impossible to say anything about possible excitation of $f_0(1500)$ in $K\overline{K}$ decays.  
\vspace*{5pt}

\noindent \textbf{The tensor $\bm{f_2'(1525)}$ in $\bm{\psi(2S)}$ decay} \\[5pt]
\noindent
No evidence for the $\pi\pi$ decay of $f_2'(1525)$ was found in the radiative decay of $J/\psi$,  and none is found in the radiative decay $\psi(2S)\to\gamma\pi\pi$, either. In $K^+K^-$ decay strong excitation of $f_2'(1525)$ is observed, and the fit to the $K^+K^-$ spectrum leads to $M(f_2'(1525)) = 1552(9)$~MeV and $\mathcal{B}_1\times\mathcal{B}_2 (K^+K^-) \times10^5= 1.3(3)$. The fit to the $K_S^0K_S^0$ spectrum leads to $\mathcal{B}_1\times\mathcal{B}_2 (K_S^0K_S^0) \times10^5= 0.9(3)$.  The total branching fraction is $\mathcal{B}_1\times\mathcal{B}_2 (K\overline{K}) \times10^5 = 2.9(6)$.
\vspace*{5pt}

\begin{table*}[!tb]
\caption{Upper limits at 90\% confidence level for the production of the presumed tensor glueball candidate $\xi(2230)$ in the reactions $J/\psi,\psi(2S)\to\gamma\mathrm{R},~\mathrm{R}\to\mathrm{PS,PS}$.}
\begin{center}
\begin{tabular}{llc|cc|c}
\hline\hline
  & & & \multicolumn{2}{c|}{ 90\% CL UL } & \\
  & & $\Gamma(\xi)$, MeV & N & $\mathcal{B}_1\times\mathcal{B}_2$ & $\mathcal{B}_1\times\mathcal{B}_2$ (ref.) \\
  & & &  & $\times10^5$ & $\times10^5$ \\
\hline
$J/\psi$ & $\to\gamma(\pi^+\pi^-)$ & 20/50 & $23.0/45.1$ & $2.6/5.2$ & $5.6(^{27}_{26})$~\cite{besii} \\
         & $\to\gamma(\pi^0\pi^0)$ & 20/50 & $16.0/23.2$ & $1.3/1.9$ & --- \\
         & $\to\gamma(K^+K^-)$ & 20/50 & $25.4/44.6$ & $1.7/3.1$ & $3.3(^{16}_{13})$~\cite{besii}, $4.2(^{19}_{16})$~\cite{markiii} \\ 
         &                     &       &         &        & $2.3$ (95\% CL)~\cite{dm2} \\
         & $\to\gamma(K_S^0K_S^0)$ & 20/50 & $11.5/19.1$ & $1.2/2.0$ & $3.1(^{17}_{15})$~\cite{besii}, $2.7(^{14}_{12})$~\cite{markiii} \\ 
         &                     &       &         &        & $1.6$ (95\% CL)~\cite{dm2} \\
$\psi(2S)$&$\to\gamma(\pi^+\pi^-)$ & 20/50 & $22.9/31.1$ & $0.32/0.43$ & --- \\
         & $\to\gamma(\pi^0\pi^0)$ & 20/50 & $15.7/23.6$ & $0.26/0.40$ & --- \\
         & $\to\gamma(K^+K^-)$ & 20/50 & $18.9/38.0$ & $0.21/0.43$ & --- \\
         & $\to\gamma(K_S^0K_S^0)$ & 20/50 & $10.7/15.6$ & $0.37/0.55$ & --- \\
\hline\hline
\end{tabular}
\end{center}
\label{tbl:xiuls}
\end{table*}

\begin{table*}[!tb]

\caption{Summary of results for $J/\psi$ and $\psi(2S)$ radiative decays to light quark resonances.  For entries with two uncertainties, the first listed uncertainties are statistical and the second are systematic as described in Sec.~VIII and listed in Tables~VIII and IX.}

\begin{center}
\begin{tabular}{l@{\hspace*{10pt}}cc@{\hspace*{20pt}}cc@{\hspace*{20pt}}c}
\hline\hline
  & \multicolumn{2}{c}{\textbf{From} $\bm{J/\psi}$\hspace*{2em}} 
  & \multicolumn{2}{c}{\textbf{From} $\bm{\psi(2S)}$\hspace*{1em}} 
  & \parbox{2.5cm}{$$\bm{\frac{\mathcal{B}_2(\psi(2S))}{\mathcal{B}_2(J/\psi)}}$$} \\
 & $M$ (MeV) & $\mathcal{B}_1\times\mathcal{B}_2\times10^5$  
 & $M$ (MeV) & $\mathcal{B}_1\times\mathcal{B}_2\times10^5$ & (\%) \\ 
\hline
$f_2(1270)\to\pi\pi$ & 1259(4)(4) & 174.4(52)(122) & 1267(4)(3) & 23.9(9)(9) & 13.7(7)(11) \\

$f_0(1370)\to K\overline{K}$ & 1360(31)(28) & 41.9(73)(134) & 1350(48)(15) & 3.1(10)(14) & 7.4(27)(41) \\
$f_0(1500)\to\pi\pi$ & 1447(16)(13) & 12.1(29)(24) & 1442(9)(4) & 3.2(6)(2) & 26.4(80)(55) \\

$f_2'(1525)\to K\overline{K}$ & 1532(3)(6) & 70.9(46)(67) & 1557(9)(3) & 2.9(6)(3) & 4.1(9)(6) \\

$f_0(1710)\to\pi\pi$ & 1744(7)(5) & 37.2(30)(43) & 1705(11)(5) & 3.6(4)(5) & 9.7(13)(18) \\

\multicolumn{1}{r}{$\to K\overline{K}$}  & 1706(4)(5) & 117.6(54)(94) & 1690(8)(3) & 6.7(6)(6) & 5.7(6)(7) \\

$f_0(2100)\to\pi\pi$ & 2090(10)(6) & 62.4(48)(87) & 2099(17)(8) & 4.8(5)(9) & 7.7(10)(18) \\

$f_0(2200)\to K\overline{K}$ & 2206(12)(8) & 58.6(49)(120) & 2188(17)(16) & 3.2(6)(8) & 5.5(11)(18) \\

\hline \hline
\end{tabular}
\end{center}
\label{tbl:compare}

\end{table*}

\noindent \textbf{The scalar $\bm{f_0(1710)}$ in $\bm{\psi(2S)}$ decay} \\[5pt]
\noindent
The scalar $f_0(1710)$ is strongly excited in both $\pi\pi$ and $K\overline{K}$ decays.
Fits to the $\pi\pi$ spectra lead to $M(f_0(1710)) = 1705(11)$~MeV,  $\mathcal{B}_1\times\mathcal{B}_2(\pi^+\pi^-)\times10^5  = 2.8(4)$, and $\mathcal{B}_1\times\mathcal{B}_2(\pi^0\pi^0)\times10^5  = 0.8(2)$,  and the sum $\mathcal{B}_1\times\mathcal{B}_2 (\pi\pi) \times10^5 = 3.6(4)$.
 Fits to the $K\overline{K}$ spectra lead to $M(f_0(1710)) = 1690(8)$~MeV,  $\mathcal{B}_1\times\mathcal{B}_2 (K^+K^-) \times10^5= 4.1(3)$,  $\mathcal{B}_1\times\mathcal{B}_2 (K_S^0K_S^0) \times10^5= 0.9(3)$,  and the total $\mathcal{B}_1\times\mathcal{B}_2 (K\overline{K}) \times10^5 = 6.7(6)$.
\vspace*{5pt}

\noindent \textbf{The $\bm{M = 1800 -2500}$ MeV region in $\bm{\psi(2S)}$ decay} \\[5pt]
\noindent
As mentioned for $J/\psi$ decays, in this region backgrounds are large and only a large, wide peak is observed in both $\pi\pi$ and $K\overline{K}$ spectra. Each of them may be composite of the excitation of more than one resonance. If analyzed as single resonances, fits to the $\pi\pi$ spectra lead to $M(f_0(2100)) = 2099(17)$~MeV,  $\mathcal{B}_1\times\mathcal{B}_2(\pi^+\pi^-)\times10^5  = 3.4(4)$, $\mathcal{B}_1\times\mathcal{B}_2(\pi^0\pi^0)\times10^5  = 1.4(3)$,  and the sum $\mathcal{B}_1\times\mathcal{B}_2 (\pi\pi) \times10^5 = 4.8(5)$. Anisovich and Uman have reported the observation of a resonance with this mass in $p\bar{p}$ annihilations~\cite{pdg}.  
Fits to the $K\overline{K}$ spectra lead to $M(f_0(2200)) = 2188(17)$~MeV,  $\mathcal{B}_1\times\mathcal{B}_2 (K^+K^-) \times10^5= 2.3(3)$,  $\mathcal{B}_1\times\mathcal{B}_2 (K_S^0K_S^0) \times10^5= 0.1(3)$, and the total $\mathcal{B}_1\times\mathcal{B}_2 (K\overline{K}) \times10^5 = 3.2(6)$.  Augustin~\cite{dm2} has reported excitation of a resonance with this mass in radiative decay of $J/\psi$.

\subsection{The $\xi(2230)$ Tensor Glueball Candidate}

While the main interest in glueball searches has been in the scalars, special interest has existed in the claim for a narrow tensor state with mass $\sim2230$~MeV, and width $\Gamma\approx20-30$~MeV.  This claim has a checkered history.  Originally claimed by Mark~III~\cite{markiii} in the radiative decay of $J/\psi$, and the decay of $\xi(2230)$ into $K^+K^-$ and $K_S^0K_S^0$, it has gone through several claimed observations~\cite{markiii,besii,itep}, and a number of non-observations~\cite{dm2,besiii,cb2,cleoxi,babarxi}.  The latest of these consist of its observation by ITEP in $\pi^-p$ formation and decay into $K_S^0K_S^0$~\cite{itep}, and its non-observation by BaBar in the radiative decay of $J/\psi$ into $K^+K^-$ and $K_S^0K_S^0$~\cite{babarxi}, and its non-observation by BES~III in the radiative decay of $J/\psi$ and decay into $\eta\eta$~\cite{besiii}.  The present measurement allow us to search for it in the radiative decays of both $J/\psi$ and $\psi(2S)$, with its decays into $\pi\pi$, $K\overline{K}$, and $\eta\eta$ pseudoscalar pairs.  In the analyses of the invariant mass spectra as described above, we add a resonance with $M=2230$~MeV, and widths of 20 and 50~MeV, and determine upper limits at 90\% confidence level  for the product branching fractions for the existence of $\xi(2230)$.  The results are listed in Table~\ref{tbl:xiuls}.  They can be summarized as upper limits of $\le5\times10^{-5}$ in $J/\psi$ decays and $<6\times10^{-6}$ in $\psi(2S)$ decays.  Our limits in $J/\psi$ decays are consistent with $\le 2.3 \times10^{-5}$ reported by DM2~\cite{dm2} for $K^+K^-$ and $K_S^0K_S^0$ decays, and $\le4\times10^{-5}$ reported by BaBar~\cite{babarxi} in $K^+K^-$ and $K_S^0K_S^0$ decays. No evidence for its existence was found by BES~III in radiative decay of a very large sample of $2.25\times10^8$~$J/\psi$ to $\eta\eta$~\cite{besiii}.

\begin{figure}[!tb]
\begin{center}
\includegraphics[width=2.35in]{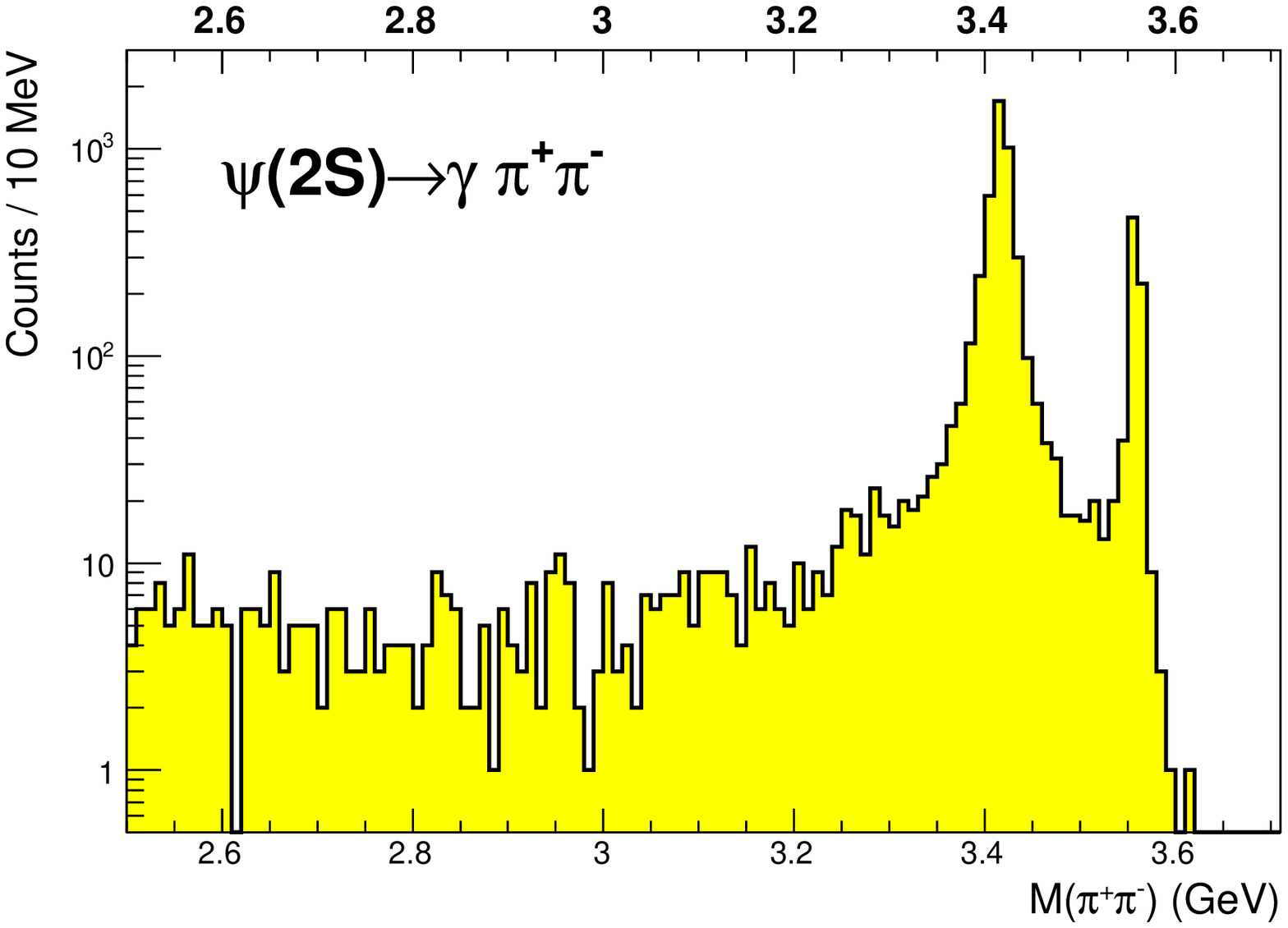}

\includegraphics[width=2.35in]{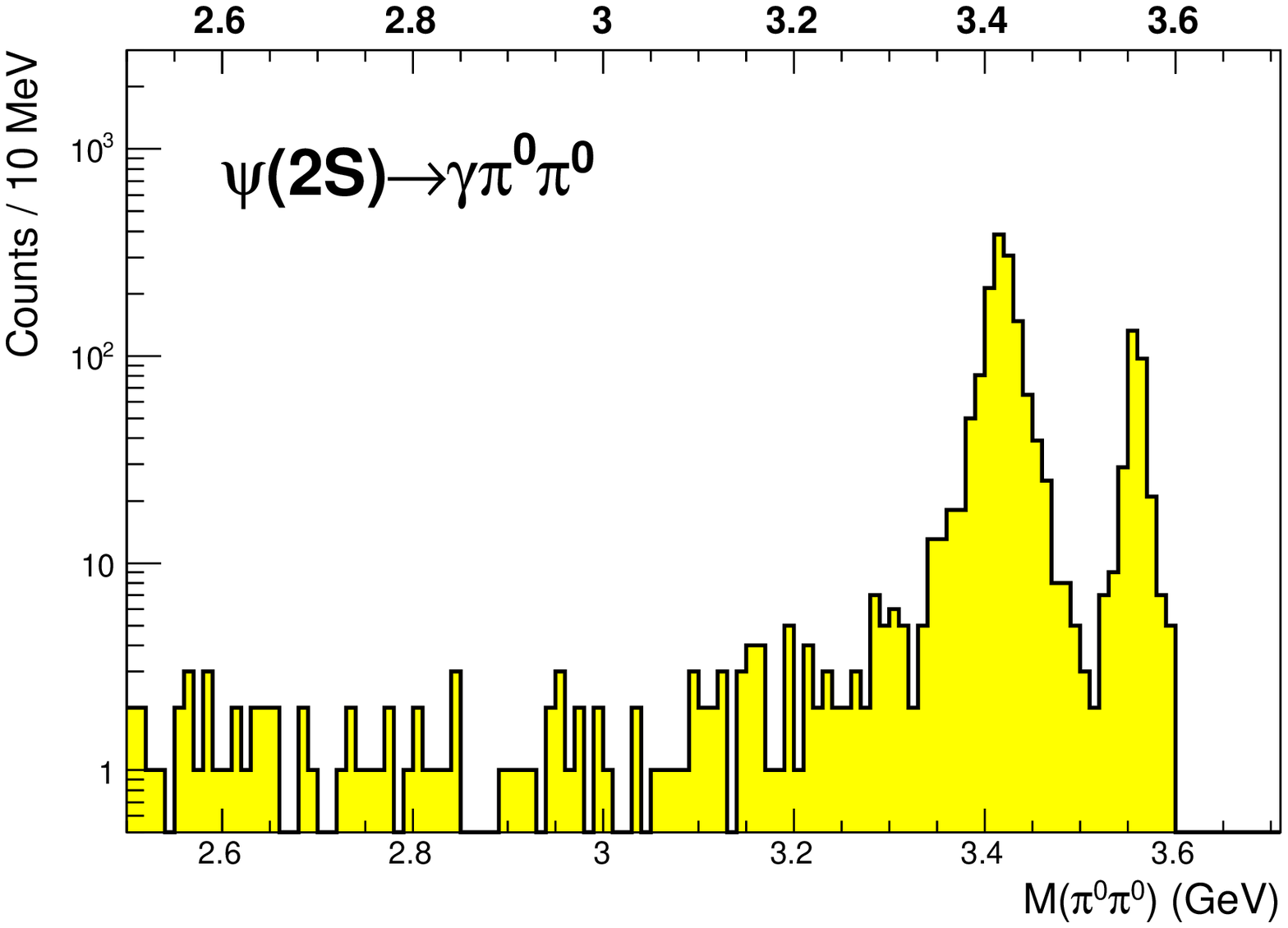}
\
\includegraphics[width=2.35in]{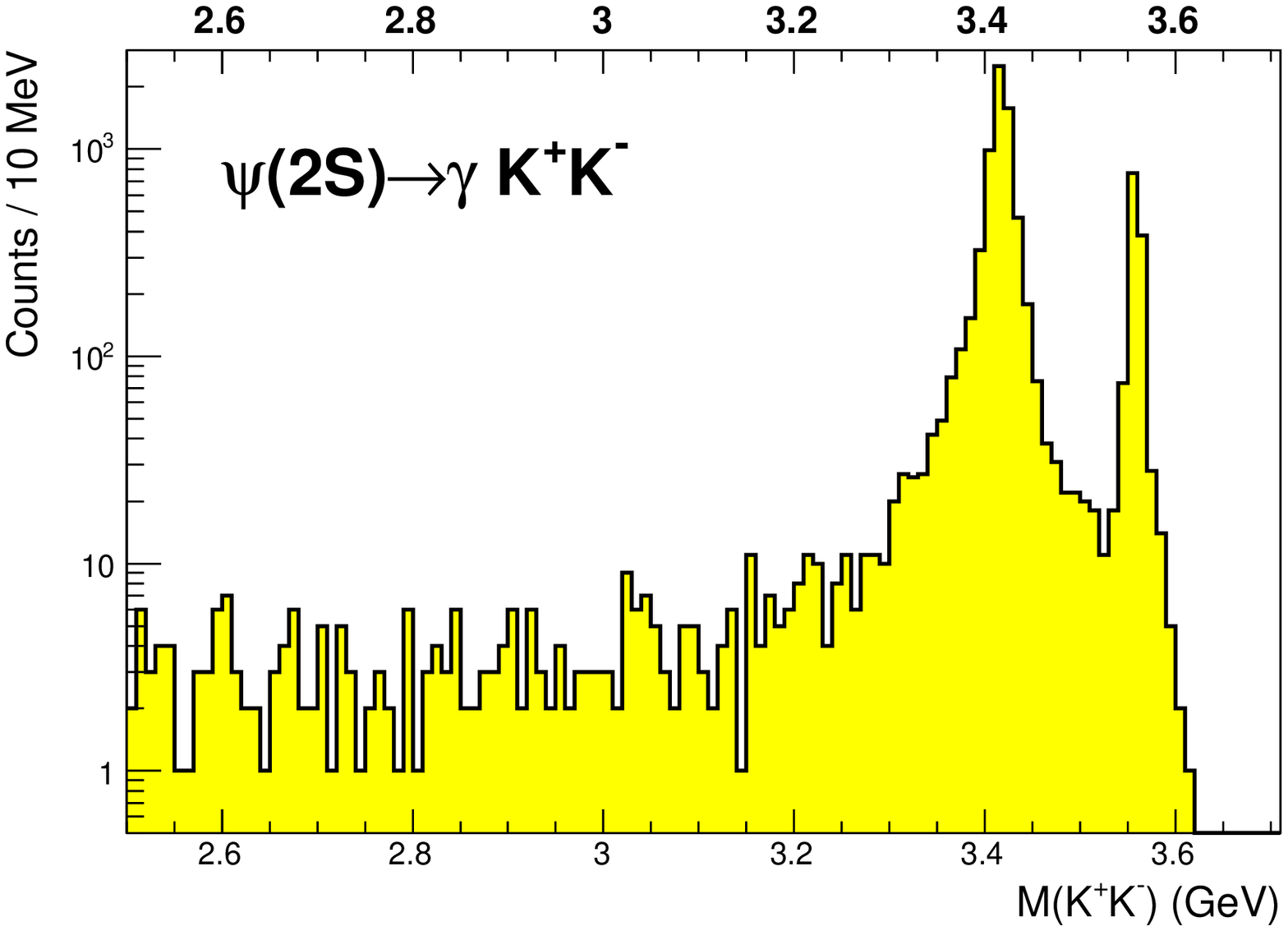}

\includegraphics[width=2.35in]{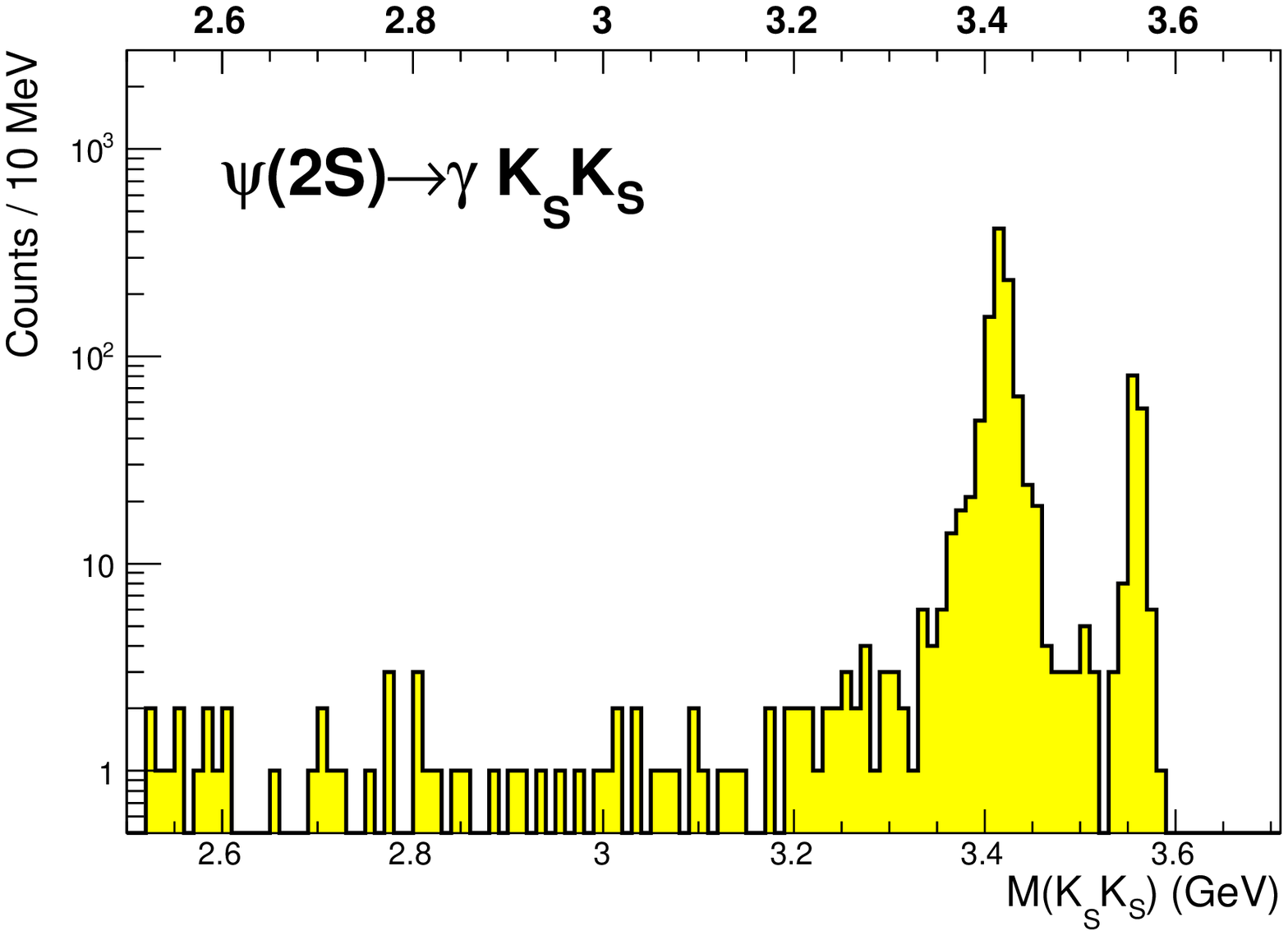}

\includegraphics[width=2.35in]{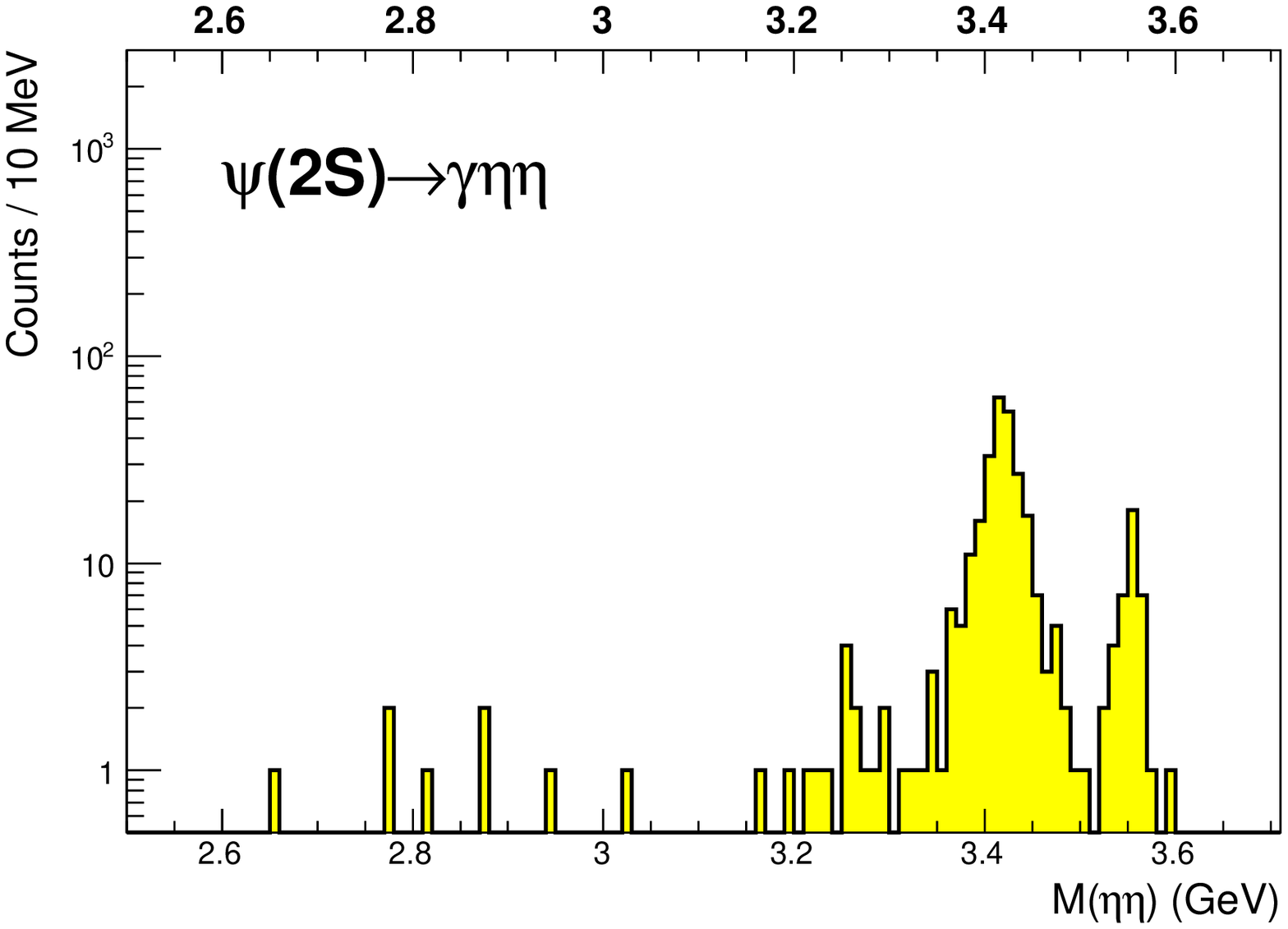}
\end{center}
\caption{Two particle invariant mass distributions for selected $\psi(2S)\to\gamma \mathrm{PS,PS}$, $(\mathrm{PS}\equiv \pi^\pm,\pi^0,K^\pm,K_S^0,\eta)$ decays, for the region $M>2.5$~GeV.  The $\chi_{c0}$ and $\chi_{c2}$ states are clearly seen.   No evidence for any other states is seen.}
\label{fig:himass}
\end{figure}

\begin{figure*}[!tb]
\begin{center}
\includegraphics[width=3.in]{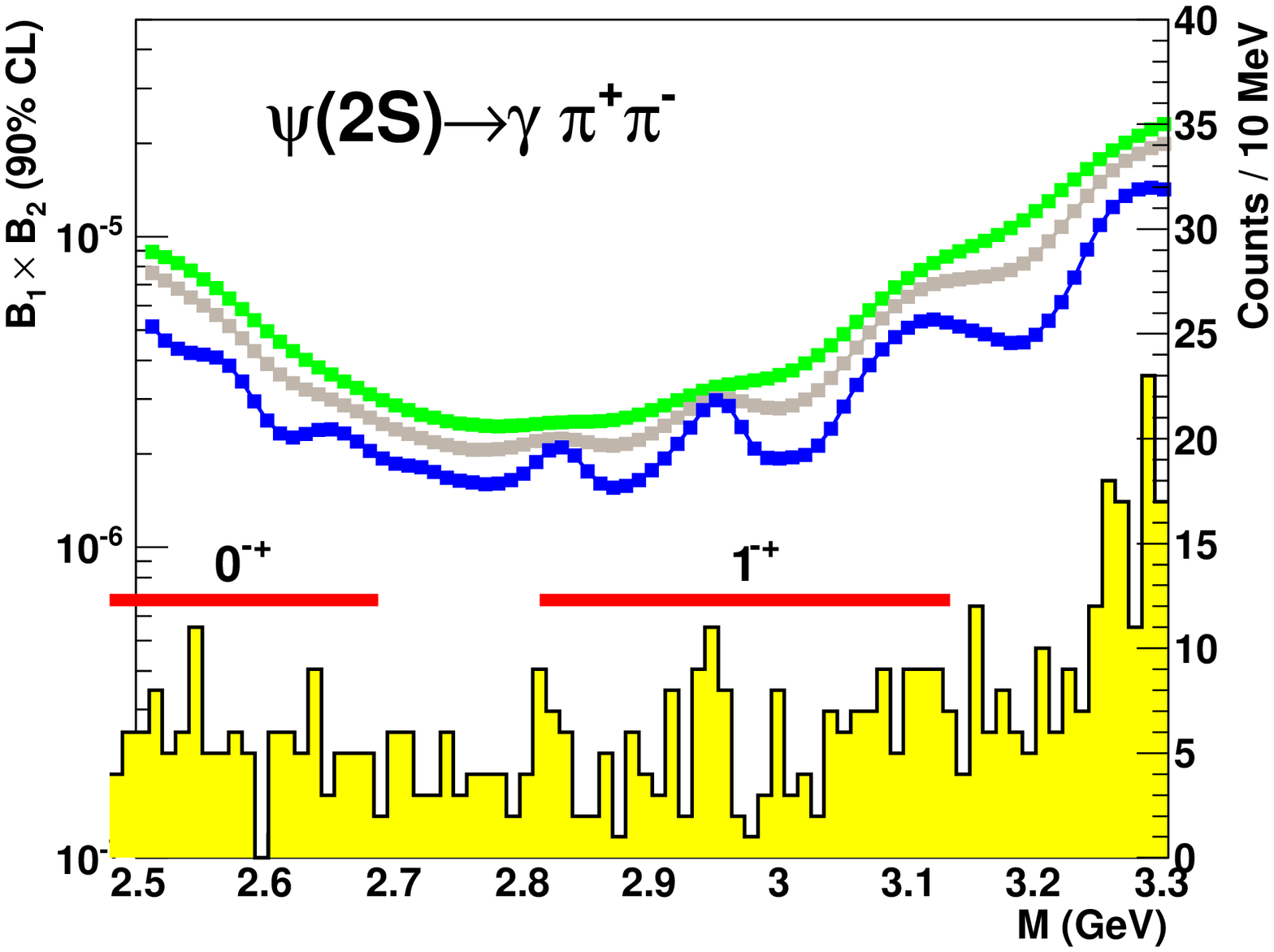}
\includegraphics[width=3.in]{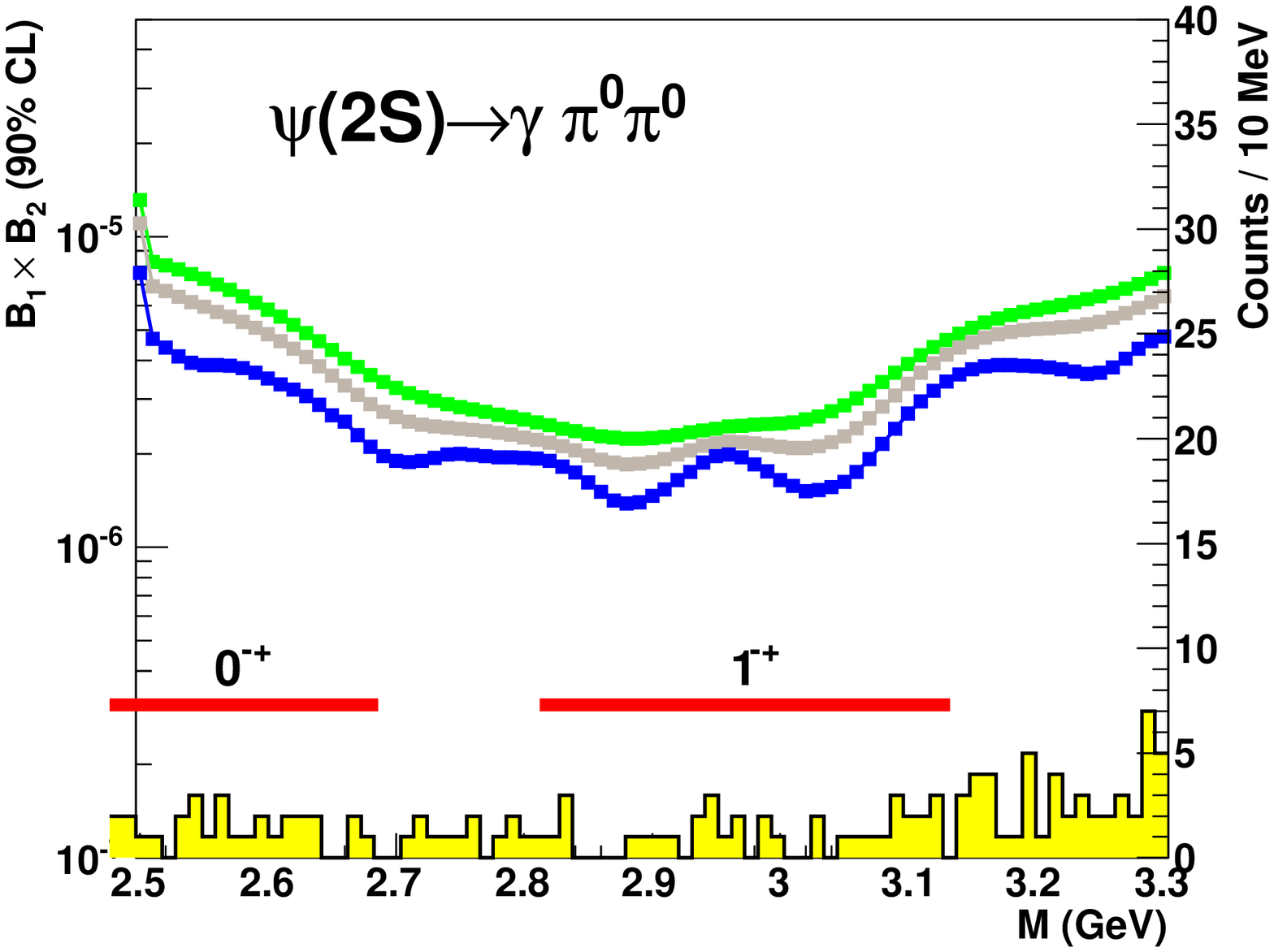}

\includegraphics[width=3.in]{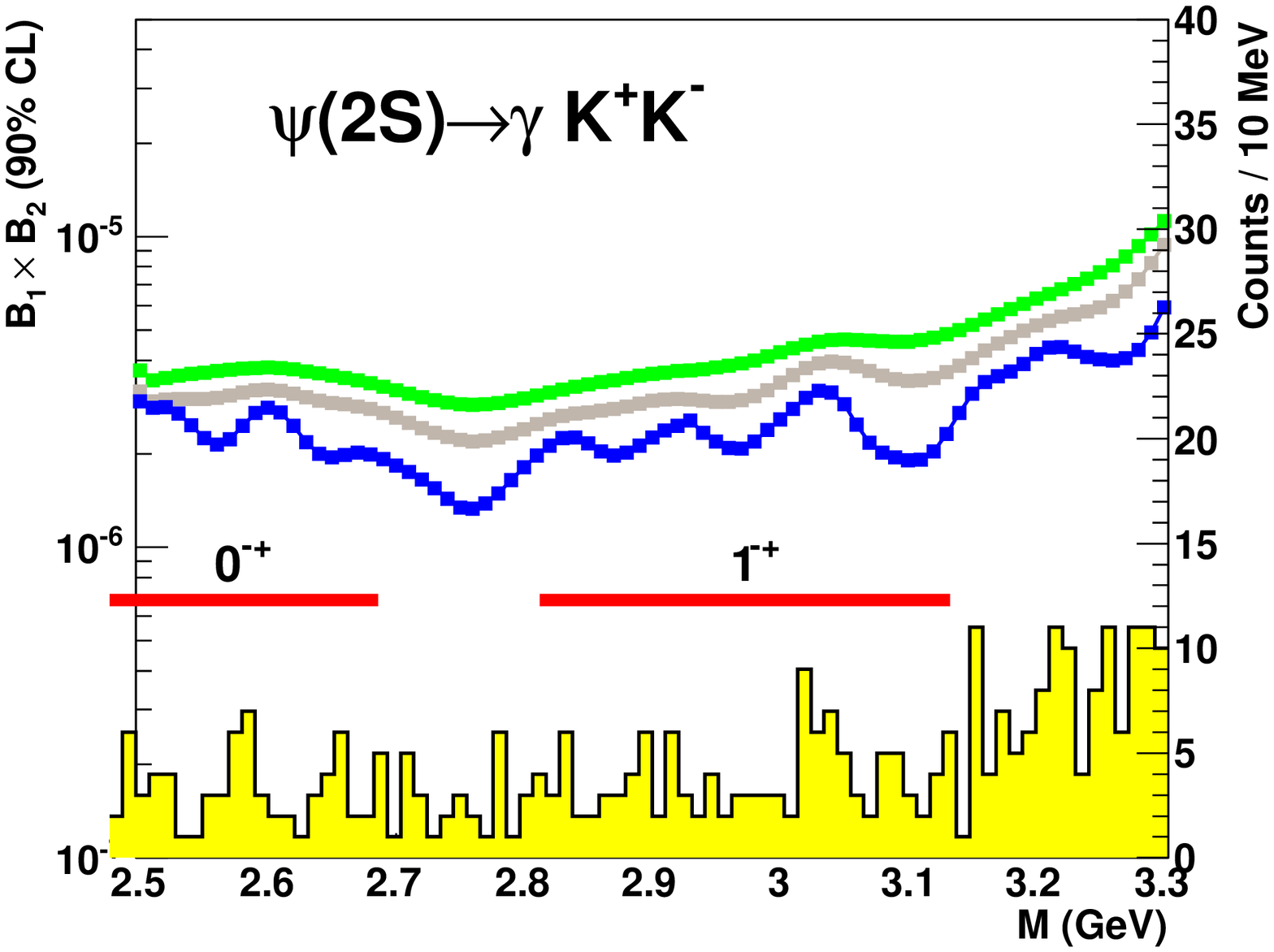}
\includegraphics[width=3.in]{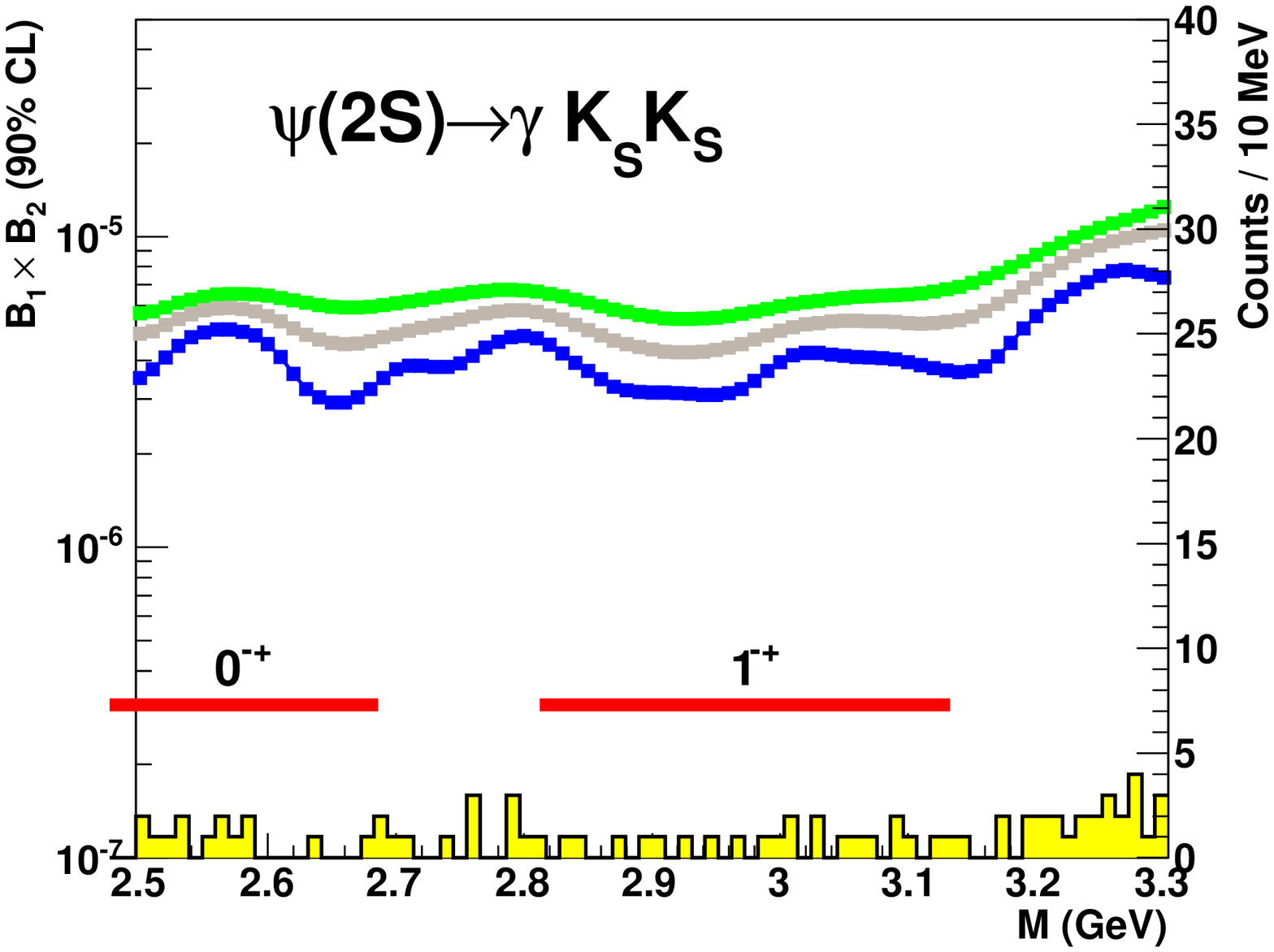}

\includegraphics[width=3.in]{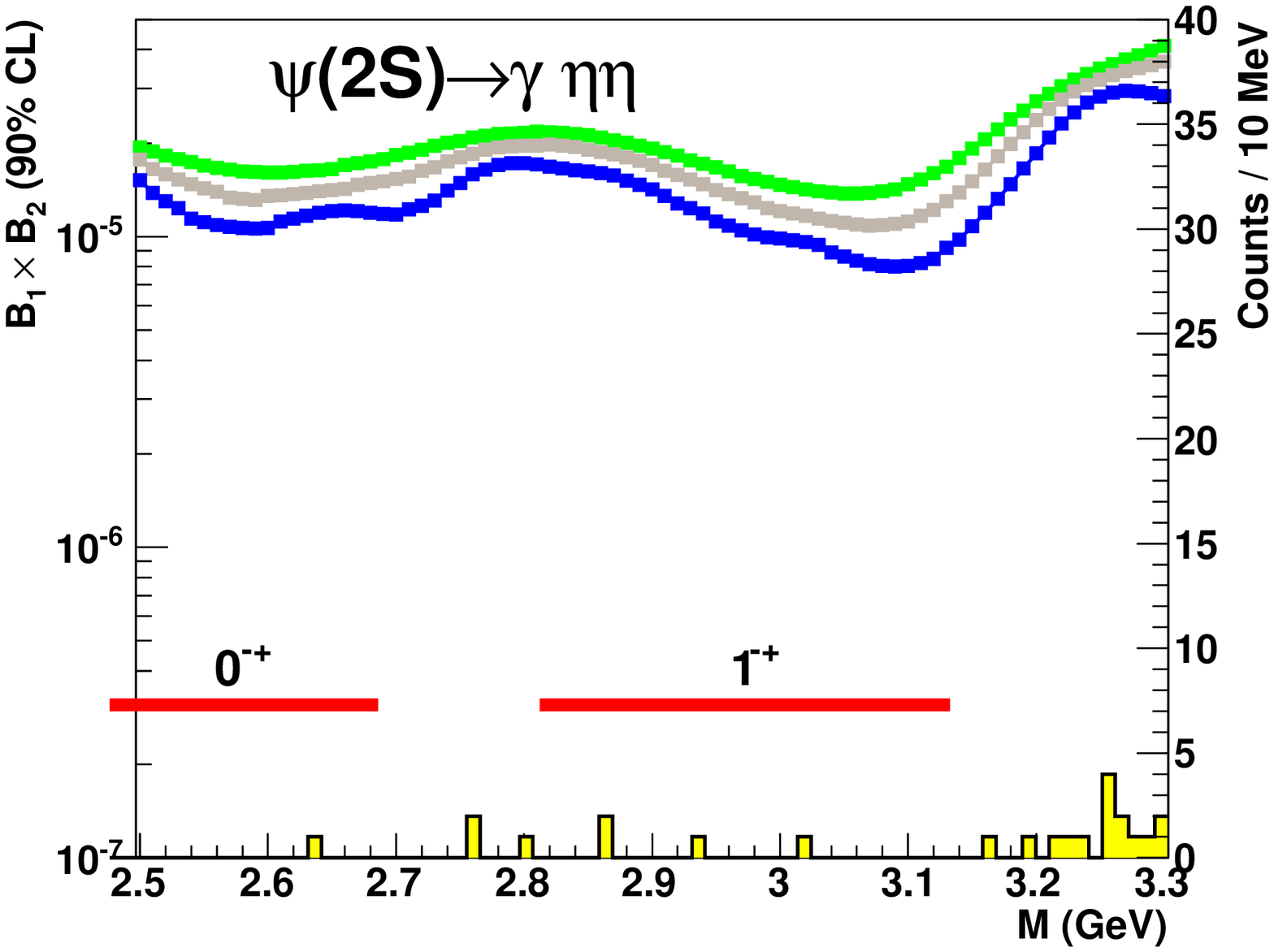}
\end{center}
\caption{Upper limits at 90\% confidence level of $\mathcal{B}_1(\psi(2S)\to R)\times\mathcal{B}_2(R\to\text{PS,PS})$ as function of resonance mass, $M$.  The differently colored curves, from top to bottom, correspond to upper limits for assumed widths $\Gamma(\mathrm{R})=150,~100~\mathrm{and}~50$~MeV.  The two-pseudoscalar invariant mass spectra are shown as the shaded histograms, with the scale of the observed counts on the right ordinate.  The thick horizontal lines illustrate the mass ranges predicated in Ref.~\cite{lattice4} for the glueball candidates with $J^{PC}=0^{-+}$ and $1^{-+}$.}
\label{fig:hiuls}
\end{figure*}

\begin{table}[tb]
\begin{center}

\caption{Product branching fractions for $\chi_{c0,2}\to\text{PS,PS}$ decays (based on Ref.~\cite{cleoprev}).}

\begin{tabular}{lcc}
\hline\hline
$\mathcal{B}_1\times\mathcal{B}_2\times10^5$  & $\chi_{c0}$ & $\chi_{c2}$ \\
\hline

$\chi_{cJ}\to\pi^+\pi^-$  & 58.8(30) & 14.8(8) \\
$\chi_{cJ}\to\pi^0\pi^0$   & 27.1(30) &  6.3(7) \\
$\chi_{cJ}\to\pi\pi$       & 85.9(42) & 21.1(11)\\
$\chi_{cJ}\to K^+K^-$     & 58.5(32) & 10.6(6) \\
$\chi_{cJ}\to K_S^0K_S^0$ & 32.2(18) & 4.9(4)  \\
$\chi_{cJ}\to K\overline{K}$ & 120.9(49) & 20.7(10) \\
$\chi_{cJ}\to\eta\eta$    & 29.2(31) & 4.8(7)  \\
$\chi_{cJ}\to\eta'\eta'$  & 19.4(20) & 0.53(31)\\
$\chi_{cJ}\to\eta\eta'$   & 1.46(56) & 0.13(29)\\

\hline\hline
\end{tabular}
\end{center}

\label{tbl:chis}
\end{table}

\section{Summary and Comparison of Results for Radiative Decays of $J/\psi$ and $\psi(2S)$ to Resonances with Masses $M<2500$~MeV}

In Table~\ref{tbl:compare} we compare the results for all resonances observed in the radiative decays of $J/\psi$ and $\psi(2S)$ to pairs of charged and neutral pseudoscalars.  Several features are notable in Table~\ref{tbl:compare}.

As mentioned in Sec.~V.B, the results for $J/\psi$ and $\psi(2S)$ decays are very similar for both charged and neutral decays. The masses are generally in agreement well within their statistical errors. Branching fraction results from BES, when available, are generally not in good agreement with ours.  The BES results are based on directly produced $J/\psi$, and as mentioned before, the BES invariant mass spectra contain large backgrounds due to ISR contributions.
\vspace*{5pt}

\noindent
\textit{The So-called 13\% Rule:} The ``13\% rule'' mentioned earlier is generally invoked for hadronic decays of vector resonances of charmonium $J/\psi$, $\psi(2S)$, \ldots $\psi(nS)$ because both their leptonic decays via one photon and their hadronic decays via three gluons are proportional to the wave function at origin.  Actually, the radiative decays to hadrons, via a photon plus two gluons have the same proportionality to the wave function at origin, so that we expect, for example, the ratio of radiative decays to pseudoscalar pairs
\begin{eqnarray*}
\mathcal{B}(\psi(2S) \to \gamma\text{PS,PS}) / \mathcal{B}(J/\psi \to \gamma\text{PS,PS}) \hspace*{1.5cm} \\
\nonumber  = \mathcal{B}(\psi(2S) \to e^+e^-) / \mathcal{B}(J/\psi \to e^+e^-) \\
\nonumber = 0.789\times10^{-2} / 5.971\times10^{-2} = 13.44(30) \%
\end{eqnarray*}
In Table~\ref{tbl:compare} we list this ratio for the various resonances we observe.  The ratio is found to vary between 3.6(8)\% and 26.2(84)\%.  This is similar to the variation which has been observed for the pure hadronic decays\cite{rule13percent}.

\begin{table*}[!tb]

\caption{Summary of systematic uncertainties on branching fractions (in units of percent).}

\begin{center}
\begin{tabular}{lcccc|cccc}
\hline\hline

 & \multicolumn{4}{c|}{$J/\psi\to\gamma\pi^+\pi^-/\gamma\pi^0\pi^0$} & \multicolumn{4}{c}{$\psi(2S)\to\gamma\pi^+\pi^-/\gamma\pi^0\pi^0$} \\
 & $f_2(1270)$ & $f_0(1500)$ & $f_0(1710)$ & $f_0(2100)$
 & $f_2(1270)$ & $f_0(1500)$ & $f_0(1710)$ & $f_0(2100)$ \\
\hline
Reconstruction \& PID & 5/5 & 5/5 & 5/5 & 5/5 & 4/5 & 4/5 & 4/5 & 4/5 \\
Kinematic Fit         & 1/2 &10/6 & 1/8 & 4/12& 1/1 & 5/29& 3/3 &23/2 \\
B-W Widths            & 1/1 & 6/8 & 3/3 & 7/5 & 1/1 & 4/10&10/4 & 7/8 \\
Background            & 1/1 & 3/13& 6/5 & 7/6 & 1/2 & 2/2 & 5/9 &10/11 \\
Binning               & 1/1 & 2/5 & 2/1 & 2/2 & 1/1 & 1/29& 2/1 & 2/1 \\
$\rho\pi$ Rejection   & 5/-- & 16/-- & 8/-- & 5/-- & --- & --- & --- & --- \\ 
Number of $\psi(2S)$  & 2/2 & 2/2 & 2/2 & 2/2 & 2/2 & 2/2 & 2/1 & 2/2 \\
&  &  &  &  &  &  &  &  \\
\textbf{Total}        & 8/6 &21/18& 12/11&13/15& 5/6 & 8/43&13/12&26/15\\

\hline

 & \multicolumn{4}{c|}{$J/\psi\to\gamma K^+K^-/\gamma K_S^0K_S^0$} & \multicolumn{4}{c}{$\psi(2S)\to\gamma K^+K^-/\gamma K_S^0K_S^0$} \\
 & $f_0(1370)$ & $f_2(1525)$ & $f_0(1710)$ & $f_0(2200)$
 & $f_0(1370)$ & $f_2(1525)$ & $f_0(1710)$ & $f_0(2200)$ \\
\hline
Reconstruction \& PID & 6/5 & 6/5 & 6/5 & 6/5 & 5/5  & 5/5 & 5/5 & 5/5 \\
Kinematic Fit         &28/3 & 4/5 & 2/3 & 5/2 &10/7  & 8/4 & 2/10& 1/30 \\
B-W Widths            &20/11& 5/4 & 3/3 & 8/16&35/24 & 7/6 & 4/4 & 13/3\\
Background            &11/6 & 1/1 & 1/2 & 2/11&26/100& 6/5 & 7/10& 22/100\\
Binning               & 3/4 & 1/3 & 1/1 & 1/2 &24/15 & 6/1 & 3/2 & 2/40\\
$K^*K$ Rejection      &28/-- & 5/-- & 6/-- & 18/-- & ---  & --- & --- & --- \\ 
Number of $\psi(2S)$  & 2/2 & 2/2 & 2/2 & 2/2 & 2/2  & 2/2 & 2/2 & 2/2 \\
&  &  &  &  &  &  &  &  \\
\textbf{Total}        &46/18& 10/9 & 9/7 &21/20&51/104&15/10&10/16& 26/112\\

\hline \hline
\end{tabular}
\end{center}

\label{tbl:systbr}
\end{table*}

\section{RESONANCES WITH MASSES $> 2500$ MeV}

\subsection{Search for Glueballs with Masses in the $2.5-3.3$~GeV Region}

One of the main reasons for our studying $\psi(2S)$ radiative decays in addition to those of $J/\psi$ was to extend our glueball searches to masses in the region $2.5-3.3$~GeV.  As stated earlier, and illustrated in Fig.~\ref{lqcd-prediction}, lattice calculations predict a $J^{PC}=0^{-+}$ glueball with mass $M(0^{-+})=2560(35)(120)$~MeV, and a $J^{PC}=2^{-+}$ glueball with mass $M(2^{-+}) = 3040(40)(150)$~MeV.  These $0^{-+}$ and $2^{-+}$ positive charge conjugation states can be excited in radiative decays of $\psi(2S)$, and we have searched for them and any other enhancements in the entire region, $M=2.5-3.3$~GeV, in their decays into pseudoscalar pairs, assuming three different values of their widths, $\Gamma(X)=50$~MeV, 100~MeV and 150~MeV. The results are displayed in Fig.~\ref{fig:hiuls}.  No notable enhancement is found anywhere in the mass region, and upper limits for their branching fractions at 90\% confidence level have been determined.  The upper limits for all product branching fractions $\mathcal{B}_1\times\mathcal{B}_2$ are determined to be all $<1\times10^{-5}$, except for decay to $\eta\eta$ for which they become as large as $3\times10^{-5}$ because of the very small number of observed counts and small $\mathcal{B}(\eta\to\gamma\gamma)\approx0.4$.

\subsection{The $\chi_{c0}$ and $\chi_{c2}$ resonances of charmonium}

As shown in the Dalitz plot of Fig.~\ref{fig:dalitz2}, and invariant mass projections in Fig.~\ref{fig:himass} for $\psi(2S)$ radiative decay, we observe strongly populated $\chi_{c0}$ and $\chi_{c2}$ states of charmonium in their decay info $\pi\pi$, $K\overline{K}$, and $\eta\eta$.  The branching fractions for these decays were reported by us earlier in the CLEO publication of Ref.~\cite{cleoprev}.  The $\mathcal{B}_2(\chi_{c0,c2}\to\text{PS,PS})$ branching fractions reported there, multiplied by $\mathcal{B}_1(\psi(2S)\to\gamma\chi_{c0})=9.22\%$ and $\mathcal{B}_2(\psi(2S)\to\gamma\chi_{c2})=9.33\%$ used there, are quoted as $\mathcal{B}_1\times\mathcal{B}_2\times10^5$ in Table~VII.

Ochs~\cite{review2} has emphasized that while it is difficult to study the flavor-independent decay of two-gluon glueball candidates to pseudoscalar pairs, the corresponding decays of $\chi_{c0}(J=0)$ and $\chi_{c2}(J=2)$, which are mediated by two gluons, provide  an excellent opportunity to test the flavor independence ansatz. 
The values for $\pi\pi$, $K\overline{K}$, and $\eta\eta$ decays listed in Table~VII  
lead to the ratios below:
\begin{widetext}
\[ \begin{array}{rccccc}
\text{Pseudoscalar~pairs} & \quad ~  \pi\pi & K\overline{K} & \eta \eta & \eta'\eta' & \eta\eta' \\[4pt]
\text{Two-gluon~decays~(theoretical)} & = 1 & 1.33 & 0.33 & 0.33 & 0 \\[4pt]

\chi_{c0}~\text{decays}  & = 1.00(5) & 1.41(6) & 0.34(4) & 0.23(3) & 0.02(1)  \\
\chi_{c2}~\text{decays}  & = 1.00(5) & 0.98(6) & 0.23(4) & 0.025(15) & 0.006(15) \\[4pt]

\chi_{c0}~\text{decays~(divided~by~phase~space~factor}~p)  & = 1.00(5) & 1.47(9) & 0.36(4) & 0.27(3) & 0.02(1)  \\
\chi_{c2}~\text{decays~(divided~by~phase~space~factor}~p^5) & = 1.00(5) & 1.18(8) & 0.28(4) & 0.058(34) & 0.010(22) 

\end{array} \]

\end{widetext}
The branching fraction ratios for $\chi_{c0}$ decays are seen to agree within experimental uncertainties with the theoretical expectations for 2--gluon decays.  The $\chi_{c2}$ ratios also agree with these expectations, except for decays to $\eta'\eta'$.

\begin{table*}[!tb]

\caption{Summary of systematic uncertainties on resonance masses (in units of MeV).}

\begin{center}
\begin{tabular}{lcccc|cccc}
\hline\hline

 & \multicolumn{4}{c|}{$J/\psi\to\gamma\pi^+\pi^-$} & \multicolumn{4}{c}{$\psi(2S)\to\gamma\pi^+\pi^-$} \\
 & $f_2(1270)$ & $f_0(1500)$ & $f_0(1710)$ & $f_0(2100)$
 & $f_2(1270)$ & $f_0(1500)$ & $f_0(1710)$ & $f_0(2100)$ \\
\hline
Mass Calibration & 2 & 2 & 2 & 2 & 2 & 2 & 2 & 2 \\
Kinematic Fit  & 1 & 7 & 2 & 5 & 1 & 1 & 3 & 5 \\
B-W Widths     & 1 & 2 & 1 & 1 & 1 & 2 & 2 & 3 \\
Background     & 1 & 2 & 1 & 1 & 1 & 1 & 1 & 4 \\
$\rho\pi$ Rejection   & 3 & 9 & 4 & 1 & --- & --- & --- & --- \\ 
Binning        & 1 & 5 & 1 & 3 & 1 & 2 & 2 & 3 \\
&  &  &  &  &  &  &  &  \\
\textbf{Total} & 4 & 13 & 5 & 6 & 3 & 4 & 5 & 8 \\

\hline

 & \multicolumn{4}{c|}{$J/\psi\to\gamma K^+K^-$} & \multicolumn{4}{c}{$\psi(2S)\to\gamma K^+K^-$} \\
 & $f_0(1370)$ & $f_2(1525)$ & $f_0(1710)$ & $f_0(2200)$
 & $f_0(1370)$ & $f_2(1525)$ & $f_0(1710)$ & $f_0(2200)$ \\
\hline
Mass Calibration & 2 & 2 & 2 & 2 & 2 & 2 & 2 & 2 \\
Kinematic Fit  & 2 & 1 & 1 & 3 & 10& 2 & 2 & 3 \\
B-W Widths     & 25& 2 & 1 & 3 & 8 & 1 & 1 & 3 \\
Background     & 5 & 1 & 1 & 1 & 5 & 1 & 1 & 5 \\
$K^*K$ Rejection   & 9 & 4 & 4 & 6 & ---  & --- & --- & --- \\ 
Binning        & 7 & 2 & 1 & 1 & 5 & 1 & 1 & 3 \\
&  &  &  &  &  &  &  &  \\
\textbf{Total} & 28 & 6 & 5 & 8 & 24& 3 & 3 & 6\\

\hline \hline
\end{tabular}
\end{center}

\label{tbl:systmass}
\end{table*}

\section{Systematic Uncertainties}

Our estimates of systematic uncertainties in product branching fractions $J/\psi\to\gamma\pi\pi$, $\gamma K\overline{K}$, and $\psi(2S)\to\gamma\pi\pi$, $\gamma K\overline{K}$, are listed in Table~\ref{tbl:systbr}.  
We estimate uncertainties due to reconstruction and particle identification as follows. We find a reconstruction uncertainty of 1\% per charged particle, with an additional 1\% and 2\% per identified $\pi^\pm$ and $K^\pm$, respectively, for the use of $dE/dx$ and RICH information. We find a $K_S^0\to\pi^+\pi^-$ identification uncertainty of 1\%.  We also find uncertainties of 2\% per $\pi^0$ and 2\% for the radiative photon.  We add the contributions for each mode in quadrature to obtain the systematic uncertainties for this source. 
We estimate the uncertainty in our requirement that the kinematic fit have $\chi^2<25$ by analyzing data with the requirements $\chi^2<20$ and $\chi^2<30$, and we take the largest variations as the systematic uncertainty due to this source.
In resonance fits, our fixed parameters are the Breit-Wigner widths of the resonances and the MC-determined mass resolution.  To estimate the uncertainties due to the fixed resonance widths, we refit the data, varying the widths by $\pm1\sigma$ of their PDG values~\cite{pdg},  and we take the largest variation for each resonance as the systematic uncertainty due to this source.  We also refit the data with the MC-determined mass resolution varied by $\pm10\%$, and find that this has no measurable effect our results, since the resonances in our fits are all much wider than the mass resolutions.
The uncertainty due to the backgrounds in the fits is obtained by varying background normalization by $\pm1\sigma$ around the central value obtained from the fits.  The uncertainty due to the 25~MeV binning of the invariant mass spectra is obtained by analyzing the data with 15~MeV and 5~MeV bins and taking the largest variation as the systematic uncertainty due to this source.
The uncertainty in the number of $\psi(2S)$ is 2\%.  

We estimate the uncertainty in our mass calibration by fitting the $\chi_{c0}$ and $\chi_{c2}$ resonances observed in the $\psi(2S)$ data.  We find that the $\chi_{c0}$ and $\chi_{c2}$ masses in all decays agree with their known values~\cite{pdg} within $\pm2$ MeV, and we take $\pm2$~MeV as the systematic uncertainty in our mass calibration.  Other contributions to uncertainties in mass determinations from the sources described above are also listed in Table~\ref{tbl:systmass}.

The sums in quadrature of all systematic uncertainties are also listed in Tables \ref{tbl:systbr} and \ref{tbl:systmass}.  For the upper limits on product branching fractions reported in Table~\ref{tbl:xiuls}, we take the largest upper limits obtained from these variations as our final upper limit results.

\section{Discussion of Results and Conclusions}

\subsection{Branching Fractions for Light Quark Resonances}

Our final results for light quark scalar and tensor resonances in the mass region $1100-2200$~MeV are summarized in Table~\ref{tbl:compare}.  As mentioned earlier, for $J/\psi$ radiative decays, results from BES measurements often differ significantly from our results.  The differences in the results can be attributed mainly to the intrinsic differences between the methods of analysis.  Our analyses are based on Breit-Wigner resonance formalism.  BES results are based on partial-wave analyses, which leads to large uncertainties depending on the number of resonances used in the analysis.  The large ISR backgrounds in the BES invariant mass spectra for $J/\psi$ radiative decay also contributes to the differences, particularly for weakly excited states.  Our spectra for $J/\psi$ radiative decay are free from ISR background contributions.

A resonance with mass $\sim1800$~MeV has been suggested in hadronic decays of $J/\psi$~\cite{review}.  We do not find evidence for such a state in either $J/\psi$ or $\psi(2S)$ radiative decays. 

Our results for $\psi(2S)$ and $J/\psi$ decays are generally in agreement.  The ratios of $\psi(2S)/J/\psi$ branching fractions vary between $\sim5\%$ to 15\%.  Similar variations from the 13\% pQCD rule have been observed before for other decays~\cite{rule13percent}.

\subsection{Results Relating to the Glueballs}

\subsubsection{The $2^{++}$ Glueball $\xi(2230)$}

We find no evidence for the claimed narrow tensor state $\xi(2230)$, and establish limits for its existence for assumed widths of 20 and 50~MeV in the different decay modes in terms of the product branching fraction at 90\% confidence level of $\mathcal{B}(J/\psi \to \gamma\xi \to \gamma(\text{PS,PS})) < 5 \times 10^{-5}$ and $\mathcal{B}(\psi(2S) \to \gamma\xi \to \gamma(\text{PS,PS})) < 0.6\times10^{-5}$.

\subsubsection{The $0^{-+}$ and $1^{-+}$ Glueballs}

As shown in Fig.~\ref{fig:hiuls}, we do not find any resonance enhancement in $\pi\pi$, $K\overline{K}$, or $\eta\eta$ decay in radiative decays $\psi(2S)$ in the mass region $2.5-3.3$~GeV in which $0^{-+}$ and $1^{-+}$ glueballs are predicted by lattice calculations~\cite{lattice4}.  At 90\% confidence level, upper limits for the product branching fractions between $10^{-5}$ and $10^{-6}$ are obtained for assumed widths $50-150$~MeV.

\subsubsection{The Scalar $f_0(1370)$}

We find convincing evidence for the existence of $f_0(1370)$, although we are able to identify it only in its decay to $K\overline{K}$.  Our inability to observe it in its $\pi\pi$ decay is due to the strong excitation of $f_2(1270)$ in its vicinity.  The existence of $f_0(1370)$ has been considered to be the crucial evidence for the supernumery nature of scalars in this mass region, and therefore of the existence of a scalar glueball, albeit mixed with other scalars.

\subsubsection{Flavor Blind Decays}

As mentioned in the introduction, the relative decay rates to pseudoscalar pairs $\pi\pi:K\overline{K}:\eta\eta:\eta'\eta'$ of $3:4:1:1$ is considered to be the defining signature of a \textit{pure}, unmixed glueball.  Our measurements of the $\pi\pi$ and $K\overline{K}$ decays, allows us to examine this conjecture only for $f_0(1710)$ which is the only resonance for which we have successfully measured branching fractions for both decays.  We obtain:
\begin{equation}
R \equiv \frac{ \mathcal{B}(f_0(1710)\to K\overline{K}) }{ \mathcal{B}(f_0(1710)\to \pi\pi) } = \frac{117.6(98)}{37.2(40)} = 3.2(4)
\end{equation}
The corresponding BES result for $f_0(1710)$ is $R(\text{BES})=2.8(9)$~\cite{besii2,besii3}.  As quoted in Ref.~\cite{review},  the Mark~III reanalysis result is $R(\text{Mark~III})=3.7(^{16}_{23})$.

Our result for the ratio $R=3.2(4)$ differs from 1.33 expected for a pure glueball by $4.3\sigma$, and can be considered as clear evidence that $f_0(1710)$ is not a \textit{pure} glueball, although it may well contain appreciable mixture of a scalar glueball in its wave function, as many mixing models suggest~(see also \cite{final}).

To summarize, our measurements have confirmed the supernumary nature of light quark scalars, and we find that $f_0(1710)$ is not a ``pure'' scalar glueball.  Further, we find no evidence for the tensor glueball candidate $\xi(2230)$.

We wish to thank Professors E. Klempt and W. Ochs for valuable comments.
This investigation was done using CLEO data, and as members of the former CLEO Collaboration we thank it for this privilege.  This research was supported by the U.S. Department of Energy.


\begin{thebibliography}{99}
\bibitem{final}  It has been pointed out to us that our measured ratio, $R=3.2\pm0.4$ agrees with the prediction of $R=3.0$ (without phase space correction) and $R=3.7$ (with phase space correction) for $f_0(1710)$ as a member of the $SU(3)$ octet.
\end{thebibliography}
\end{document}